\documentclass[aps,prd,10pt,preprint,tightenlines,preprintnumbers,showpacs,superscriptaddress,notitlepage,nofootinbib,eqsecnum,floatfix,longbibliography]{revtex4-2}
\pdfoutput=1

\usepackage{amsmath,amssymb,color,graphicx,amscd,amsfonts}
\usepackage{latexsym}
\usepackage{graphicx}
\usepackage{bbm}
\usepackage[normalem]{ulem}
\usepackage[usenames,dvipsnames,svgnames,table]{xcolor}
\usepackage{booktabs,multirow,tabularx}
\usepackage[utf8]{inputenc}
\usepackage[bookmarks,linktocpage,colorlinks]{hyperref}
\usepackage{enumitem}

\DeclareGraphicsExtensions{.pdf}
\graphicspath{{./}}

\DeclareMathOperator {\Tr}{Tr}

\graphicspath{{./}}
\newcommand \TZeroGammaStarNfEight {1.0830(11)}
\newcommand \TZeroGammaStarNfTwelve {0.4395(20)}
\newcommand \ChiralLimitF {0.02120(127)({}^{252}_{15})}
\newcommand \ChiralLimitPsibarPsi {0.000200(20)}
\newcommand \ChiralLimitMassAOneOverMassRho {1.59(14)({}^{0}_{19})}
\newcommand \ChiralLimitCZero {0.0526(184)({}^{26}_{787})}
\newcommand \ChiralLimitMSigmaOverFOverRootTwo {3.51(1.25)({}^{17}_{5.27})}

\newcommand \ChiralLimitFSquaredOverTwo {0.0002247(270)({}^{377}_{22})}
\newcommand \ChiralLimitSqrtDOneNfTwelve {0.844(136)({}^{21}_{24})}
\newcommand \ChiralLimitFourOverSqrtDOneNfEight {3.97(40)({}^{96}_{63})}

\newcommand \ChiralLimitDOneNfFour {3.5(1.1)}
\newcommand \GammaMFromDOneNfFour {2.11(21)}
\newcommand \NewEnsembleMSigma {0.112(17)({}^{0}_{33})}
\newcommand \NewEnsembleLMsigma {5.38(79)({}^{0}_{1.56})}
\newcommand \NewEnsembleChi {4.388(528)({}^{43}_{737})}

\begin{document}
%
\preprint{RIKEN-iTHEMS-Report-22, UTHEP-804, UTCCS-P-167, KEK-TH-2721}

\title{A novel view of the flavor-singlet spectrum from multi-flavor QCD on the lattice}

\author{Yasumichi Aoki}
\affiliation{RIKEN Center for Computational Science, Kobe 650-0047, Japan}

\author{Tatsumi Aoyama}
\affiliation{Institute for Solid State Physics, University of Tokyo, Kashiwa, 277-8581, Japan}

\author{Ed Bennett}
\affiliation{Swansea Academy of Advanced Computing, Swansea  University, Bay Campus, Swansea, SA1 8EN, UK}

\author{\\ Toshihide Maskawa\footnote{Deceased, July 23, 2021.}}
\affiliation{Kobayashi-Maskawa Institute for the Origin of Particles and the Universe, Nagoya University, Nagoya 464-8602, Japan}

\author{Kohtaroh Miura}
\affiliation{Institute of Particle and Nuclear Studies,
High Energy Accelerator Research Organization (KEK), Tsukuba 305-0801, Japan}

\author{Hiroshi Ohki}
\affiliation{Department of Physics, Nara Women’s University, Nara 630-8506, Japan}

\author{Enrico Rinaldi}
\affiliation{Interdisciplinary Theoretical \& Mathematical Science (iTHEMS) Program, RIKEN, Wako, Saitama 351-0198, Japan}

\author{Akihiro Shibata}
\affiliation{Computing Research Center, High Energy Accelerator Research Organization (KEK), Tsukuba 305-0801, Japan}

\author{Koichi Yamawaki}
\affiliation{Kobayashi-Maskawa Institute for the Origin of Particles and the Universe, Nagoya University, Nagoya 464-8602, Japan}

\author{Takeshi Yamazaki}
\affiliation{Institute of Pure and Applied Sciences, University of Tsukuba, Tsukuba, Ibaraki 305-8571, Japan}
\affiliation{Center for Computational Sciences, University of Tsukuba, Tsukuba, Ibaraki 305-8577, Japan}

\collaboration{LatKMI Collaboration}
\noaffiliation
\begin{abstract}
SU(3) gauge theories with increasing number of light fermions are the templates of strongly interacting sectors and studying their low-energy dynamics and spectrum is important, both for understanding the strong dynamics of QCD itself, but also for discovering viable UV completions of beyond the Standard Model physics.
In order to contrast many-flavors strongly interacting theories with QCD on a quantitative footing, we use Lattice Field Theory simulations.
We focus on the study of the flavor-singlet spectrum in the scalar and pseudoscalar channels: this is an interesting probe of the dynamics of the strongly interacting sector, as reminded by the QCD case with the $f_0(500)$ ($\sigma$) and $\eta^\prime$ mesons.
The hierarchy of the spectrum of a strongly coupled new gauge sector of the Standard Model defines the potential reach of future colliders for new physics discoveries.
In addition to a novel hierarchy with light scalars, introducing many light flavors at fixed number of colors can influence the dynamics of the lightest flavor-singlet pseudoscalar.
We present a complete lattice study of both these flavor-singlet channels on high-statistics gauge ensembles generated by the LatKMI collaboration with 4, 8, and 12 copies of light mass-degenerate fermions.
We also present other hadron masses on the lightest ensemble for $N_f=8$ generated by the LatKMI collaboration and discuss the chiral extrapolation of the spectrum in this particular theory.
We contrast the results to $N_f=4$ simulations and previous results of $N_f=12$ simulations.
\end{abstract}
\maketitle

\section{Introduction}\label{intro}

Since the discovery in 2012 of the Higgs boson~\cite{ATLAS:2012yve,CMS:2012qbp}, the Standard Model (SM) has been so successful that there has been little hint of new physics beyond the SM\@. However, the origin of mass remains a central mystery of the SM in the form of the \emph{naturalness problem} (see Ref.~\cite{Craig:2022uua} for a recent review), which is suggestive that there should be some underlying theory to account for it. Walking technicolor~\cite{Yamawaki:1985zg,Bando:1986bg}
is a candidate for such a theory;
it is
characterized by a chiral condensate with a large anomalous dimension
$\gamma_m \simeq 1$, and by an approximate scale symmetry which is broken spontaneously and explicitly (the scale anomaly) by the same chiral condensate.\footnote{
Walking technicolor has also been
advocated without the notion of scale symmetry breaking
giving a dilaton or a large anomalous dimension~\cite{Holdom:1984sk,Akiba:1985rr,Appelquist:1986an}.}
The latter of these
predicts
the Higgs as a composite light pseudo-dilaton (``technidilaton''),
a pseudo Nambu-Goldstone (NG) boson of the broken scale symmetry.
Such a theory would be strongly coupled,
with a slowly running (``walking'')
gauge coupling.

Such a walking gauge theory may be realized by an asymptotically free $SU(N_c)$ gauge theory
with large number of massless flavors $N_f (\gg N_c)$~\cite{Appelquist:1996dq},
which we may refer to as ``large-$N_f$ QCD''.
It is inspired by the near scale-invariant/conformal structure around the Caswell-Banks-Zaks (CBZ) infrared (IR) fixed point $\alpha_*$
in the two-loop beta function
at large $N_f$ ($8 < N_f \leq 16$ for $N_c=3$)~\cite{Caswell:1974gg,Banks:1981nn}.

The walking regime is anticipated
to be at a value of the IR strong coupling $\alpha_*$ larger than a critical value, $\alpha_* \gtrsim \alpha_{\rm cr}$,
such that $1\ll  n_f\equiv N_f/N_c < n_f^{\rm cr},$
where $n_f=n_f^{\rm cr}$ at $\alpha_* =\alpha_{\rm cr}$.
We expect a significant separation between the UV scale $\Lambda_{\mathrm{UV}}$
and the scale of the chiral condensate with dynamical mass $m_{D} \ll \Lambda_{\rm UV}$, 
where $\Lambda_{\mathrm{UV}}$ is identified with $\Lambda_{\mathrm{QCD}}$ generated by the regularization in perturbation theory.
This is in contrast to QCD where $\Lambda_{\mathrm{QCD}} \approx m_{D}$. 
In the infrared region $\mu< \Lambda_{\rm IR}=O(m_D)$ the coupling $\alpha$  blows up with the fermion loop decoupled, i.e. the perturbative IR fixed point $\alpha_*$ is  washed out. For $\Lambda_{\rm IR} < \mu < \Lambda_{\rm UV}$, it starts running slowly in units of $\Lambda_{\rm IR}$ as $\alpha(\mu) \searrow \alpha_{\rm cr}$ and 
$\gamma_m(\alpha(\mu)) \ne 0$, with $\alpha_{\rm cr}$ now being the UV fixed point; this gives rise to a non-perturbative trace anomaly of order ${\cal O}(m_D^4)$ 
in contrast to the perturbative one of order ${\cal O}(\Lambda_{\rm UV}^4)$.

On the other hand, the region $\alpha_* < \alpha_{\rm cr}$
($n_f>n_f^{\rm cr}$) is the conformal window, with $m_D\equiv 0$.

While experiments provide direct access to this spectrum in the case of QCD with light flavors $N_f=2+1$, the case $N_f \gg 2$ has not been observed in nature, and so can only be studied theoretically.
Analytic approaches involve making
an approximation, such as the ladder approximation
(see e.g.~\cite{Matsuzaki:2015sya} and references therein),
which introduces somewhat large theoretical uncertainties in our understanding of the spectrum.
Lattice simulations fill this gap, providing a non-perturbative first-principles
approach relatively free of large or uncontrolled approximations, and
providing us with data
which can be used to
benchmark phenomenological models.

Several lattice groups, including the LatKMI collaboration, have searched for a candidate walking theory within the space of large $N_f$ QCD ($N_c=3$) theories on the lattice. The cases $N_f=8$ and $12$ have drawn particular attention.
The spectrum of the theory toward the chiral limit gives one avenue to probing
the viability of such a theory as a candidate walking theory.
Among other observables, we have particularly focused on the possible large anomalous dimension visible in the spectra of bound states, with a special focus on a flavor-singlet scalar $\sigma$, 
which would be
a candidate for the light pseudo-dilaton~\cite{Aoki:2012eq,Aoki:2013xza,Aoki:2013zsa,Aoki:2014oha,Aoki:2016wnc}.
In order to shed light on signals of near-conformal dynamics, we have checked the hyperscaling and otherwise compared the behavior of the $\sigma$ mass from lattice simulations of $N_f=4$, 8 and 12 QCD with simulations that have used the same lattice setup to minimize systematic differences between results from the three theories.

Our prior lattice numerical simulations indicated that these three different values of $N_f$ correspond to rather different types of dynamics, implying that $n_f^{cr} \lesssim 4$.

\begin{itemize}
  \item $N_f=4$
        shows~\cite{Aoki:2013xza,Aoki:2016wnc} essentially the same features as of $N_f=2+1$ QCD, specifically spontaneous chiral symmetry breaking, and neither hyperscaling nor a light flavor-singlet scalar state $\sigma$ 
        are observed.

  \item The $N_f=12$ theory appears to show~\cite{Aoki:2012eq, Aoki:2016wnc} conformal dynamics, and hence lie in the conformal window, implying $m_D\equiv 0$. The two-body hadronic mass $M_{\rm H}$ shows typical universal hyperscaling of the (renormalized) fermion mass $m_f^{(R)}$, $M_{\rm H}/2 \sim m_D + m_f^{(R)} =  m_f^{(R)} \sim (m_f)^{1/(1+\gamma_m(\alpha_*))}$~\cite{Miransky:1998dh,DelDebbio:2010ze,DelDebbio:2010jy}, with $\gamma_m (\alpha_*) \simeq 0.4  - 0.5$\footnote{
        This result is roughly consistent with the ladder Schwinger--Dyson (SD) equation in the conformal phase, with $\gamma_m = 1 - \sqrt{1- \alpha/\alpha_{\rm cr}} \sim 0.8$ at $\alpha=\alpha_*$.
        Including non-leading terms for
        the
        large mass and finite size gives $\gamma_m \sim 0.5-0.6$ when fitted by a simple hyperscaling form (with only the leading term) as in the lattice analyses in~\cite{Aoki:2012ve}.\label{gammaSD}}.
        The ratios of various bound state mass scales remain
        constant toward the chiral limit.
        We subsequently found a light $\sigma$~\cite{Aoki:2013zsa},
        even lighter than the pion $\pi$, an observation to which we shall return in this work.

        In the conformal window,
        the presence of an explicit fermion mass $m_f$
        means that all bound states are in fact non-relativistic ``unparticles'',
        due to the conformal phase transition~\cite{Miransky:1996pd}, 
        since there are no bound states in the chiral limit.
        Nevertheless,
        $\sigma$ is still a pseudo-NG boson,
        while $\pi$ is not:
        chiral symmetry is only broken explicitly by $m_f$,
        while scale symmetry is further broken spontaneously by the gluon condensate
        (triggered in turn by the same $m_f$).
        We will return to this discussion in Sec.~\ref{sec:antiveneziano}.

  \item Most excitingly, we found~\cite{Aoki:2012zwc,Aoki:2013xza,Aoki:2016wnc} that $N_f=8$ is a candidate for a walking theory.
        The pion mass decreases $M_\pi \rightarrow 0$ towards the chiral limit  $m_f\rightarrow 0$,
        while other (flavor non-singlet) bound states mass $M_{\rm H}$ do not; i.e. $M_{\rm H}/M_\pi \rightarrow \infty$.
        This signals that the $\pi$ is a pseudo NG boson of a spontaneously broken symmetry.
        At the same time the theory shows signals of being in the near conformal phase: specifically, most states in the mass spectrum  (including $F_\pi$ and the string tension),
        dominated by the explicit chiral symmetry breaking $m_f^{(R)} \gtrsim m_D$, obey the approximate hyperscaling relation
        with a large anomalous dimension
        \begin{eqnarray}
          \gamma_m \simeq 1\;.
          \label{gamma8}
        \end{eqnarray}
        The exception to this is the pion, as we expect:
        since the pion is a pseudo-NG boson,
        the hyperscaling form is expected to fit best near the chiral limit,
        while for other states the scaling breaks in this region.
        The would-be anomalous dimension that we see for the pion is ``$\gamma_m$'' $\simeq 0.6$ in the mass region we study~\cite{Aoki:2016wnc};
        we refer to this as ``non-universal'' hyperscaling.
        This has been found to be consistent with other groups' results~\cite{Cheng:2013eu, Appelquist:2014zsa, LSD:2023uzj}.\footnote{
See also Refs.~\cite{Hasenfratz:2022qan, Witzel:2024bly} for recent and different approaches,  
that suggest that the $N_f=8$ QCD theory exhibits a fixed-point existing at much stronger coupling region than previous studies, 
consistent with being near onset of the conformal window. 
}
\end{itemize}

Moreover,
for the $N_f=8$ theory we reported~\cite{Aoki:2014oha, Aoki:2016wnc} a signal that
the flavor-singlet scalar state $\sigma$ may be a pseudo-dilaton.
Specifically,
it was observed to be as light as the pseudo NG boson pion $\pi$ all the way down to
the vicinity of
the chiral limit,
\begin{eqnarray}
  M_\sigma \simeq M_\pi \ll M_\rho\,,
  \label{sigmamass0}
\end{eqnarray}
and displayed
``$\gamma_m$''
$\approx 0.5$
(again, as far as $\gamma_m$ can be considered meaningful for a pseudo-NG boson).
These observations
contrast sharply with QCD with smaller number of flavors.
By implementing the proposal in Ref.~\cite{Matsuzaki:2013eva}, we also observed
that the ratio of decay constants of the $\sigma$ and $\pi$~\footnote{
Decay constants of the pion and $\sigma$ are defined as: $\langle 0 | D_\mu(0) |\sigma(q_\mu)\rangle =
  i F_\sigma q_\mu$,  or equivalently $\langle 0|\theta_{\mu\nu}|\sigma(q_\mu)\rangle= - F_\sigma (q_\mu q_\nu-g_{\mu\nu} q^2)/3$,
   $\langle 0 |
  A^\alpha_\mu(0) |\pi^\beta (q_\mu)\rangle$
  $ =
   i \delta^{\alpha\beta} (F_\pi/\sqrt{2}) \cdot q_\mu$, where
  our $F_\pi$ corresponds to 130 MeV for the usual QCD pion.
  }:
\begin{equation}
  \frac{F_\sigma}{F_\pi/\sqrt{2}} \approx 4 \gg 1\;,
\end{equation}
which, with Eq.~(\ref{sigmamass0}), suggests
that $\sigma$  in this theory may be identified as a candidate for the Higgs.

We obtained the latter result by fitting lattice data using the mass formula~\cite{Matsuzaki:2013eva}:
  \begin{equation}
M_\sigma^2\equiv d_0 + d_1 \cdot M_\pi^2\;.\label{msigma}
\end{equation}
This relation may be derived\footnote{This was first derived in Ref.~\cite{Matsuzaki:2013eva} in the context of dilaton chiral perturbation theory, giving the same expressions for $d_{0}$ and $d_{1}$.} via the scale and axial Ward-Takahashi (WT) identities. 
Since its original derivation via these routes, there have been further developments in a variant of this mass formula based on a low-energy dilaton effective field theory description~\cite{
Golterman:2016lsd, Appelquist:2017wcg, Golterman:2018mfm, Appelquist:2019lgk, Fodor:2019vmw, Fodor:2020niv, Golterman:2020utm, Appelquist:2022mjb, Freeman:2023ket, Golterman:2024vra}.
To preempt the results we will present in Sec.~\ref{sec:antiveneziano}, we will later demonstrate that $d_1\approx 1$ in this theory, implying $\gamma_m \approx 1$, and hence Eq.~\eqref{msigma}.
This allows a direct comparison between the three theories we consider here, regardless of whether they are conformal or chirally broken, which we will employ in this work.
Here $d_0 = \mathcal{O}(m_D^2) (\ll \Lambda_{\mathrm UV}^2)$, 
which is due to the nonpertubative trace anomaly, and will be non-zero even in the chiral limit.

We have also observed~\cite{Aoki:2014oha} that this coincides with linear sigma model-based and holographic calculations,
a result that was further confirmed by subsequent lattice computations of the scalar decay constant $F_S$~\cite{LatKmi:2015non,Aoki:2016wnc}.
 The light $\sigma$ has been confirmed by
 lattice studies by other groups~\cite{Appelquist:2016viq,Appelquist:2018yqe}.
 (For more recent lattice studies see e.g., Refs.~\cite{Witzel:2019jbe, Svetitsky:2017xqk,Pica:2016ejc,LatticeStrongDynamics:2023bqp}.)
The value we obtain for $F_\sigma$ is roughly consistent with more recent analyses of data
 from the Lattice Strong Dynamics (LSD) collaboration, $F_\sigma/(F_\pi/\sqrt{2})\simeq 3.4$
 ~\cite{Appelquist:2017vyy,
 Appelquist:2019lgk}
 (see also
Refs.~\cite{Fodor:2019vmw, Golterman:2020tdq, LatticeStrongDynamics:2023bqp}).

We have also reported preliminary results for the flavor-singlet pseudoscalar $\eta^\prime$~\cite{Aoki:2016fxd,Aoki:2017fnr},
which has a mass even in the chiral limit due to the chiral
anomaly  in the anomalous chiral WT identity:
\begin{equation}
M_{\eta^\prime}^2= \left(M_{\eta^\prime}^2\right)^{\textnormal{(anomalous)}}
  +M_\pi^2.
  \label{etaprimemass}
  \end{equation}
While for $M_\sigma$, $d_0$ and $d_1$ depend on $N_f$ and $N_c$ only implicitly via $\gamma_m$,
we found~\cite{Aoki:2017fnr} that when normalised by an appropriate IR scale $t_0$,
the $\eta^\prime$ mass depends directly on these as~\cite{Matsuzaki:2015sya}
\begin{eqnarray}
8 t_0 \cdot M^2_{\eta^\prime} \simeq
8t_0 \cdot \left(
M_{\eta^\prime}^2\right)^{\textnormal{(anomalous)}}
 \sim n_f^2,
 \label{antiVenezianolimit}
\end{eqnarray}
where $n_f=N_f/N_c$.
These flavor-singlet bound states have additional mass due to
anomalies---the trace anomaly for $\sigma$,
and the chiral anomaly for $\eta^\prime$---which
gives extra symmetry breaking beyond the fermionic mass term.
We will revisit this in more detail in Sec.~\ref{sec:antiveneziano}.

Extending our previous study of the flavor-singlet scalar $\sigma$~\cite{Aoki:2013zsa,Aoki:2014oha,Aoki:2016wnc} and also the preliminary studies of the flavor-singlet pseudoscalar 
$\eta^\prime$~\cite{Aoki:2016fxd,Aoki:2017fnr}, in this work
we perform lattice field theory simulations of QCD with different $N_f$
to study the spectrum of bound states. We highlight the differences, and similarities, between QCD theories with different numbers of light (or massless) flavors, in a quantitative way, with new data.
In particular, we present an extensive study of
$\eta^\prime$, as well as an updated study of the spectra of $\sigma$
and other bound states, all of which are consistent with our previous results mentioned above.

This comparison is informed by the mass formulae above,
Eqs.~\eqref{msigma} and~\eqref{etaprimemass},
discussed in more detail in Sec.~\ref{sec:antiveneziano},
which are valid independently of the chiral phase
(spontaneously broken or conformal),
and hence may be applied equally to all values of $N_f$ we consider.

We also measure the gradient flow energy scale $1/\sqrt{8 t_0}$
for $N_f=4,8,12$, with the same systematics, including a study of the hyperscaling of $t_0$ for $N_f=8,12$.

The remainder of this paper is organized as follows:

In Sec.~\ref{sec:antiveneziano}
we motivate the use of the gradient flow scale $1/\sqrt{8t_{0}}$ as an infrared scale
 and derive mass formulae from anomalous WT identities
for the scale symmetry (for $\sigma$) and for the $U(1)_{A}$ symmetry (for $\eta^{\prime}$),
which will be used to contextualize the lattice data in subsequent sections.

In Sec.~\ref{sec:lattice-setup} we describe our setup of the lattice measurements. We then present an update of the flavor-non-singlet spectra:
\begin{enumerate}
  \item
        We first confirm our previous results on the characteristic $m_f$ dependence of $M_\rho/M_\pi$ toward the chiral limit
        for  $N_f=4, 8, 12$, as deeply broken, walking, and conformal phases, respectively.

  \item
        We present measurements of the gradient flow energy scale $1/\sqrt{8 t_0}$ for $N_f=4,8,12$, which provide the basis for comparing the lattice results for different $N_f$.
        We show that it obeys hyperscaling for $N_f=8$ and $12$,
        while not for $N_f=4$,
        reinforcing previous indications that the former theories are in a (near-)conformal phase,
        while the latter has a deeply spontaneously broken chiral symmetry.

  \item
        Then  the updated results for the  non-singlet  $N_f=8$ spectra  are given, which are all consistent with our previously-published data~\cite{Aoki:2016wnc}.
\end{enumerate}

Then in Sec.~\ref{sec:scalar} we present the data for $\sigma$ mass, giving an update to our previous data for $N_f=8$, which are analyzed according to the formula Eq.~(\ref{msigma}) and found to give $d_0, d_1$ consistent with the previous publications~\cite{Aoki:2014oha, Aoki:2016wnc}.
We also present new data for $N_f=4$, and further compare $d_0, d_1$ among $N_f=4, 8$ and $N_f=12$.
We also present the $\sigma$ mass and the non-singlet masses both in units of $F_\pi/\sqrt{2}$ and in units of
$1/\sqrt{8 t_0}$, and highlight characteristic features of $N_f =4, 8, 12$.

Section~\ref{sec:pseudoscalar} is the core part of the present paper, presenting completed results for the $\eta^\prime$ in comparison with other spectra.  We show  the results
for the mass both in terms of UV and IR scales; the latter indicate
good agreement with the prediction of the $n_f$ dependence
in the anti-Veneziano limit, Eq.~(\ref{antiVenezianolimit}), thus confirming the previous
results~\cite{Aoki:2016fxd,Aoki:2017fnr}.

Section~\ref{sec:summary} is devoted to the summary.

We also include more detailed data and analyses in subsequent Appendices.
In particular, Appendix~\ref{sec:app-taste} compares our results with those of the LSD Collaboration in units of $1/\sqrt{8t_0}$.

\clearpage
\newpage
\section{Meson masses in the anti-Veneziano limit}\label{sec:antiveneziano}

In the following sections we will present new lattice data
for $N_{f}=4,8,12$ QCD.
In particular,
we will present results for the masses of
the flavor-singlet scalar $\sigma$
and the pseudoscalar $\eta^{\prime}$,
each of which has characteristic contributions from the anomaly,
the scale,
and the $U(1)$ axial anomaly.
In order to understand and contextualise the numerical results we will present,
we must have a theoretical functional form which
 to compare them.
In this section we will discuss the theoretical issues
surrounding the characteristic $ n_f  (\equiv N_{f}/N_{c})$  dependence
of the anomalous parts of the two mesons we will be focusing on,
in the limit
\begin{equation}
 N_{c}\rightarrow\infty \,\,
{\rm with}\,\,  N_{c}\alpha=  {\rm fixed},
 n_f  =N_{f}/N_{c}=
 {\rm fixed} \gg 1\,.
 \label{antiVeneziano}
 \end{equation}
We refer to this as the \emph{anti-Veneziano limit}~\cite{Matsuzaki:2015sya},
by analogy with the related limit
proposed by Witten and Veneziano~\cite{Witten:1979vv,Veneziano:1979ec}
in which $ n_f$ is fixed at a small value,
referred to as the Veneziano limit.
The anti-Veneziano limit is of interest in our setup
as in the theories we are considering,
$ n_f \in[\frac{4}{3},4]$;
this range is well away from the Veneziano limit,
and we may expect the upper end to be better-represented
by the anti-Veneziano limit
than the lower end.

\subsection{An IR scale from the gradient flow}
\label{sec:gradientflow}
In the discussions that follow,
we will normalise spectral quantities using an IR scale $\Lambda_{\rm IR}=\Lambda_{\rm IR}|_{N_c,  N_f}$;
that is,
a scale where the coupling becomes strong,
allowing the formation of bound states from the nonperturbative dynamics. This scale is intrinsic to
the theory specified by the combination of $N_{c}$ and $N_{f}$.
To allow comparisons between theories with different $N_{f}$,
bound state masses
such as those of the $\eta^{\prime}$ and $\sigma$
should be measured in terms of such a scale.

In the broken phase,
$\Lambda_{\mathrm{IR}}$ is essentially the confinement scale
and/or the spontaneous chiral breaking scale  $m_D$
(the ``dynamical mass'').
More properly this can be called the
``consitutent quark mass''
$m_{f}^{\textnormal{(constituent)}}$;
it is related to the mass of the ground state bound state
\begin{equation}
  \Lambda_{\mathrm{IR}}\simeq 2 m_{f}^{\textnormal{(constituent)}}\simeq 2  m_D+ 2 m_{f}^{(R)}\simeq M_{\rho}\;,
  \label{eq:lambda-ir-mf}
\end{equation}
where $m_{f}^{(R)}$ is the ``current quark mass'',
defined as the renormalized, on-shell mass of the fermion: $m_{f}^{(R)}=m_{f}^{(R)}(\mu)|_{\mu=m_{f}^{\textnormal{(constituent)}}}$.\footnote{
  Note that another ground state mass,
  that of the $\pi$,
  is the NG boson mass in the broken phase,
  and so has a different dependence on  $m_D$ and $m_{f}^{(R)}$.
}

However,
defining $\Lambda_{\rm IR}$ via Eq.~\eqref{eq:lambda-ir-mf} is not suitable for theories in the conformal phase, where
bound states exist only
in the presence of a non-zero explicit fermion mass $m_{f}^{(R)}\ne 0$.
Ordinary fermionic bound states
(except for the flavor-singlet scalar $\sigma$ discussed in a later subsection)
are formed via the weakly-coupled Coulomb force
instead of the confining force.
The masses of two-body hadrons then become
\begin{equation}
  M_{\mathrm{H}}\sim 2m_{f}^{(R)}
  =2m_{f}\cdot Z_m^{-1}
  =2m_{f}\left(\frac{\Lambda_{\mathrm{UV}}}{m_{f}^{(R)}}\right)^{\gamma_{m}}
  =2\Lambda_{\mathrm{UV}}\left(\frac{m_{f}}{\Lambda_{\mathrm{UV}}}\right)^{\frac{1}{1+\gamma_{m}}}\;,
  \label{eq:coulomb-bs}
\end{equation}
where $\gamma_m=\gamma_m(\alpha(\mu))|_{\mu=m_{f}^{(R)}}$ and
$\Lambda_{\mathrm{UV}}$ a corresponding UV scale (the intrinsic scale denoted in QCD as $\Lambda_{\rm QCD}$), 
which characterizes the asymptotically free running of the coupling.
Specifically, for scales $\mu\gg\Lambda_{\mathrm{UV}}$,
then $\gamma_m(\alpha(\mu))\sim \alpha(\mu)\sim1/\ln(\mu/\Lambda_{\mathrm{UV}})\ll1$.
On the lattice, 
we may identify this UV scale 
with the inverse lattice spacing $\Lambda_{\mathrm{UV}}=1/a$,
then this is nothing but the hyperscaling relation~\cite{Miransky:1998dh,DelDebbio:2010ze,DelDebbio:2010jy}
\begin{equation}
  aM_{\mathrm{H }}\sim 2am_{f}^{(R)}=2(am_{f})^{\frac{1}{1+\gamma_{m}}}\;.
\end{equation}

This explicit fermion mass also introduces a new IR scale, $\Lambda_{\mathrm{YM}}$,
below which the fermions decouple, leaving a confined pure Yang--Mills theory.
(Although the IR fixed point $\alpha_*$ is washed out, there still remains a remnant $\alpha(\mu) \simeq \alpha_*$ for
$m_{f}^{(R)}< \mu <\Lambda_{\rm UV}$.)
This is given by \cite{Miransky:1998dh}
\begin{equation}
  \Lambda_{\mathrm{YM}}=m_{f}^{(R)} \exp\left(-\frac{1}{b_{0}\alpha_{*}}\right)=\Lambda_{\mathrm{UV}}\left(\frac{m_{f}}{\Lambda_{\mathrm{UV}}}\right)^{\frac{1}{1+\gamma_{m}}}\exp\left(-\frac{1}{b_{0}\alpha_{*}}\right)\;,\label{eq:conformalir}
\end{equation}
where $ b_{0} = \frac{11N_{c}}{6\pi}$
is the coefficient of the one-loop beta function of pure Yang-Mills,
and
$\alpha_{*}$ the IR fixed point in two-loop perturbation theory.

For the specific case of $N_{c}=3,N_{f}=12$,
then $b_{0}=\frac{11}{2\pi}$, $\alpha_{*}\simeq0.754$,
giving $\exp\left(-\frac{1}{b_{0}\alpha_{*}}\right)\simeq0.47$.
On the lattice,
Eq.~\eqref{eq:conformalir} then becomes
\begin{equation}
  a\Lambda_{\mathrm{YM}}
  =am_{f}^{(R)}\exp\left(-\frac{1}{b_{0}\alpha_{*}}\right)
  \simeq0.47 am_{f}^{(R)} < am_{f}^{(R)}\;. \label{eq:conformalir2}
  \end{equation}
Comparing Eq.~\eqref{eq:coulomb-bs} with Eq.~\eqref{eq:conformalir2}, 
we see that in the conformal phase, 
\begin{equation}
  \Lambda_{\mathrm{YM}} < M_{\mathrm{H }}. 
\end{equation}

When comparing theories in different phases
including inside and outside the conformal window,
we may choose to use for $\Lambda_{\rm IR}$ the gradient flow scale~\cite{Luscher:2010iy}
\begin{equation}
\Lambda_{\mathrm{IR}}=\frac{1}{\sqrt{8t_{0}}}\,,
\label{eq:gradientflow}
\end{equation}
as an alternative to using $M_\rho$ as in Eq.~\eqref{eq:lambda-ir-mf}.
As a gluonic operator,
we expect this to be
relatively independent of explicit
chiral breaking effects, 
and allow a more robust comparison between theories in different phases.
On the lattice, we may compare $\Lambda_{\mathrm{IR}}$ with a corresponding UV scale, $\Lambda_{\mathrm{UV}}=1/a$.
Analogous to Eq.~\eqref{eq:lambda-ir-mf}, 
we expect the scale $1/\sqrt{8t_{0}}$ to consist of two contributions: 
the chiral limit value from the dynamics of the theory $\left.1/\sqrt{8t_{0}}\right|_{m_{f}=0}$,
and a deformation due to the fermion mass, proportional to $m_{f}^{(R)}$.
This leads us to anticipate the following behavior for the ratio of the two scales: 
\begin{align}
  \frac{\Lambda_{\mathrm{IR}}}{\Lambda_{\mathrm{UV}}} = \frac{a}{\sqrt{8t_{0}}}
  \simeq 2a  m_D+C_{t_0} am_{f}^{(R)}
  \simeq \left.\frac{a}{\sqrt{8t_{0}}}\right|_{m_{f}=0}  + C_{t_0}(am_{f})^{\frac{1}{1+\gamma_{m}}}
\;.
  \label{gradientflowasIR}
\end{align}
The first term is expected to scale with the dynamical mass $m_D$ generated by the chiral condensate, 
$\left.a/\sqrt{8t_{0}}\right|_{m_{f}=0}\sim a\cdot 2 m_D$;
this then tends to zero as the conformal window is approached from below,
while in the conformal window,
there is no chiral condensate,
and $\left.a/\sqrt{8t_{0}}\right|_{m_{f}=0}\equiv 0$.
We also introduce the coefficient $C_{t_0}$, which depends on the flow time.
Since this is a gluonic operator, $C_{t_0}$ is expected to be smaller than 
the coefficients associated with two-body hadron masses in Eq.~\eqref{eq:coulomb-bs}. 
Therefore, in the broken phase, we expect $C_{t_0} \leq 2$, 
while in the conformal phase, Eq.~\eqref{eq:conformalir2} implies $C_{t_0} < 1$.

More concretely,
as we will demonstrate in Sec.~\ref{sec:lattice-setup}, 
these expectations---or even the stronger condition $C_{t_0} \leq 1$ for $N_f=4,8$---are typically satisfied for our chosen values of the flow time.
This implies 
\begin{align}
  \textnormal{Broken (near-conformal):}&&M_{\rho}\sqrt{8t_{0}}&\simeq\frac{2 m_D+2m_{f}^{(R)}}{2 m_D+C_{t_0}m_{f}^{(R)}}\rightarrow \begin{cases}\gtrsim 2 & (m_{f}^{(R)}\gg  m_D)\\ \simeq 1 &
  (
  m_{f}^{(R)}\ll  m_D)\end{cases}, \nonumber\\
   \textnormal{Conformal:}&& M_{\rho}\sqrt{8t_{0}}  & ( \gtrsim  M_{\pi}\sqrt{8t_{0}} ) >  2\;.
   \label{MrhovsTzero}
\end{align}

\subsection{$\eta^\prime$ mass in the anti-Veneziano limit}\label{sec:aVlimit}

We will consider first the $\eta^{\prime}$ meson.
In QCD with $N_{f}$ degenerate fermions,
the flavor-singlet axial vector current
\begin{equation}
  A_{\mu}^{0}(x)=\sum_{i=1}^{N_{f}}\bar{\psi}_{i}(x)\gamma_{\mu}\gamma_{5}\psi_{i}(x)
\end{equation}
has divergence
\begin{equation}
  \partial^\mu A^0_\mu(x) = 2 m_f \sum^{N_f}_{i=1} \bar \psi_i i \gamma_5 \psi_i + 2 N_f \frac{\alpha}{8\pi} G^{\mu\nu} {\tilde G}_{\mu\nu} (x)\;,
\end{equation}
where the second term is the chiral anomaly.
We may then define the $\eta^{\prime}$ decay constant via
\begin{align}
  \langle 0| A_\mu^0(x)|\eta^\prime(q_\mu)\rangle &= i \sqrt{2 N_f}\,
   F_{\eta^\prime\, }\,
    q_\mu\, e^{- i qx}\,,\\
  \langle 0| \partial^\mu A_\mu^0(x)|\eta^\prime(q_\mu)\rangle &=\sqrt{N_f} \, F_\pi\,
 M^2_{\eta^\prime}\,  e^{- i qx}\,,
\end{align}
which then gives $F_{\eta^\prime}=F_\pi/\sqrt{2}$;
this is independent of whether the theory is in the broken or conformal phase.

 \begin{figure}
  \begin{center}
    \includegraphics[scale=0.7,clip]{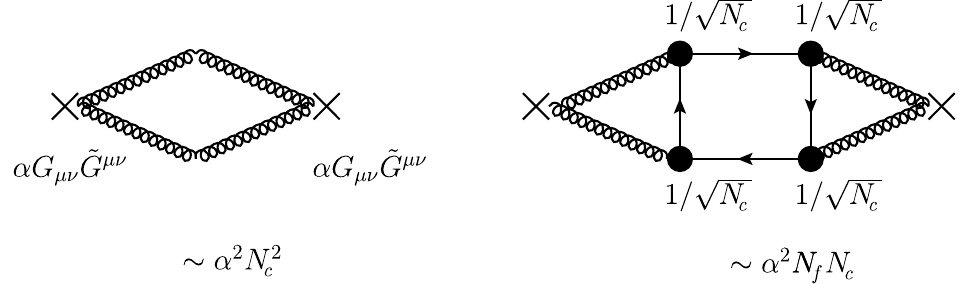}
  \end{center}
\caption{
 The loop diagrams contributing to the correlation function of $\alpha G_{\mu\nu} \tilde{G}^{\mu\nu}$ coming from
 the gluon loop (left panel) and fermion loop (right panel).
 The large $N_c$ and $N_f$ scalings have also been specified.
}
\label{GGtil-loops}
\end{figure}

The relevant anomalous chiral Ward--Takahashi (WT) then takes the form
  \begin{align}
 0& = \lim_{q^\mu\rightarrow 0} i q^\mu \int d^4x  e^{-i qx} \left< T(A^0_\mu (x) \cdot \partial^\mu A^0_\mu(0)\right>
  \nonumber \\
  &=
\left\langle [i Q_5^0, \partial^\mu A^0_\mu(x)] \right\rangle  + i  {\cal F.T.} \left\langle T\left(\partial^\mu A^0_\mu(x) \cdot \partial^\mu A^0_\mu(0)\right)\right\rangle\Big|_{q_\mu\rightarrow 0}, \\
-\left\langle \big[i Q_5^0, \partial^\mu A^0_\mu(x)\big] \right\rangle &=  i
{\cal F.T.} \left\langle T\left(\partial^\mu A^0_\mu(x) \cdot \partial^\mu A^0_\mu(0)\right)\right\rangle\Big|_{q_\mu\rightarrow 0} \nonumber \\
&=\frac{ N_f F_\pi^2
M_{\eta^\prime}^{4}
}{M^2_{\eta^\prime} -q^2}\Big|_{q_\mu\rightarrow 0}\\  \nonumber
&= N_f  F_\pi^2 M_{\eta^\prime}^2\nonumber\\
&= N_f  F_\pi^2 \left[
\left(M_{\eta^\prime}^2\right)^{\textnormal{(non-anomalous)}}
+  \left(M_{\eta^\prime}^2\right)^{\textnormal{(anomalous)}}
\right]\,,
\label{WTforetaprime}
\end{align}
where $\mathcal{F.T.}$ refers to the Fourier transform of the correlator,
which the anti-Veneziano (i.e.~large $N_c$) limit
allows us to evaluate using a single pole.
  
The non-anomalous and anomalous contributions are then given by
\begin{align}
\left(M_{\eta^\prime}^2\right)^{\textnormal{(non-anomalous)}}
  &= \frac{1}{ N_f F_\pi^2 }
 \left\langle \big[-i Q_5^0, 2 m_f \sum^{N_f}_{i=1} \bar \psi_i i \gamma_5 \psi_i \big] \right\rangle
 \,,\nonumber\\
&=\frac{1}{N_f F_\pi^2} \cdot 2   m_f \sum^{N_f}_{i=1} - 2 \langle \bar \psi_i  \psi_i \rangle = M_\pi^2 \rightarrow 0 \,\,(m_f\rightarrow0) \,,
\label{non}
\\
\left(M_{\eta^\prime}^2\right)^{\textnormal{(anomalous)}}
  &= M_{\eta^\prime}^2- M_\pi^2
= \frac{1}{ N_f F_\pi^2 } i {\cal F.T.} \left\langle T\left(
2N_f \frac{\alpha}{8\pi} G^{\mu\nu} {\tilde G}_{\mu\nu} (x)
\cdot
 2N_f \frac{\alpha}{8\pi} G^{\mu\nu} {\tilde G}_{\mu\nu}(0)
\right)\right\rangle\Big|_{q_\mu\rightarrow 0}
\nonumber\\
 &\sim\frac{N_f}{
  F_\pi^2 } \Lambda_{\rm IR}^4\left[ (\textnormal{gluon loop})  +
 \frac{N_f}{N_c}  \,\ (\textnormal{fermion loop})
 \right]\,\nonumber \\
&\sim
\frac{N_f}{N_c}
 \Lambda_{\rm IR}^2\left[ (\textnormal{gluon loop})  +
 \frac{N_f}{N_c}  \,\ (\textnormal{fermion loop})\right]
 \,.
  \label{WTI}
\end{align}
Here ``gluon loop'' and ``fermion loop'' refer to
the contributions from the diagrams
in the left and right panels of Fig.~\ref{GGtil-loops}, respectively~\cite{Matsuzaki:2015sya}.

In the Veneziano limit $ n_f \equiv N_f/N_c \ll 1$,
the theory is in the broken phase,
and the gluonic loop contribution in Eq.~\eqref{WTI} dominates over the fermionic one.
In this case,
the anomalous mass of the $\eta^{\prime}$ becomes
\begin{equation}
  \frac{\left(M_{\eta^\prime}^2\right)^{\textnormal{(anomalous)}}}{\Lambda^2_{\rm IR}} \sim  n_f \ll 1\;.
\end{equation}
This matches expectations that in this limit,
the $\eta^{\prime}$ meson is a pseudo-NG boson with parametrically vanishing mass.\footnote{
  In $N_{f}=N_{c}=3$ QCD,
  as observed in experiments,
  we see a large value of $M_{\eta^{\prime}}$,
  comparable with $M_{\rho}$,
  which does not align with this prediction of a parametrically vanishing mass.
  This is because $N_f=N_c=3$ is far from
  the idealised Veneziano limit of $N_{f}\ll N_{c}$.
}

In the anti-Veneziano limit as defined in Eq.~(\ref{antiVeneziano}), however,
the fermion loop contribution in Eq.~\eqref{WTI} dominates over the gluon one.
This then gives
\begin{equation}
  \left(M^2_{\eta^\prime}\right)^{\textnormal{(anomalous)}} \sim \frac{N_f \alpha^2}{
   F_\pi^2 } \left(N_c^3 N_f\alpha^2\Lambda_{\rm IR}^{ 4}
\right)
\sim  n_f
^2\cdot \Lambda_{\rm IR}^2
\,.
\label{eq:anomalous-eta-prime-nf-dependence}
\end{equation}
The non-anomalous contribution
proportional to the pion mass
is negligible in this limit,
giving
\begin{equation}
  \frac{M_{\rm \eta^{\prime}}^{2}}{\Lambda_{\mathrm{IR}}^{2}} \simeq \frac{{M_{\eta^{\prime}}^{2}}^{\textnormal(\rm anomalous)}}{\Lambda_{\mathrm{IR}}^{2}}\; .
\end{equation}
We therefore expect
to observe
the normalised $\eta^{\prime}$ mass to be approximately independent of $m_{f}$
in the broken phase.
Meanwhile,
in the conformal phase,
the only scale is provided by the explicit deforming fermion mass,
meaning ${M_{\eta^{\prime}}^{2}}^{\textnormal{(anomalous)}}$ obeys the same hyperscaling relations as
$M_{\pi}^{2}$ and $\Lambda_{\mathrm{IR}}^{2}  =1/(8t_0)$,
and so the normalised $\eta^{\prime}$ mass should be fully independent of $m_{f}$.

We may now define the ratio
\begin{equation}
  R^{2}_{N_{f}}\equiv
 8 t_0 \cdot \left(M_{\eta^{\prime}}^{\textnormal{(anomalous)}} \right)^{2}_{N_f} =
   \left(\frac{M_{\eta^{\prime}}^{\textnormal{(anomalous)}}}{\Lambda_{\mathrm{IR}}}\right)^{2}_{N_{f}}
  = \left(\frac{M_{\eta^{\prime}}^{2}-M_{\pi}^{2}}{\Lambda_{\mathrm{IR}}^{2}}\right)_{N_{f}}
  \sim  n_f
  ^{2}
  \label{eq:R-definition}
  \end{equation}
in the anti-Veneziano limit.
Thus for the values of $N_{f}$ we consider,
we expect
\begin{equation}
  R_{12}^{2}:R_{8}^{2}:R_{4}^{2} \simeq 9:4:1\;.
\end{equation}
These will be shown to be consistent with our lattice data in Sec.~\ref{sec:pseudoscalar}.

\subsection{$\sigma$ mass in the conformal and broken phases}
\label{sigmamass}

The flavor-singlet scalar $\sigma$ has a mass obeying
the WT identity for the scale symmetry,
such that
\begin{equation}
  \theta^{\mu}_{\mu}=\partial_{\mu}D^{\mu}=\frac{\beta^{(NP)}(\alpha)}{4\alpha}G_{\mu\nu}^{2}+(1+\gamma_{m}) m_f
   \sum^{N_f}_{i=1}
   \bar{\psi}^{i}\psi^{i}
   \;,
   \label{dilatation}
\end{equation}
where $\frac{\beta^{(NP)}(\alpha)}{4\alpha}G_{\mu\nu}^{2}$ is
the nonperturbative component of the trace anomaly,
in which
$G_{\mu\nu}$ is the gluon field strength,
and $\beta^{(NP)}(\alpha)$ is the non-perturbative beta function
for the non-perturbative running (or walking) of the coupling.
The mass anomalous dimension $\gamma_{m}$ here is the non-perturbative infrared
value, $\gamma_m=\gamma_m(\alpha(\mu))|_{\mu=\Lambda_{\mathrm{IR}}}$.


In the broken phase,
both the non-perturbative beta function
and the non-perturbative infrared mass anomalous dimension
may be estimated via the ladder Schwinger--Dyson (SD) equation
(which conceptually aligns with the anti-Veneziano limit),
giving
\begin{align}
m_f &\equiv m_f^{(R)}\cdot Z_m  \sim m_f^{(R)}\cdot \frac{m_D}{\Lambda}  \,\,\,
\,, \nonumber\\
\beta^{(NP)}(\alpha)&=
\frac{\partial \alpha(\Lambda)}{\partial \ln \Lambda}\Big|_{m_D ={\rm fixed}}= -\frac{2 \pi^2 \alpha_{\rm cr}}{\ln^3 (\frac{4 \Lambda}{m_D} )}=-\frac{2 \alpha_{\rm cr}} {\pi} \left(\frac{\alpha}{\alpha_{\rm cr}} -1\right)^{3/2}\,,\nonumber\\
\gamma_m(\alpha)
&=\frac{\partial \ln Z_m^{-1}}{\partial \ln \Lambda}\Big|_{m_D ={\rm fixed}}
 =1\,,
\label{laddermD}
\end{align}
and
\begin{equation}
  m_D = 4 \Lambda  \exp\left(-\frac{\pi}{\sqrt{\alpha/\alpha_{\mathrm{cr}}-1}}\right)\,; \\
\end{equation}
then $m_D /\Lambda=m_D/\Lambda_{\rm UV}\rightarrow 0$ near the conformal edge
in an essential singularity,
and cannot be power expanded.
(Fuller details are presented in Ref.~\cite{Matsuzaki:2015sya} and references therein.)


We then expect bound state masses to have a hierarchy of order
$
\mathcal{O}(\Lambda_{\mathrm{IR}})=\mathcal{O}( 2 m_D)\ll\Lambda_{\mathrm{UV}}$
   as $\alpha\rightarrow \alpha_{\rm cr}$.
Noting that
$n_f=n_f^{\rm cr}$ at $\alpha=\alpha_*=\alpha_{\rm cr}$,  we have \begin{align}
  \frac{\Lambda_{\mathrm{IR}}}{\Lambda_{\mathrm{UV}}}
  &\sim \frac{2  m_D}{\Lambda_{\mathrm{UV}}} \nonumber\\
  &\sim \exp\left(-\frac{\pi}{\sqrt{\alpha/\alpha_{\mathrm{cr}}-1}}\right) 
  \sim
   \exp\left(-\frac{A}{\sqrt{n_f^{\rm cr} -n_f}}
  \right)\nonumber \\
  &\rightarrow 0\;\textnormal{ as }  n_f\equiv  \frac{N_{f}}{N_{c}}\rightarrow   n_f^{\rm cr}
  \equiv  \left(\frac{N_{f}}{N_{c}}\right)^{\rm cr}
  \;.
  \label{essentialsingularity}
\end{align}

At this point, at the two-loop level,
\begin{equation}
  \alpha(\mu^{2}=\Lambda_{\mathrm{UV}}^{2})=\frac{\alpha_{*}}{1+W(e^{-1})}\simeq0.78\alpha_{*}\;,
\end{equation}
where the IR fixed point is at
\begin{equation}
  \alpha_{*}\simeq\frac{4\pi}{N_{c}}\cdot\frac{11-2 n_f
  }{13 n_f
  -34}\;,
\end{equation}
and the ladder critical coupling is
\begin{equation}
   \alpha_{\mathrm{cr}}
 =\frac{\pi}{3} \cdot \frac{1}{C_2}
  =\frac{\pi}{3}\frac{2N_{c}}{N_{c}^{2}-1}   \,\, \gtrsim  \frac{2\pi}{3N_c}\;.
\end{equation}
The two-loop values are then
\begin{align}
   A \simeq
  \frac{3\sqrt{2}\pi}{5}\simeq2.7,\quad
   n_f^{\rm cr}\lesssim  4\;.
\end{align}
This suggests that $N_f=12$ is barely inside the conformal window, although the two-loop value for $\alpha_{*}$ is not expected to be precise,
and hence neither is the value of $n_f^{\rm cr}$.
Fuller details may be found in Ref.~\cite{Matsuzaki:2015sya}.

Turning our attention back to the conformal phase,
we may at first sight expect that
since we introduce a non-zero fermion mass,
both the scale symmetry and chiral symmetry
are broken explicitly but not spontaneously.
In this case,
all fermionic bound states,
including the $\pi$ and $\sigma$,
would be Coulombic bound states as discussed in Eq.~\eqref{eq:coulomb-bs} above.
This is what is observed in heavy quarkonia,
for similar reasoning.

However,
a key difference of the conformal phase from the heavy quarkonia
is that the degenerate massive fermions giving rise to confinement decouple,
with the induced confinement scale---the
gluon condensate---totally
governed by the fermion mass $m_{f}^{(R)}$ as described by Eq.~\eqref{eq:conformalir}.
The gluon condensate breaks the scale symmetry not only explicitly,
but also spontaneously.
Specifically,
the $\sigma$ is a pseudo NG boson of the scale symmetry
(although
the explicit breaking scale $m_{f}^{(R)}$
is much larger than the spontaneous breaking scale
$ \Lambda_{\mathrm{YM}} \simeq m_{f}^{(R)}/2$ in the $N_{f}=12$ case).
The gluon condensate does not however break chiral symmetry,
which as such is only explicitly broken;
the $\pi$ therefore remains a non-NG boson.

Given this,
we may try to estimate the $\sigma$ mass  not only in the broken phase, but also in
 the conformal phase,
 where the scale symmetry is broken
both explicitly and spontaneously
by the same infrared mass scale of the fermion mass $m_{f}^{(R)}$.
(It is of course also explicitly broken in the UV by the regularization;
this makes no difference to the formation of bound states.)
Following the argument of Ref.~\cite{Matsuzaki:2013eva},
we may start with the dilatation current given in Eq.~(\ref{dilatation}),
whose WT identity  (with single-pole dominance as in the case for $\eta^\prime$ in Eq.~(\ref{WTforetaprime}))
is
\begin{align}
  M_\sigma^2 F_\sigma^2&= i {\cal F.T.}
 \left.\langle T
  \left(\partial_\mu D^\mu(x)\cdot  \partial_\mu D^\mu(0)\right)
\rangle\right|_{q_\mu\rightarrow 0} =\langle [ -i Q_D, \partial_\mu D^\mu(0)]\rangle 
  \label{sigmamassformula}  \\
&    =-  4 \cdot \frac{\beta^{(NP)}(\alpha)}{4\alpha}
\langle G_{\mu\nu}^2\rangle
 - \left(3-\gamma_m\right) \cdot \left(1+\gamma_m\right) m_f
  \sum^{N_f}_{i=1}  \langle  \bar{\psi}^{i}\psi^{i}\rangle.
  \label{sigmamassformula2}
 \end{align}
Similarly,
the single-pole-dominated WT identity for the non-singlet axial-vector current
$A_\mu^\alpha\,(\alpha=1,2,3)$ for each doublet  $\psi^i \,(i=1,2)$ gives the Gell-Mann--Oakes--Renner (GMOR) relation:
\begin{equation}
\left(\frac{F_\pi}{\sqrt{2}}\right)^2 M_\pi^2\cdot  \delta^{\alpha\beta} = \langle [ -i Q_5^\alpha,  \partial^\mu A_\mu^\beta(0)]\rangle = - m_f  \sum^{2}_{i=1}\langle  \bar \psi^i  \psi^i \rangle \cdot  \delta^{\alpha\beta}.
\label{pionmass}
\end{equation}
While Eqs.~\eqref{sigmamassformula} and~\eqref{pionmass}
are usually derived using the soft pion theorem,
they are in fact based simply on the pole dominance,
and hence are valid in both the broken and conformal phases.
This then gives~\cite{Matsuzaki:2013eva}
\begin{align}
  M_\sigma^2 &= (M_\sigma^{\textnormal{(anomalous)}})^2
          +\frac{ (3-\gamma_m)\cdot (1+\gamma_m)\frac{N_f}{2}
          (\frac{F_\pi}{\sqrt{2}})^2}{F_\sigma^2}\cdot M_\pi^2 \nonumber
          \\
        &= d_0 + d_1 \cdot M_\pi^2\,.
        \label{eq:msigma-summary}
\end{align}

This may then be used as a fit form for lattice data.

When combined with an estimate of the anomalous dimension $\gamma_m$, 
the results of the fit for $d_1$ may be used to estimate
the ratio
$F_\sigma/(F_\pi/\sqrt{2})=\sqrt{N_D} F_\sigma/v_{\rm EW}$
(where $N_D$ is the number of electroweak doublets).
From Eq.~\eqref{eq:msigma-summary},
\begin{equation}
 d_{1}=\left(\sqrt{(3-\gamma_{m})(1+\gamma_{m})\frac{N_f}{2}}
  \cdot \frac{F_\pi/\sqrt{2}}{F_{\sigma}}\right)^2\equiv \left(C_{\gamma_m}\cdot \frac{F_\pi/\sqrt{2}}{F_{\sigma}}\right)^2\,.
  \label{d_1}
\end{equation}

Linear sigma model and holographic QCD calculations
in the near-conformal broken and conformal phases
suggest that
(via Footnote~18 of~\cite{Matsuzaki:2015sya} and references therein)
\begin{equation}
  F_{\sigma}^{2}= d_\sigma^2 \cdot \frac{N_f}{2} \left(\frac{F_\pi}{\sqrt{2}}\right)^2
  =(3-\gamma_{m})^{2}\cdot
 \frac{N_{f}}{2}
\left (\frac{F_\pi}{\sqrt{2}}\right)^{2}\;,
  \label{linearsigmarel}
\end{equation}
 with $d_\sigma=3-\gamma_m$  the scale dimension of  $\sigma\sim \bar \psi \psi$\footnote{ For $N_f=2$
this becomes the standard linear sigma model relation $F_\sigma=F_\pi/\sqrt{2}=\langle \sigma \rangle$ for $d_\sigma=1$, or equivalently the Nambu--Jona-Lasinio (NJL) model with $\gamma_m=2$.
Note also that the kinetic term (the scale-invariant part) of the $N_f=2$ linear sigma model (or the Standard Model Higgs Lagrangian) is rewritten into
the dChPT Lagrangian, Eq.~(2) of Ref.~\cite{Matsuzaki:2013eva}, through polar decomposition, with its mass from the explicit scale-symmetry breaking potential, $M^2 =2 \lambda \langle \sigma \rangle^2$, as a pseudo-dilaton for $\lambda\ll 1$. The interested reader is referred to
Ref.~\cite{Yamawaki:2018jvy}
and references cited therein.
\label{linearsigmaNf2}}.
If this is the case throughout the near-conformal and conformal phases,
then  from Eq.~(\ref{d_1}) we have 
\begin{equation}
  d_{1}=\frac{1+\gamma_{m}}{3-\gamma_{m}}\;.
  \label{eq:d1-slope}
\end{equation}

We may also use this result to make predictions
informing potential lattice studies of $12<N_f<16.5$:
in this case, the SD equation gives
\begin{align}
  \gamma_m &\simeq 1-\sqrt{1-\alpha_*/\alpha_{\rm cr}} \\
      & \searrow 0\;\textnormal{ as }\alpha_* \searrow 0\, (N_f \nearrow  16.5)\;.
\end{align}
This in turn would predict
\begin{align}
  d_1 &\searrow 1/3 \quad\textnormal{as } M_\sigma \searrow  M_\pi/\sqrt{3}\;,\textnormal{ and } \\
  \frac{F_\sigma}{F_\pi/\sqrt{2}} &\nearrow 8.6 \quad\textnormal{as } M_\sigma^w \simeq d_1 M_\pi^2 (\gg d_0)\;.
\end{align}

On the other hand,
\begin{equation}
  d_0 = {M_\sigma^2}^{(\mathrm{anomalous})}
       = - \frac{4 \cdot \langle \theta^{\mu}_{\mu}\rangle
    }{F_\sigma^2}
      =- \frac{\beta^{(NP)}(\alpha)}{\alpha} \frac{\langle G_{\mu\nu}^2\rangle}{F_\sigma^2}
    \end{equation}
comes from the non-perturbative trace anomaly.
In the broken phase it is non-zero even in the chiral limit,
as $m_D$ violates the scale symmetry,  both spontaneously and explicitly. Thus the chiral limit $\sigma$ mass $d_{0}$ is a key observable to understand whether a theory is a suitable candidate for a walking technicolor model with a composite Higgs of appropriate mass.

In the anti-Veneziano limit,
the fermion loop dominates over the gluon loop in
the computation of the anomaly.
Similarly as for the $\eta^\prime$ in Fig.~\ref{GGtil-loops},
this then gives from Eq.~\eqref{sigmamassformula}:
\begin{align}
  {M_\sigma^2}^{(\mathrm{anomalous})} \cdot F_\sigma^2
  &= i {\cal F.T.} \left.\left\langle T
    \left(
    \frac{\beta^{(NP)}(\alpha)}{4 \alpha} G_{\mu\nu}^2(x)
    \cdot
    \frac{\beta^{(NP)}(\alpha)}{4 \alpha} G_{\mu\nu}^2(0)
    \right)
    \right\rangle\right|_{q_\mu\rightarrow 0} \\
  &\sim N_f N_c m_D^4\;,
\end{align}
with
\begin{equation}
  F_\sigma^2\sim (3-\gamma_m)^2 N_f N_c m_D^2
  \label{eq:fsigma}
\end{equation}
up to a numerical factor roughly independent of  $N_f$ and $N_c$.
An implicit $N_{f}$ dependence
 comes from $F_{\sigma}^{2}\sim (3-\gamma_{m})^{2}$,
via the details of the dynamics as in the linear sigma model/holography discussed above.

This then implies
\begin{equation}
 \left.M_{\sigma}^{2}\right|_{m_{f}=0}=d_{0}
  = \mathcal{O}
  \left(
  \left( \frac{ m_D}{
3-\gamma_m}\right)^{2}
\right)   \;,\quad\forall N_{f}, \; .
\label{sigmamassbroken}
\end{equation}
As the conformal window is approached from below,
we expect the anomalous dimension $\gamma_m \rightarrow 1$,
and so this vanishes:
\begin{equation}
 \left.M_{\sigma}^{2}
\right|_{m_{f}=0} = {\cal O} \left( \left(\frac{ m_D}{2}\right)^{2} \right)
\rightarrow 0  \quad
(n_f \nearrow n_f^{\rm cr}).
\end{equation}This is similar to other states in the spectrum,
which obey
\begin{equation}
  \left.
  M^{2}_{\rm H}\right|_{m_{f}=0}
 =\mathcal{O}\left(\left(2m_D\right)^{2}\right) \rightarrow 0 \quad   (n_f \nearrow n_f^{\rm cr})   \;;
\end{equation}
with no additional suppression,
contrary to the popular assumption~\cite{Bando:1986bg, Appelquist:2010gy, Golterman:2016lsd,DelDebbio:2021xwu}
that the vanishing beta function suppresses $M_\sigma^2$ relative to other states.\footnote{An explicit ladder computation~\cite{Hashimoto:2010nw,Matsuzaki:2015sya}
shows that the vanishing beta function in Eq.~(\ref{laddermD})
is precisely cancelled by the {\it diverging} $ \langle G_{\mu\nu}^2\rangle$
 $ \sim N_f N_c m_D^4\cdot \ln^3(\Lambda/m_D)$
 $ \rightarrow \infty$, such that
 $- \beta^{(NP)}(\alpha)/\alpha \cdot  \langle G_{\mu\nu}^2\rangle$  $\sim  N_f N_c m_D^4$ is independent of
 $\Lambda$,
with the result precisely the same as an independent computation of $\langle \theta^\mu_\mu\rangle $ through the effective potential.
 Such a complete cancellation might be avoided only by  including non-ladder effects, such as in the holographic model
~\cite{Matsuzaki:2012xx}.}
Thus in the near-conformal region,  $\sigma$ as a pseudo NG boson  has a small but non-zero ratio:
\begin{equation}
  \left.\frac{M^2_{\sigma}}{M^2_{\rm H}}\right|_{m_{f}=0} \rightarrow \textnormal{const.} (\ll 1) \quad   (n_f \nearrow n_f^{\rm cr})\; .
\end{equation}

Note that
$\left.M_{\sigma}^{2}\right|_{m_{f}=0}=d_{0}
\sim {m_D}^{2}$
has no additional dependence on $N_{f}$
when expressed in units of $\Lambda_{\mathrm{IR}} =
1/\sqrt{8 t_0} $
(aside from the above-mentioned
 implicit $N_{f}$ dependence of
$F_{\sigma}^{2}$ through $\gamma_m$).
Similarly,
$d_{1}$ also has no explicit dependence on $N_{f}$
in the anti-Veneziano limit
(but again has an implicit $N_{f}$ dependence from $\gamma_{m}$).
This contrasts the result
in Eq.~\eqref{eq:R-definition} that
$ M_{\eta^{\prime}}^{2}\cdot 8t_0=
M_{\eta^{\prime}}^{2}/\Lambda_{\mathrm{IR}}^{2}\sim N_{f}^{2}$.

Eq.~\eqref{eq:msigma-summary}, here based on the WT identity for  the scale symmetry was originally derived
 in the broken phase  via
dilaton chiral perturbation theory
(dChPT)~\cite{Matsuzaki:2013eva} at tree level.
It has
further been shown explicitly~\cite{Matsuzaki:2013eva} that there exists
a
 possible deviation from Eq.~\eqref{eq:msigma-summary}  due to the chiral log of the pion loop effects
 (as in the standard ChPT) near the chiral limit.
For the loop expansion parameter
\begin{equation}
 \chi =M_\pi^2/\Lambda_\chi^2 \ll 1\;,
\end{equation}
where
\begin{align}
  \Lambda_\chi^2 &=(4\pi\cdot (F/\sqrt{2}))^2/N_f\;, & F &\equiv F_\pi|_{m_f=0}\;,
\end{align}
as well as $M_\pi^2 \ll M_\sigma^2$.
Then from Eq.~(10) of Ref.~\cite{Matsuzaki:2013eva}:
\begin{align}
\frac{M_\sigma^2}{\Lambda_\chi^2} &\simeq  \frac{d_0}{\Lambda_\chi^2} \left(1+ \frac{d_1}{2}  \chi \ln \chi\right)
+ d_1 \left(1+ \frac{9}{4} \chi \ln \chi\right)\cdot \chi
 \quad (N_f\gg 1).
 \label{chirallog}
 \end{align}
While this expansion is relatively flat for $0 < \chi < 1$,
as $\chi$ increases past this region,
the chiral log present in dChPT no longer makes sense,
and Eq.~\eqref{eq:msigma-summary} should be used instead.
This is the region where our numerical results lie,
with $\chi \gg 1$ and (for $N_f = 8$)
$M_\sigma^2 \lesssim M_\pi^2$.
Further,
since $\Lambda_\chi^2 \simeq (N_c/N_f)\cdot  (2 m_D)^2$,
simulations at much smaller values of $m_f$ would be needed
in order to observe the loop effects.

This implies that the value obtained for $d_0$ by fitting using Eq.~\eqref{eq:msigma-summary}
is effectively shifted to a smaller value than the true chiral limit value of $d_0=M_\sigma^2|_{m_f=0}={M_\sigma^2}^{(\mathrm{anomalous})}$,
as illustrated in Fig.~2 of Ref.~\cite{Matsuzaki:2013eva}.
On the other hand, the fit value of $d_1$ is the same as $d_1$ in the chiral limit $\chi \rightarrow 0$ where chiral log effects disappear, and so is the value $\frac{F_\sigma}{F_\pi/\sqrt{2}}$ in Eq.~(\ref{sigmamassdecayconst}).
We shall return to this point in Sec.~\ref{sec:scalar} where we discuss the new data of this paper on the mass of the $\sigma$ as a candidate for the  Higgs. 

While dChPT is valid in the near-conformal broken phase,
the alternative derivation~\cite{Matsuzaki:2013eva}
presented here shows that
Eq.~\eqref{eq:msigma-summary} is also valid
not only deep in the broken phase  (again with the $d_0$ fit value up to chiral log)
in the absence of any remnant of scale symmetry
(and where we do not expect to observe hyperscaling),
but also inside the conformal window  where dChPT is irrelevant.
Inside the conformal window,
both $d_{0}$ and $\beta^{(NP)}(\alpha)$ are identically zero in the chiral limit,
where no bound states exist.
As such there is a ``conformal phase transition'',
where the order parameter $m_D$
continuously changes going from the broken to the conformal phase
in the essential singularity form,
while bound states do not~\cite{Miransky:1996pd}.

When we introduce an explicit fermion mass $m_f$, the scale symmetry is broken explicitly and spontaneously as described in Sec.~\ref{sec:gradientflow}.
We expect that similarly to Eq.~\eqref{eq:conformalir},
the anomalous component of the $\sigma$ mass
is of the order the confining scale of pure Yang--Mills $\Lambda_{\mathrm{YM}}\simeq m_f^{(R)}/2
=1/(2\sqrt{8 t_0})$ instead of $\Lambda_{\rm IR}=1/\sqrt{8 t_0}$---in contrast to
that of  $\eta^{\prime}$ (Eq.~(\ref{eq:R-definition})), which is irrelevant to the spontaneous scale symmetry breaking.
Specifically,
\begin{equation}
  d_0= \left(M_{\sigma}^{\textnormal{(anomalous)}}\right)^{2}
  \simeq \frac{\Lambda_{\mathrm{YM}}^{2}}{(3-\gamma_m)^2}
  \simeq \frac{\left(m_{f}^{(R)}\right)^{2}}{4   (3-\gamma_m)^2}
  \simeq \frac{M_{\pi}^{2}}{16   (3-\gamma_m)^2}\;,
\label{sigmamassconformal}
\end{equation}
where we have again relied on relation Eq.~(\ref{linearsigmarel}).
Putting this back into Eq.~\eqref{eq:msigma-summary} gives
\begin{equation}
  M_{\sigma}^{2} =
  \left(\frac{1}{16
 (3-\gamma_m)^2}
  +d_{1}\right)M_{\pi}^{2}.
  \label{sigmamass12}
\end{equation}

Let us now briefly explore the compatibility of the predictions above with our previous work.
Considering Eq.~\eqref{eq:msigma-summary},
for $N_f=8$ we found in our previous work~\cite{Aoki:2016wnc}\footnote{A consistent set of coefficients was obtained in the first lattice result for $M_\sigma$ in $N_f=8$ QCD \cite{Aoki:2014oha}.}:
\begin{align}
d_0 &= - 0.0028(98)(^{36}_{354}),&d_1&=0.89(26)(^{75}_{11}).
\label{d_0d_1previous}
\end{align}
Here,
$d_0 \lesssim 0$ would suggest that
\begin{equation}
M_\sigma^2 \lesssim d_1  M_\pi^2\simeq M_\pi^2\;,
\label{d_0Nf8}
\end{equation}
suggesting that
if the $N_f=8$ theory is in the broken phase,
then our data are far from the chiral limit,
where we expect $M_\sigma^2> M_\pi^2=0$.

Our previous observations for $\gamma_*$ from hyperscaling would then indicate that
for $N_f=8$, $\gamma_m\simeq1$.
Considering Eq.~\eqref{d_1},
this then implies
$\Rightarrow C_{\gamma_m} \simeq 4$,
and so\footnote{
A consistent result was also obtained through the measurement of the the scalar density decay constant $F_S$ which is independent of the $d_1$ measurement---see Fig.~50 in Ref. \cite{Aoki:2016wnc}.
}
\begin{eqnarray}
  \frac{\sqrt{N_D} F_\sigma}{v_{\rm EW}} =\frac{F_\sigma}{F_\pi/\sqrt{2}} =
C_{\gamma_m}  \cdot \frac{1}{\sqrt{d_1}
 }
  \simeq 4
 \,.
  \label{sigmamassdecayconst}
\end{eqnarray}
This is consistent with the linear sigma model prediction in Eq.~\eqref{linearsigmaNf2},
and gives $d_1 \simeq 1$. 
For $N_f=12$, $\gamma_m \simeq 0.4-0.5$ would give 
\begin{equation}
d_{1}^{(N_{f}=12)} \simeq 0.54-0.6.
  \label{eq:d1}
\end{equation}

Eq.~\eqref{sigmamass12} implies that $M_{\sigma}$ is less than $M_{\pi}$ for $N_f=12$,
since $d_{1}^{(N_{f}=12)} <1$ (see Eq.~\eqref{eq:d1}); 
i.e. the smallness of the $\sigma$ mass is related to that of its slope. 
Since there is no chiral log correction in the conformal phase, we expect $M_\sigma^2/M_\pi^2 <1$ all the way down to near the chiral limit (although not, as previously discussed, at $m_f\equiv 0$, where no bound states exist).

By directly comparing the three theories, we will later demonstrate in Sec.~\ref{sec:scalar}
how our updated lattice results for $M_\sigma$ and $M_\pi$ align with these formulas. 
Meanwhile, the relation $M_\sigma < M_\pi$ for $N_f=12$ has been numerically confirmed in our previous lattice result~\cite{Aoki:2013zsa}, independently of these formulas.

\clearpage
\newpage
\section{Setup of lattice simulations and hadron spectrum}\label{sec:lattice-setup}
The LatKMI collaboration has been systematically investigating $N_f=4$, 8, 12, and 16 QCD using a common setup for the lattice action, 
exploring different fermion masses and volume sizes, and employing a single value for $N_f=8$ 
and two values of the gauge coupling for $N_f=4$ and $N_f=12$~\cite{Aoki:2012eq,Aoki:2013xza,Aoki:2013zsa,Aoki:2014oha,Aoki:2016wnc}.

The fermionic action discretization used is the highly improved staggered quark (HISQ) action~\cite{Follana:2006rc}, while the gauge action is the tree-level improved Symanzik action~\cite{Symanzik:1983dc}.
It suppresses the effects of taste breaking that spoil flavor symmetry~\cite{Aoki:2013xza,Aoki:2016wnc}.
Moreover, for theories with a number of degenerate quarks that is a multiple of four, staggered quarks are very efficient compared to other numbers of flavors, as taking a fourth root of the fermionic determinant is not required.

The spectrum of the low-lying flavor-non-singlet states is extracted using standard techniques involving the analysis of two-point correlation functions of interpolating operators with the correct quantum numbers.
The benchmark states are the vector and the pseudoscalar, which are called the $\rho$ and $\pi$ meson respectively, by analogy with the corresponding QCD states.
Their masses, denoted by $M_\rho$ and $M_\pi$ respectively, are easy to compute and their mass ratio is a good qualitative indicator for spontaneous chiral symmetry breaking or for conformality.
In the first case, the $\pi$ mass would be expected to go to zero and $M_{\rho}/M_{\pi}$ to go to infinity in the massless fermion limit.
On the other hand, an infrared conformal spectrum in the presence of an explicit fermion mass would show a constant finite $M_{\rho}/M_{\pi}$ towards the masslesss limit, with possible corrections to this behavior at large quark masses.

\begin{figure}[thb]
  \centering  \includegraphics[width=\textwidth]{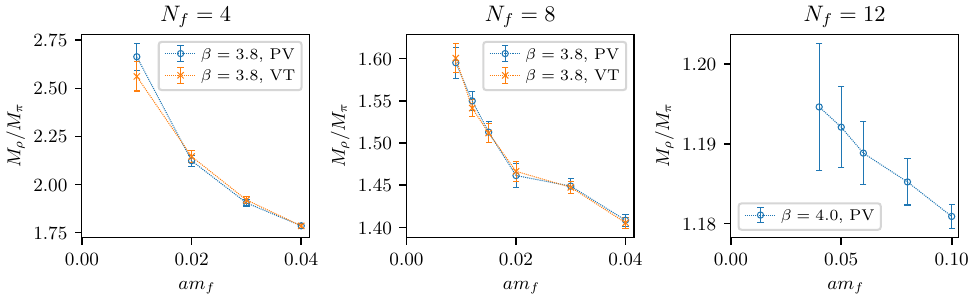}

  \caption{\label{fig:rhopiratio} The ratio $M_{\rho}/M_{\pi}$  shows indications of spontaneous chiral symmetry breaking in the massless fermion limit for $N_f=4$ and 8 QCD different from $N_f=12$ QCD. For $N_f=8$ the results using different taste operators for the $\rho$ meson are shown. The range of fermion mass considered in the three theories is similar. Lines connecting the data points are meant to guide the eye.}\label{fig:nf4-nf8-nf12-chisb}
\end{figure}

Our previous studies have shown that the SU(3) gauge theory with $N_f=4$ and 8 are in a hadronic, chirally-broken phase where the lattice simulations show a decisive increase of the ratio $M_{\rho}/M_{\pi}$ towards the massless quark limit.
On the contrary, the $N_f=12$ theory displays a constant ratio in our simulations.
We repeat this computation using an updated set of lattice simulations (discussed in more detail in the specific subsection below), and confirm that we continue to observe this effect, as can be seen in Fig.~\ref{fig:nf4-nf8-nf12-chisb}.
Note that, while the fermion mass range for the three theories is similar (in lattice units), the hadron mass ratio $M_{\rho}/M_{\pi}$ has a different scale, highlighting a large deviation from a constant in the $N_f=4$ and 8 cases.
The staggered fermion taste symmetry breaking, which is the relevant lattice discretization effect in this investigation, is briefly discussed in Appendix~\ref{sec:app-taste}; this includes for example $N_f=4$ simulations performed at different lattice spacings.
For $N_f=8$ we have already shown that these effects are small~\cite{Aoki:2016wnc}.

In the following subsections we will first define the means by which we set the scale using the gradient flow, as well as how this scale can be used to extract the anomalous dimension via finite-size hyperscaling, and then revisit the spectrum computations of our previous work. After establishing that the conclusions of our previous work still hold for $N_f=8$ with the updated simulations at smaller quark masses we will not repeat a full chiral or conformal analysis; subsequent sections will focus on the flavor-singlet spectrum.

\subsection{Gradient flow scale and hyperscaling}

\begin{figure}[thb]
  \centering
  \includegraphics[width=0.52\textwidth]{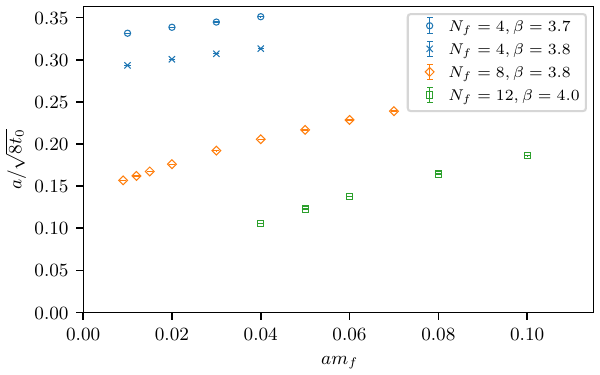}

  \caption{\label{fig:s8t0}The hadronic energy scale $1/\sqrt{8t_0}$ in lattice units as a function of the bare quark mass is plotted for all values of $N_f$.}
\end{figure}

In addition to the mass scales of the hadronic spectrum, we measure the gradient flow scale $t_0$~\cite{Luscher:2010iy}.
We use the Symanzik flow kernel to smooth the clover definition of the one-point energy operator $\langle E \rangle$.
From this,
we define the scale $t_{\mathcal{E}}$ as the value of flow time $t$ such that $t^2\langle E(t) \rangle\equiv  \mathcal{E}$,
and in particular the scale $t_0$ as the value of $t_{\mathcal{E}}$ where $\mathcal{E}= 0.3 \equiv \mathcal{E}_0$. 
This is the same setup we used in our previous publications~\cite{Aoki:2016yrm,Aoki:2016wnc}.
We calculate $t_0$ for $N_f=4$ at two $\beta$ values and four quark masses $m_f$, for $N_f=8$ at one $\beta$ and six quark masses, and for $N_f=12$ at one $\beta$ and five quark masses, 
while we have simulated at two $\beta$ values for $N_f=12$ (see Ref.~\cite{Aoki:2016wnc} for details of the mass spectra analysis).
The results are summarized in Tab.~\ref{tab:t0}.
While $t_0$ is in units of $a^2$, it is convenient to define a length scale
(associated with the smearing radius of the diffusion process to which the gradient flow is equivalent)
as $r=\sqrt{8t_0}$, which can be intuitively associated to the Sommer radius $r_0$ in QCD~\cite{Sommer:1993ce}.
We plot the energy scale corresponding to $1/\sqrt{8t_0}$ for all values of $N_f$ as a function of the bare quark mass in lattice units in Fig.~\ref{fig:s8t0}.
We observe smaller $a/\sqrt{8 t_0}$ as $N_f/N_c$ increases, which indicates that the chiral limit values would decrease towards the conformal window.

\begin{table}[ht]
  \caption{The gradient flow scale $t_0$ in lattice spacing units for the ensembles used in the spectrum analysis.}\label{tab:t0}

  \begin{ruledtabular}\begin{tabular}{ccccc}
$N_f$ & $\beta$ & $L$ & $m_f$ & $t_0$ \\
\hline
4 & 3.7 & 20 & 0.01 & 1.13807(89) \\
4 & 3.7 & 20 & 0.02 & 1.09109(73) \\
4 & 3.7 & 20 & 0.03 & 1.05237(86) \\
4 & 3.7 & 20 & 0.04 & 1.01459(73) \\
4 & 3.8 & 20 & 0.01 & 1.4535(19) \\
4 & 3.8 & 20 & 0.02 & 1.3841(20) \\
4 & 3.8 & 20 & 0.03 & 1.3262(17) \\
4 & 3.8 & 20 & 0.04 & 1.2752(15) \\
\hline
8 & 3.8 & 48 & 0.009 & 5.0762(72) \\
8 & 3.8 & 42 & 0.012 & 4.7543(56) \\
8 & 3.8 & 36 & 0.015 & 4.4556(64) \\
8 & 3.8 & 36 & 0.02 & 4.0304(49) \\
8 & 3.8 & 30 & 0.03 & 3.3842(35) \\
8 & 3.8 & 30 & 0.04 & 2.9563(39) \\
8 & 3.8 & 24 & 0.05 & 2.6599(70) \\
8 & 3.8 & 24 & 0.06 & 2.39320(77) \\
8 & 3.8 & 24 & 0.07 & 2.1885(32) \\
\hline
12 & 4.0 & 36 & 0.04 & 11.206(57) \\
12 & 4.0 & 36 & 0.05 & 8.292(26) \\
12 & 4.0 & 30 & 0.05 & 8.338(26) \\
12 & 4.0 & 30 & 0.06 & 6.603(19) \\
12 & 4.0 & 36 & 0.08 & 4.632(17) \\
12 & 4.0 & 24 & 0.08 & 4.627(12) \\
12 & 4.0 & 24 & 0.1 & 3.6133(70) \\
\end{tabular}\end{ruledtabular}

\end{table}

\begin{figure}[htb]
  \centering
  \includegraphics[clip]{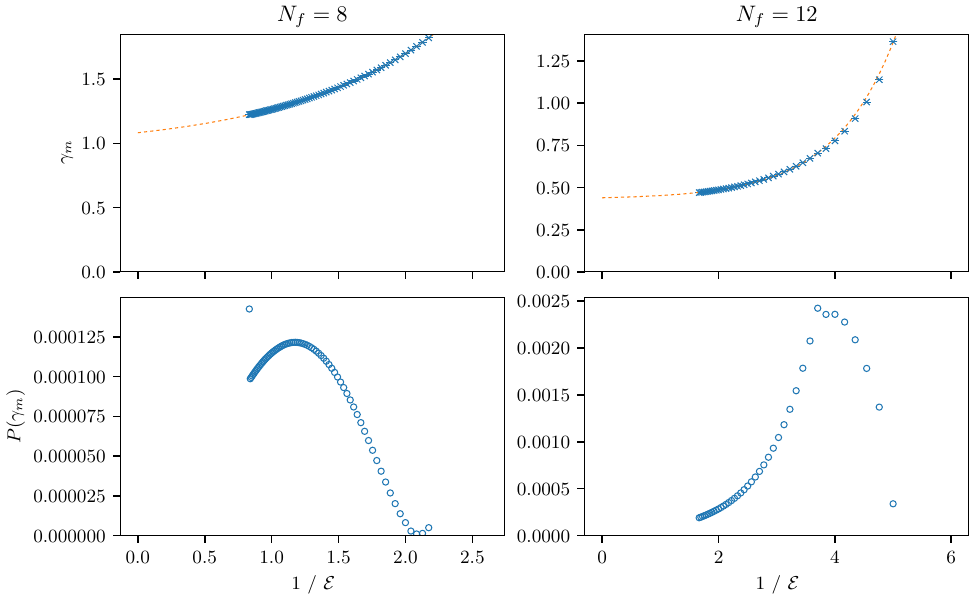}

  \caption{The result of the finite-size hyperscaling analysis for $N_f=8$ (left) and $N_f=12$ (right). In each case, the upper panel shows the value of the anomalous dimension $\gamma_m$ obtained via a fit of the energy scale $t$ as a function of the reference scale $\mathcal{E}$ (points), and the lower panel the residual $P(\gamma_m)$ (defined in Eq.~(19) of \cite{Athenodorou:2021wom}) of the fits for the points in the upper plot. The dashed line shows the exponential fit used to find the final quoted values for $\gamma_*$.}\label{fig:t0fshs}
\end{figure}

To understand the scaling behavior of $a/\sqrt{8 t_{\mathcal{E}}}$,
we fit these data with a finite-size hyperscaling (FSHS) ansatz of the form
\begin{equation}
  \frac{L}{\sqrt{8t_{\mathcal{E}}}} =  f\left(Lm_f^{\frac{1}{1+\gamma_m}}\right)\;,
\end{equation}
where $\gamma_m$ is the mass anomalous dimension,
as enters into Eq.~\eqref{eq:coulomb-bs} in the conformal phase.
This fit was performed using the piecewise-interpolating curve-collapse method proposed by Bhattacharjee~\cite{Bhattacharjee_2001} and previously deployed by DeGrand~\cite{DeGrand:2009hu}.
The value found for $\gamma_m$ shows significant dependence on $\mathcal{E}$, as shown in Fig.~\ref{fig:t0fshs}.
As the gradient flow has a smoothing effect removing ultraviolet effects, the fit result at larger values of $t$---and since $t^2 E$ increases monotonically with $t$, larger values of $\mathcal{E}$---is closer to the infrared value than that at smaller values.
Furthermore, since the residual $P(\gamma_m)$ (analogous to a $\chi^2$ measure for this type of fit) decreases as $1/\mathcal{E}$ approaches zero for $N_f=12$, 
this suggests that $a/\sqrt{8 t_{\mathcal{E}}}$ can be well described by the FSHS ansatz at larger flow times $t$. 

This result, together with the observation in Fig.~\ref{fig:s8t0}, supports our identification of a hadronic energy scale $\Lambda_{\mathrm{IR}}/\Lambda_{\mathrm{UV}}=a/\sqrt{8 t_0}$,  
as indicated in Eq.~\eqref{gradientflowasIR}, though Eq.~\eqref{gradientflowasIR} strictly applies only in the infinite volume limit.
Among various possible definitions, for comparisons across different theories, 
it is convenient to use a gluonic definition $a/\sqrt{8 t_0}$ as the reference scale $\Lambda_{\mathrm{IR}}/\Lambda_{\mathrm{UV}}$, 
since it exhibits only mild fermion mass dependence, as discussed later in Secs.~\ref{sec:scalar} and \ref{sec:pseudoscalar}.
As a result,
in the following we will often use the length scale $r=\sqrt{8t_0}$ (and its reciprocal) as a reference scale across theories with different number of flavors.

In Fig.~\ref{fig:t0fshs},
we show the result for $\gamma_m$ as a function of $1/\mathcal{E}$,
for both values of $N_f=8,12$.
We note that the region of $\mathcal{E}$ in which a stable fit is obtained
differs between the two cases.
This is due to differences in the fermion masses and lattice volumes considered in the two cases,
since there must be sufficient overlap between the data at different $L$ to allow all data to contribute to the fit;
we would anticipate that dedicated studies controlling these parameters
could allow overlapping regions of $\mathcal{E}$ to be explored.

In both cases,
the data visibly begin to level off at larger values of $\mathcal{E}$,
which we interpret as $\gamma_{*} = \lim_{\mathcal{E}\rightarrow \infty}\gamma_{m}$.
This is most pronounced in the $N_f=12$ case,
which appears to almost reach its plateau value;
in the $N_f=8$ case
extrapolation to the $\mathcal{E}\rightarrow\infty$ limit is needed.
Imposing \emph{ad hoc} an exponential form for this behaviour gives a limiting value of
$\gamma_* = \TZeroGammaStarNfEight$ for $N_f=8$ and
$\gamma_* = \TZeroGammaStarNfTwelve$ for $N_f=12$,
both of which agree closely with results previously obtained
from (basic and finite-size) hyperscaling of the mass spectrum~\cite{Aoki:2016wnc}
(specifically that obtained for states other than the $\pi$ in the $N_f = 8$ case).
\subsection{Updates of hadron spectra in $N_f = 8$ QCD}\label{sec:spectrum}

Before discussing the scalar spectrum, we will give an update of the hadron spectra and their chiral extrapolations, which include a newly added lightest mass point in $N_f=8$ QCD at $m_f = 0.009$. These new data were generated with volume $L^3\times T = 48^3 \times 64$ at $\beta = 3.8$, using the same algorithm as in our previous works~\cite{Aoki:2016wnc,Aoki:2013xza,Aoki:2014oha} where the lightest point was $m_f=0.012$ on $L^3\times T = 48^3 \times 64$ lattices.

We report details of the simulation parameters and a finite volume study in Appendix~\ref{sec:app-lattice}.

In the analyses that follow, physical quantities are extrapolated to the chiral limit by a polynomial function of $m_f$, as in the previous work~\cite{Aoki:2016wnc}, whereas effects from chiral logarithm are taken into account to estimate the systematic error of chiral extrapolation for some limited cases.

\subsubsection{$F_\pi$ and $M_\pi$}

The decay constant $F_\pi$ is fitted by a quadratic fit with the fitting range $am_f \in [0.009, 0.03]$, because including larger masses results in large fit residuals ($\chi^2/$dof). Having a new lightest mass point at $m_f=0.009$ allows us to shift our fitting ranges towards the chiral limit while keeping the same number of fit degrees of freedom.

\begin{figure}[tp]
   \centering  \includegraphics[width=\textwidth]{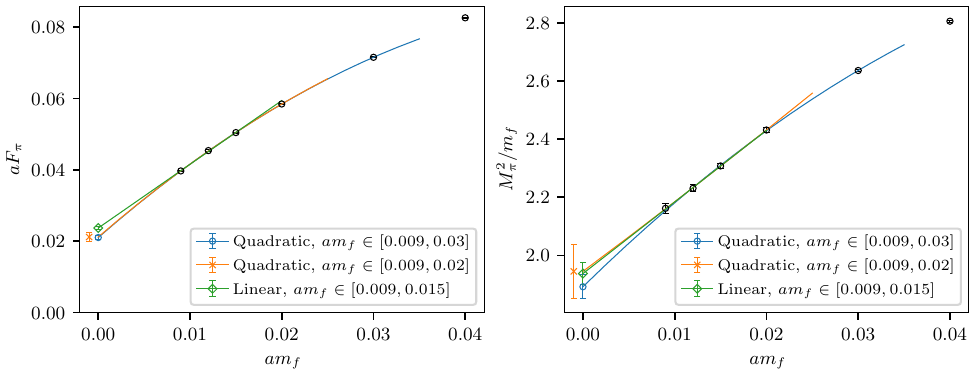}

   \caption{Left: Chiral extrapolations of $F_\pi$ in $N_f = 8$.
     The range of $m_f$ considered for each fit curve is noted in the legend.
   Right: Same as the left panel but for $M_\pi^2/m_f$.}\label{fig:spect_chiral:fpi}
\end{figure}

The chiral limit extrapolations of these fits ($F$) are shown in Table~\ref{tab:spect_chiral:pi}.
The result of $F$ with a quadratic form using the fit range $m_f \in [0.009, 0.03]$ agrees reasonably well with other fit results, which are from quadratic and linear fit forms using the data in the intervals $m_f \in [0.009, 0.02]$ and $m_f \in [0.009, 0.015]$, respectively.
These results are presented in Fig.~\ref{fig:spect_chiral:fpi}, and a comparison of these results with those from the LSD collaboration~\cite{Appelquist:2018yqe} is provided in Appendix~\ref{sec:app-taste}.

The data for $M_\pi^2/m_f$ are also fitted by the same fit forms as $F_\pi$ and shown in Fig.~\ref{fig:spect_chiral:fpi}.
The results are presented in Table~\ref{tab:spect_chiral:pi}, and are in good agreement with those in the previous paper~\cite{Aoki:2016wnc}:
$F=0.0212(12)$$(^{+49}_{-71})$
and $M_\pi^2/m_f = 1.866(57)$,  including the possible chiral log effects for $F$.

Although the updated and previous results are consistent, our data are far from the chiral limit.
At $m_f = 0.009$ the expansion parameter of chiral perturbation theory (ChPT) for $N_f = 8$ QCD~\cite{Soldate:1989fh,Chivukula:1992gi,Harada:2003jx} is still large,
\begin{equation}
\chi = M_\pi^2/\Lambda_\chi^2= 2 N_f (M_\pi/4\pi F)^2 = \NewEnsembleChi \gg 1\,.
\end{equation}
In order to obtain a more reliable value of $F$, we would need several data points at yet smaller values of $m_f$.
The systematic error of $F$ is estimated with the same procedure we used in our previous paper~\cite{Aoki:2016wnc}, and we obtain
\begin{equation}
F = 0.0210(6)(^{+27}_{-70}) ,
\label{eq:F_chiral}
\end{equation}
where the quoted uncertainties are statistical and systematic respectively.
The central value and statistical error are determined from the quadratic fit with $m_f \in [0.009, 0.03]$. The lower systematic error is estimated from the effect of the chiral log term in next-to-leading order (NLO) ChPT, and the upper from the difference between the quadratic and linear fit results tabulated in Table~\ref{tab:spect_chiral:pi}.

\subsubsection{Chiral condensate}

We estimate the chiral condensate in the chiral limit with two analyses.
The first is a direct measurement of the chiral condensate
\begin{equation}
\langle \overline{\psi}\psi\rangle = \frac{{\rm Tr}[D^{-1}_{\rm HISQ}]}{4}.
\end{equation}
The chiral extrapolation of $\langle \overline{\psi}\psi\rangle$ using a quadratic fit in the range $m_f \in [0.009, 0.03]$
is presented in Appendix~\ref{sec:app-chiral}.

\begin{figure}[tp]
  \centering
  \includegraphics[width=0.52\textwidth]{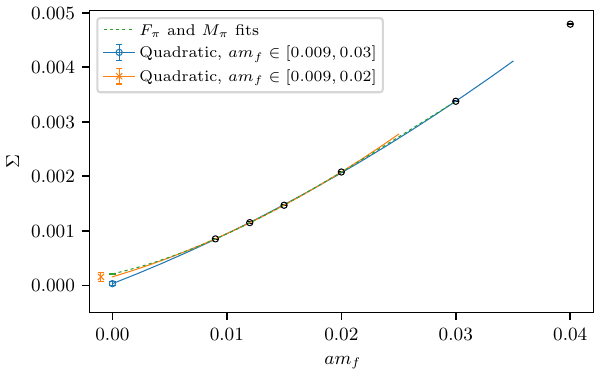}

  \caption{Chiral extrapolations of $\Sigma = F_\pi^2 M_\pi^2 / 4 m_f$ in $N_f=8$, using a quadratic fit form. The dashed line represents the estimated $m_f$ dependence of $\Sigma$ using the fit results for $F_\pi$ and $M_\pi$.}\label{fig:spect_chiral:Sigma}
\end{figure}

The second analysis is based on the Gell-Mann--Oakes--Renner (GMOR) relation,
\begin{equation}
\langle \overline{\psi}\psi\rangle|_{m_f\to 0} = \frac{F^2B}{2},
\label{eq:GMOR}
\end{equation}
where $2B = M_\pi^2/m_f$ in the chiral limit.
We calculate $\Sigma$ given by
\begin{equation}
\Sigma = \frac{F_\pi^2 M_\pi^2}{4m_f},
\label{eq:Sigma_mf}
\end{equation}
at each $m_f$.
The data in the interval $m_f \in [0.009, 0.02]$ are extrapolated to the chiral limit with a quadratic function of $m_f$, as shown in Fig.~\ref{fig:spect_chiral:Sigma}.
Comparing the fit curve with one estimated using the fit results of $F_\pi$ and $M_\pi^2/m_f$ shows a reasonably good agreement.
However, we observe a large fit range dependence of the chiral limit value, similarly to our previous study~\cite{Aoki:2016wnc}.
A fit with a wider fit range gives a smaller value in the chiral limit, as presented in Table~\ref{tab:spect_chiral:pbp}.

The fit results in the chiral limit are summarized in Table~\ref{tab:spect_chiral:pbp}.
We also tabulate the estimated value from the GMOR relation in Eq.~(\ref{eq:GMOR}), using the fit results for $F_\pi$ and $M_\pi^2/m_f$ presented in Table~\ref{tab:spect_chiral:pi}.
The fit result of the direct measurement agrees with the GMOR relation, and also with the result in the previous paper~\cite{Aoki:2016wnc} which was $\langle \overline{\psi}\psi\rangle|_{m_f \to 0} = 0.00022(4)$.
The new result, $\langle \overline{\psi}\psi\rangle|_{m_f \to 0} = \ChiralLimitPsibarPsi$, has a smaller error thanks to the wider mass range for the fit given by the addition of a precise value at a lighter fermion mass.

The result of $\Sigma$ with the shorter fit range, $m_f \in [0.009, 0.02]$, is consistent with the direct measurement, although the error is large.
It is in tension with the fit results with $\Sigma$ in our previous work~\cite{Aoki:2016wnc}, where all the extrapolated values are negative.
It is expected that the chiral extrapolation of $\Sigma$ will become more stable as we add several smaller $m_f$ data points.

\subsubsection{Masses of mesons and baryons}

\begin{figure}[tp]
   \centering
   \includegraphics{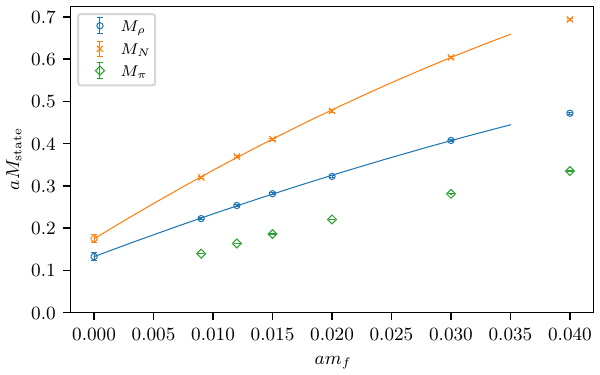}

   \caption{Chiral extrapolations of $M_\rho$ and $M_N$ in $N_f=8$, using a quadratic fit form. There is a noticeable curvature at the lightest fermion masses.}\label{fig:spect_chiral:rho-N}
\end{figure}

We extrapolate the masses of the $\rho$, $a_0$, $a_1$, $b_1$, $N$, and $N_1^*$ states to the chiral limit using a linear fit of the data for $m_f \in [0.009, 0.03]$.
The $\rho$ and $N$ states show a curvature in this range, and so are also extrapolated using a quadratic fitting form, for the same fit range, as shown in Fig.~\ref{fig:spect_chiral:rho-N}.
The results of these fits are tabulated in Table~\ref{tab:spect_chiral:hadron}; plots of the individual fits for the different mesons are included in Appendix~\ref{sec:app-chiral}.
As the $\chi^2/$ degree of freedom (dof) of the linear fit of $N$ is large, we consider only the quadratic fit result for the following discussion.
While the linear fit of $\rho$ is acceptable, we choose the quadratic fit result as the central value, and the linear fit result is used to estimate a systematic error.
The results in the chiral limit are consistent with the ones in the previous paper~\cite{Aoki:2016wnc}, except for $M_N$: this is about 20\% smaller than the previous result due to the curvature at light $m_f$.

Note that $M_\rho$ agrees with the value in the previous paper within the systematic error coming from the difference of the quadratic and linear fits.
Using the results in Table~\ref{tab:spect_chiral:hadron}, we obtain $M_{a_1}/M_\rho = \ChiralLimitMassAOneOverMassRho$, where the first and second errors are statistical and systematic ones respectively.
Here we have used the quadratic fit for $M_\rho$ and the linear fit for $M_{a_1}$ to determine the central values.
This agrees with the result in the previous paper within uncertainties\footnote{In Ref.~\cite{Aoki:2016wnc} we extrapolated linearly the ratio $M_{a_1}/M_\rho$; repeating the same analysis with our new data reproduces the same results within uncertainties.}.
Moreover, it is also compatible with the PDG value~\cite{Zyla:2020zbs} of $M_{a_1}/M_\rho = 1260/770 \approx 1.64$ corresponding to ordinary QCD.
A more detailed discussion of this ratio will be presented in a forthcoming publication~\cite{Latkmi-Sparameter}.
To obtain more stable results in the chiral limit, calculations with several lighter fermion will be necessary.
We also note that our extrapolations are not based on a rigorous effective description, but rather are meant to provide guidance for interpreting our results in the context of strongly-coupled theories near the edge of the conformal window, and as a comparison with previous results.

The ratios for the hadron masses (shown in Table~\ref{tab:spect_chiral:hadron}) to $F$ (in Eq.~(\ref{eq:F_chiral})) are presented in Table~\ref{tab:spect_chiral:ratio} and are our chiral limit predictions for the spectrum of the $N_f=8$ QCD theory.
These ratios are suggesting that this theory is in the hadronic, chirally-broken phase, but these results alone cannot establish how close the theory is to the conformal window.

Note that the central values for $\rho$ and $N$ are determined from the quadratic fit results and the systematic error quoted for $\sqrt{2}M_\rho/F$ includes the difference between the quadratic and linear fit results for $M_\rho$ shown in Table~\ref{tab:spect_chiral:hadron}.
Our result in Table~\ref{tab:spect_chiral:ratio}
\begin{equation}
\label{eq:Mrho_F_chiral}
\frac{\sqrt{2} M_\rho}{F} =8.91(66)(^{+4.44}_{-1.00}),
\end{equation}
obtained from a quadratic fit for both $M_\rho$ and $F$
is consistent with the values included in our previous works: 
$\sqrt{2} M_\rho/F= 7.7(1.5)(^{+3.8}_{-0.4})$ in \cite{Aoki:2013xza} and $\sqrt{2}M_\rho/F=10.1(0.6)(^{+5.0}_{-1.9})$ in \cite{Aoki:2016wnc}, 
where the latter was obtained from a linear extrapolation for $M_{\rho}$ and a quadratic one for $F$.

Finally, we should discuss the discretization effects and their implications for our study of the phase structure.
In the $N_f=8$ theory, the suppression of taste breaking indicates that lattice artifacts are at least reduced in the fermion sector. 
However, the overall magnitude of the discretization error is difficult to estimate, 
and with the present data we cannot definitively exclude the conformal scenario.
At the same time, our spectrum analysis provides strong evidence for chiral symmetry breaking. 
Unlike other bound states, the pion mass does not exhibit universal hyperscaling. 
Instead, the effective anomalous dimension $\gamma_m$ extracted from $M_\pi$ decreases as $m_f$ increases. 
This behavior is naturally explained by the GMOR relation,
\begin{equation}
M_\pi^2 = 4\frac{\Sigma}{F_\pi^2}m_f = \frac{4\langle \overline{\psi}\psi\rangle|_{m_f\to 0}}{F^2}m_f + \mathcal{O}(m_f^2),
\end{equation}
in which the linear term dominates near the chiral limit ($\gamma_m \simeq 1$) 
while the quadratic term controls the large-mass regime ($\gamma_m \simeq 0$). 
Our $N_f = 8$ results are fully consistent with this tendency, 
which is characteristic of NG bosons, in sharp contrast to conformal dynamics.
For $N_f=12$, by contrast, the effective $\gamma_m$ never decreases with $m_f$
and shows a universal value at small fermion masses, as confirmed for two different $\beta$ values (Fig.~33 of our previous paper~\cite{Aoki:2016wnc}.)
For the $N_f=8$ theory, it remains possible that all $\gamma_m$ values could converge near 1 at
much smaller fermion masses than we have simulated, 
making it exceedingly difficult to distinguish a chirally broken phase from a conformal one 
based solely on spectrum analysis (cfr.~\cite{LatticeStrongDynamics:2023bqp}).
Furthermore, the theory could undergo a phase transition at stronger coupling, as recently suggested in Refs.~\cite{Hasenfratz:2022qan, Witzel:2024bly}. 
Since our $N_f=8$ study is limited to a single $\beta$ and masses far from the chiral limit, 
exploring a wider range of $\beta$ values and smaller fermion masses will be essential for a more definitive conclusion, 
though such an analysis lies beyond the scope of this work.


\begin{table}[ht]
\caption{Fit results for $F$ and $M_\pi^2/m_f$ in the chiral limit of $N_f = 8$,
using a fit form $C_0 + C_1 m_f + C_2 m_f^2$.
``Linear'' and ``Quadratic'' denote
the fit forms with $C_2=0$ and $C_2 \ne 0$ respectively.
The value of $\chi^2/$dof, the number of degrees of freedom (dof) in the fits, and the fit range are also tabulated.
}\label{tab:spect_chiral:pi}

  \begin{ruledtabular}\begin{tabular}{cccccc}
 & $C_0$ & $\chi^2/$dof & dof & Fit form & $m_f$ fit range \\
\hline
$F$ & 0.02105(61) & 0.17 & 2 & Quadratic & $[0.009, 0.03]$ \\
$F$ & 0.0212(13) & 0.33 & 1 & Quadratic & $[0.009, 0.02]$ \\
$F$ & 0.02371(52) & 1.22 & 1 & Linear & $[0.009, 0.015]$ \\
$M_{\pi}^2/m_f$ & 1.892(41) & 0.24 & 2 & Quadratic & $[0.009, 0.03]$ \\
$M_{\pi}^2/m_f$ & 1.945(92) & 0.06 & 1 & Quadratic & $[0.009, 0.02]$ \\
$M_{\pi}^2/m_f$ & 1.938(39) & 0.07 & 1 & Linear & $[0.009, 0.015]$ \\
\end{tabular}\end{ruledtabular}

\end{table}

\begin{table}[ht]
\caption{Fit results for the chiral condensate
$\langle \overline{\psi}\psi\rangle$ and $\Sigma = F^2_\pi M_\pi^2/4m_f$
in chiral limit of $N_f = 8$ using a quadratic fit form
$C_0 + C_1 m_f + C_2 m_f^2$
together with the result from the GMOR relation $F^2B/2$.
The value of $\chi^2/$dof, the number of degrees of freedom (dof), and the fit range
in the fits are also tabulated.
}\label{tab:spect_chiral:pbp}

  \begin{ruledtabular}\begin{tabular}{ccccc}
 & $C_0$ & $\chi^2/$dof & dof & $m_f$ fit range \\
\hline
$\langle\bar{\psi}\psi\rangle$ & 0.000200(20) & 0.41 & 2 & $[0.009, 0.03]$ \\
$\langle\bar{\psi}\psi\rangle$ & 0.000198(42) & 0.82 & 1 & $[0.009, 0.02]$ \\
$\Sigma$ & 0.000031(39) & 1.69 & 2 & $[0.009, 0.03]$ \\
$\Sigma$ & 0.000156(79) & 0.09 & 1 & $[0.009, 0.02]$ \\
$F^2 B/2$ & 0.000210(13) & --- & --- & --- \\
\end{tabular}\end{ruledtabular}

\end{table}

\begin{table}[ht]
\caption{Fit results for hadron masses in the chiral limit of $N_f = 8$ with
a polynomial fit form $C_0 + C_1 m_f + C_2 m_f^2$ in the range $m_f \in [0.009, 0.03]$.
``Linear'' and ``Quadratic'' denote
the fit forms with $C_2=0$ and $C_2 \ne 0$ respectively.
The value of $\chi^2/$dof and the number of degrees of freedom (dof)
in the fits are also tabulated.}\label{tab:spect_chiral:hadron}

  \begin{ruledtabular}\begin{tabular}{ccccc}
 & $C_0$ & $\chi^2/$dof & dof & Fit form \\
\hline
$M_{\rho}$ & 0.1485(25) & 1.68 & 3 & Linear \\
$M_{\rho}$ & 0.1326(90) & 0.83 & 2 & Quadratic \\
$M_{a_0}$ & 0.1450(99) & 0.85 & 3 & Linear \\
$M_{a_1}$ & 0.210(12) & 1.28 & 3 & Linear \\
$M_{b_1}$ & 0.206(18) & 0.39 & 3 & Linear \\
$M_{N}$ & 0.2058(26) & 4.84 & 3 & Linear \\
$M_{N}$ & 0.1749(92) & 1.10 & 2 & Quadratic \\
$M_{N_1^*}$ & 0.273(11) & 0.00 & 3 & Linear \\
\end{tabular}\end{ruledtabular}

\end{table}

\begin{table}[ht]
\caption{Ratios of $\sqrt{2} M_H/F$ in the chiral limit.
The first and second errors are statistical and systematic errors.
$\rho$ and $N$ are evaluated from the quadratic fit results,
and others from the linear results shown in
Table~\ref{tab:spect_chiral:hadron}.
}\label{tab:spect_chiral:ratio}

  \begin{ruledtabular}\begin{tabular}{cc}
$H$ & $\sqrt{2}M_H/F$ \\
\hline
$M_{\rho}$ & $8.91(66)({}^{4.44}_{1.00})$ \\
$M_{a_0}$ & $9.74(72)({}^{4.85}_{1.09})$ \\
$M_{a_1}$ & $14.13(88)({}^{7.04}_{1.59})$ \\
$M_{b_1}$ & $13.9(1.3)({}^{6.9}_{1.6})$ \\
$M_{N}$ & $11.75(70)({}^{5.86}_{1.32})$ \\
$M_{N_1^*}$ & $18.34(92)({}^{9.14}_{2.06})$ \\
\end{tabular}\end{ruledtabular}

\end{table}


\clearpage
\newpage
\section{The flavor-singlet scalar state}\label{sec:scalar}

In this section, we report our results for the lightest flavor-singlet scalar state in the  $N_f=8$ and $N_f=4$   theories.
While the former is an update of our previous results by adding a new lightest fermion mass, the latter is new to this work. We also summarise these data, comparing both with theoretical predictions and with the non-singlet spectrum
together with the data for $N_f=12$
previously reported in
~\cite{Aoki:2012eq, Aoki:2016wnc}.

\subsection{Update for $N_f = 8$ QCD}

The mass of the flavor-singlet scalar in $N_f=8$ QCD at $\beta=3.8$, $m_f = 0.009$ is calculated in the same way as the results at larger $m_f$ presented in our previous work~\cite{Aoki:2016wnc,Aoki:2014oha,Aoki:2013zsa}.
The simulation parameters, including the number of configurations and the bin size for the jackknife analysis, are summarized in Table~\ref{tab:hadron_stat}.

\begin{figure}[tp]
   \centering
   \includegraphics[width=0.52\textwidth]{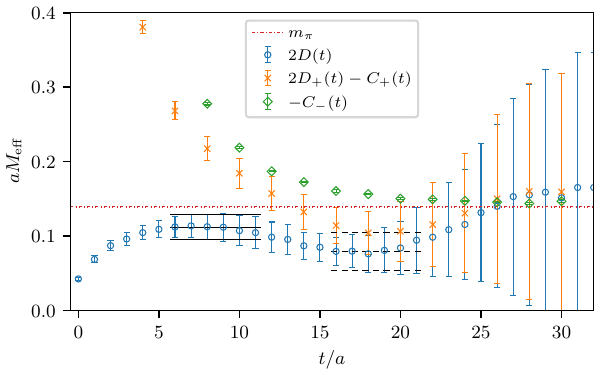}

     \caption{The effective $\sigma$ mass at $m_f = 0.009$ in $N_f = 8$. Circle and diamond symbols represent effective masses from positive parity projected and disconnected correlators, respectively. The effective mass of the connected correlator with negative parity projection is also plotted for comparison. Solid and dashed horizontal lines express fit results with one standard deviation error bands. The dot--dashed line is $M_\pi$.}\label{fig:scalar:effm}
\end{figure}

Figure~\ref{fig:scalar:effm} shows the effective mass of the $\sigma$ state at $m_f = 0.009$ evaluated from the vacuum subtracted disconnected correlator $2D(t)$.
The effective masses from the positive parity projected full correlator $2D_+(t) - C_+(t)$ and the negative parity projected connected correlator $-C_-(t)$ are also plotted, where $C_\pm(t) = 2C(t) \pm C(t+1) \pm C(t-1)$ at even $t$.
We estimate $M_\sigma$ separately from two fit ranges, $t \in [6, 11]$ and $t \in [16, 21]$.
Both results are statistically consistent with each other and also with the effective $\sigma$ mass of the full correlator.
We obtain $M_\sigma = \NewEnsembleMSigma$ as shown in Table~\ref{tab:spect48}.
The central value is determined from the result with the fit range at smaller $t$.
The result from the larger-$t$ fit range is used to estimate the systematic error, as in our previous work~\cite{Aoki:2016wnc,Aoki:2014oha,Aoki:2013zsa}.
At this $m_f$, finite volume effects are expected to be small, because the value of $L M_\sigma = \NewEnsembleLMsigma$ statistically satisfies the criterion for negligible finite volume effects, $L M_\sigma \ge 6$~\cite{Aoki:2016wnc}.

\begin{figure}[tp]
   \centering
  \includegraphics[width=0.52\textwidth]{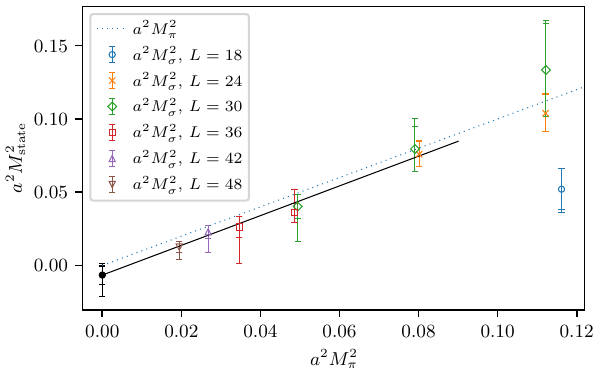}

  \caption{\label{fig:Sigmafit} The chiral extrapolation of $M_\sigma$ in $N_f = 8$, fitted with $M_\sigma^2 = d_0 + d_1 M_\pi^2$. The inner error is statistical, while the outer error represents the combined error, where statistical and systematic errors are added in quadrature. The dotted line shows the case where $M_\sigma^2 = M_\pi^2$.
  }
   \label{fig:scalar:fit_dchpt}
\end{figure}

\begin{figure}[tp]
   \centering
  \includegraphics[width=0.52\textwidth]{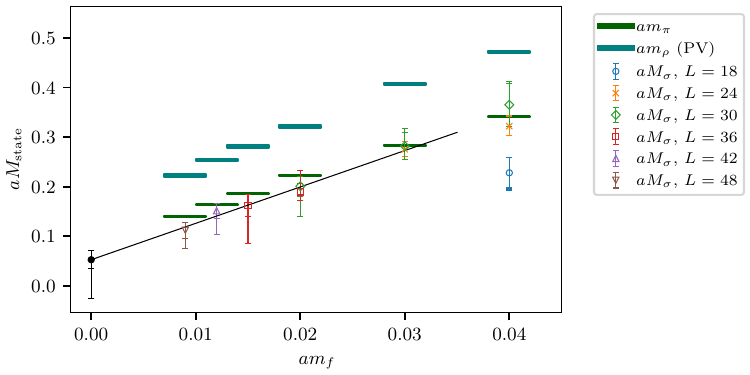}

  \caption{The chiral extrapolation of $M_\sigma$ in $N_f = 8$, fitted with $M_\sigma = c_0 + c_1 m_f$. The inner error is statistical, while the outer error represents the combined error, where statistical and systematic errors are added in quadrature. $M_\pi$ and $M_\rho$ are plotted for comparison.}\label{fig:scalar:fit_lin}
\end{figure}

We have updated chiral extrapolations of the $\sigma$ mass with this new data point.
The data are fitted using the form  from the WT identity/dChPT
 discussed in Sec.~\ref{sec:antiveneziano},
Eq.~\eqref{eq:msigma-summary},
re-presented here for convenience,
\begin{equation}
  M_\sigma^2 = d_0 + d_1 M_\pi^2\;,
  \label{eq:msigma-recap}
\end{equation}
and with an empirical linear fit form,
\begin{equation}
  M_\sigma = c_0 + c_1 m_f\;.
\end{equation}
The results of these fits are tabulated in Tab.~\ref{tab:spect_chiral:scalar},  and  plotted in Figs.~\ref{fig:scalar:fit_dchpt} and~\ref{fig:scalar:fit_lin}.
As emphasized in Sec.~\ref{sec:antiveneziano},
any possible deviations of the dChPT from the WT identity Eq.~(\ref{eq:msigma-summary})
that may arise near the chiral limit
(due to the $\pi$ loop~\cite{Matsuzaki:2013eva})
are not relevant to our lattice data,
lying as they do far from the chiral limit.

These results are consistent with the those presented previously~\cite{Aoki:2014oha, Aoki:2016wnc} with a good overlap, while the lower systematic errors are improved by more than a factor of two.

The linear fit result $c_0 = \ChiralLimitCZero$, combined with our chiral fit value of $F=F_\pi|_{m_f=0}=\ChiralLimitF$, Eq.~(\ref{eq:F_chiral}), gives $M_\sigma/(F/\sqrt{2}) =\ChiralLimitMSigmaOverFOverRootTwo$.
This is slightly smaller than,
but consistent within uncertainties,
with our previous result omitting the new lightest point~\cite{Aoki:2016wnc},
$M_\sigma/(F/\sqrt{2}) =4.2 (2.0)$$(^{1.4}_{9.5})$.
The new value
would give  $M_\sigma/v_{\rm EW} = (M_\sigma/F) \sqrt{2/N_D} \sim 1.75$ for $N_D=4$ (the  one-family model),
which remains consistent
(within uncertainties, which remain large)
with the possibility to identify $\sigma$ with the Higgs, with $M_\sigma/v_{\rm EW}\simeq \frac{1}{2}$.

The fit results for $d_0$ and $d_1$ from the WT identity and dChPt,
presented in Table \ref{newfitNf8},
also remain consistent with the previous ones~\cite{Aoki:2014oha, Aoki:2016wnc},
discussed in Eq.~(\ref{d_0d_1previous}).
As such,
the logic of Eqs.~(\ref{d_0Nf8})--(\ref{sigmamassdecayconst}) holds:
\begin{eqnarray}
M_\sigma^2 &\lesssim
&d_1 M_\pi^2\simeq M_\pi^2,
\nonumber\\
 \frac{F_\sigma}{F_\pi/\sqrt{2}}
&=&C_{\gamma_m}\cdot  \frac{1}{\sqrt{d_1}} = \ChiralLimitFourOverSqrtDOneNfEight \quad {\rm for}\quad C_{\gamma_m}=4
\;.
\label{sigmadecayconst2}
\end{eqnarray}
$d_0$ is consistent
(up to a large uncertainty, and the possible chiral log effects discussed in Sec. II)
with the value
\begin{equation}
  d_0 = M_\sigma^2|_{m_f=0} \simeq \frac{v_{\rm EW}^2}{4}= \frac{N_D}{4} \cdot \frac{F^2}{2} \simeq  \frac{N_D}{4} \cdot \ChiralLimitFSquaredOverTwo
\end{equation}
that would be required for $\sigma$ to be identified with the Higgs in the TC model with $N_D$ electro-weak doublets.

Our data for $M_\sigma^2$ vs $M_\pi^2$ (and $\gamma_m \simeq 1$)
 are also
consistent with those of the LSD Collaboration~\cite{Appelquist:2018yqe};
we provide details of the comparison in Appendix~\ref{sec:app-taste}.
As such,
 the result  Eq.~(\ref{sigmadecayconst2}) is also consistent with $F_\sigma/(F_\pi/\sqrt{2}) \simeq 3.4$ in Ref.~\cite{Appelquist:2019lgk}\footnote{
 Their fit formula, Eq.~(14) combined with Eq.~(15) in \cite{Appelquist:2019lgk}, is the same as our Eq.~(\ref{sigmamassdecayconst}), with the parameters corresponding to $C_{\gamma_m} =\sqrt{(N_f/2) y (4-y)}=4.00$ (from  $y(=3-\gamma_m)=2.06 \pm 0.05$ in their Eq.~(16)), and
 $\sqrt{d_1}=M_\sigma/M_\pi \gtrsim 1$ from their Fig.~1, thus with $F_\sigma/(F_\pi/\sqrt{2}) $ slightly smaller than ours.
 },  fit from the LSD Collaboration's data.
(See also similar estimates in Refs.~\cite{Fodor:2019vmw,Golterman:2020tdq}.)

\begin{table}[ht]
\caption{\label{tab:Sigmafit} Results of the chiral extrapolation of $M_\sigma$ in $N_f = 8$ with the fit form from
 the anomalous WT identity/dChPT, Eq.~(\ref{eq:msigma-recap}),
  and the linear fit form.
  The fit range in both cases is $m_f \in [0.009, 0.03]$. The first and second errors are statistical and systematic errors, respectively.
The value of $\chi^2/$dof and the number of degrees of freedom (dof) in the fits are also tabulated.\label{tab:spect_chiral:scalar}}

  \begin{ruledtabular}\begin{tabular}{ccc}
Fit form & $M_\sigma^2 = d_0 + d_1 M_\pi^2$ & $M_\sigma = c_0 + c_1 m_f$ \\
\hline
 & $d_0=-0.0066(62)({}^{81}_{144})$ & $c_0=0.0526(184)({}^{26}_{787})$ \\
 & $d_1=1.01(20)({}^{32}_{49})$ & $c_1=7.35(96)({}^{2.95}_{39})$ \\
$\chi^2/$dof & 0.21 & 0.34 \\
dof & 3 & 3 \\
\end{tabular}\end{ruledtabular}

\label{newfitNf8}
\end{table}

\subsection{Results for $N_f = 4$ QCD}

The $\sigma$ mass in $N_f = 4$ at $\beta = 3.8$ on $L^3\times T = 20^3\times 30$ is calculated using the same action as in $N_f = 8$.
Fermion masses are chosen to be $m_f = 0.01$, 0.02, and 0.03 in lattice units.

\begin{figure}[tp]
  \centering
  \includegraphics[width=0.52\textwidth]{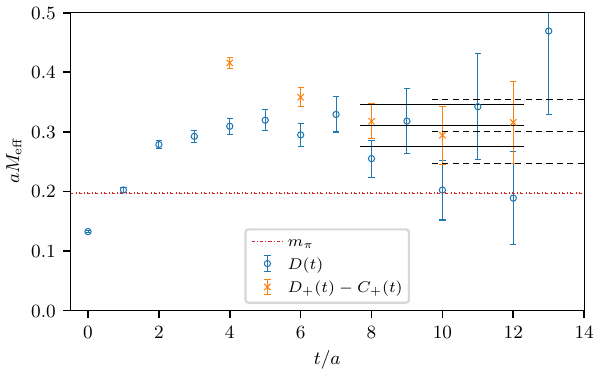}

  \caption{Effective masses for the $D_+(t) - C_+(t)$ and $D(t)$ correlators
at $m_f = 0.01$ and $\beta=3.8$ for $N_f = 4$. The dot-dashed line represents
$M_\pi$. The solid lines express the fit result of $D_+(t) - C_+(t)$ with
the one standard deviation band.}\label{fig:nf4:mf0.01:sigma}
\end{figure}

In contrast to $N_f=8$, the effective $\sigma$ mass of $D(t)$ in $N_f = 4$ has a strong oscillating behavior.
(A typical example is the data at $m_f = 0.01$, as plotted in Fig.~\ref{fig:nf4:mf0.01:sigma}.)
This is because, despite using the HISQ action, the taste symmetry breaking of the pseudoscalar masses is larger in $N_f = 4$ QCD (cfr. Fig.~\ref{fig:nf4-tastesb})~\cite{Aoki:2016wnc}.

Due to this oscillating behavior, we do not determine $M_\sigma$ from $D(t)$; instead, we calculate $M_\sigma$ from the positive parity projected correlator $D_+(t)-C_+(t)$, whose effective mass and fit result are plotted in Fig.~\ref{fig:nf4:mf0.01:sigma}\footnote{We have confirmed that the same $M_\sigma$ is obtained from the positive parity projected correlator after subtracting the contribution of the parity partner of $\sigma$.}.
Since the signal of the effective $\sigma$ mass is limited in the $N_f = 4$ case compared to the one in $N_f = 8$, we do not estimate the systematic error coming from the choice of the fit range.
In contrast to $N_f = 8$, the figure shows that $M_\sigma > M_\pi$ in $N_f = 4$, which is similar to the usual QCD~\cite{Fu:2012gf,Kunihiro:2003yj,Briceno:2016mjc,Briceno:2017qmb} and to the $N_f = 4$ QCD result reported by the LSD Collaboration~\cite{Appelquist:2018yqe} with a different choice of lattice regularization.

\begin{figure}[tp]
   \centering
  \includegraphics[width=0.52\textwidth]{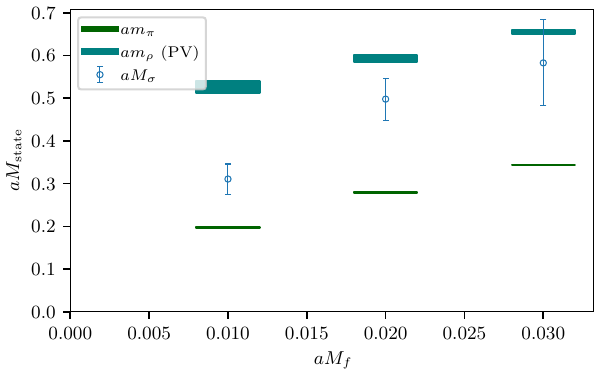}

  \caption{
   The fermion mass dependence of $M_\sigma$ in $N_f = 4$ at $\beta=3.8$, compared with $M_\pi$ and $M_\rho$.}
 \label{fig:nf4:mf_dep:sigma}
\end{figure}

We also obtain $M_\sigma$ in $m_f = 0.02$ and 0.03 using the same method. The results are tabulated in Table~\ref{tab:nf04:scalar} together with the masses for $\pi$ and $\rho$.
The extraction of $M_\sigma$ becomes harder as $m_f$ increases, as the correlator of $\sigma$ rapidly degrades as $m_f$ increases, as can be seen in Fig.~\ref{fig:nf4:mf_dep:sigma} where we plot the data in lattice units.
Similarly to the $N_f=8$ case, we extrapolate the $\sigma$ mass to the chiral limit using Eq.~\eqref{eq:msigma-recap}.

\begin{table}[ht]
\caption{Results for $M_\sigma$, $M_\pi$, $F_\pi$, and $M_\rho$
at each $m_f$ in $N_f = 4$ at $\beta=3.8$. The number of configurations in the ensemble analysed ($N_{\rm conf}$)
and the bin size ($N_{\rm bin}$) for the flavor-singlet scalar calculation
are also tabulated.
}\label{tab:nf04:scalar}

  \begin{ruledtabular}\begin{tabular}{ccccccc}
$am_f$ & $aM_{\sigma}$ & $N_{\mathrm{conf}}$ & $N_{\mathrm{bin}}$ & $aM_\pi$ & $aF_\pi$ & $aM_\rho$ \\
\hline
0.01 & $0.311(35)$ & 3750 & 150 & $0.19731(26)$ & $0.09039(17)$ & $0.526(14)$ \\
0.02 & $0.498(49)$ & 6250 & 250 & $0.27927(26)$ & $0.10487(13)$ & $0.5931(78)$ \\
0.03 & $0.58(10)$ & 6250 & 250 & $0.34384(24)$ & $0.11555(12)$ & $0.6549(56)$ \\
0.04 & $\cdots$ & $\cdots$ & $\cdots$ & $0.40030(20)$ & $0.124636(96)$ & $0.7146(46)$ \\
\end{tabular}\end{ruledtabular}

\end{table}


\subsection{Comparison of theories}
\label{subC}

\begin{figure}[hp]
  \begin{center}
  \includegraphics[width=0.52\textwidth]{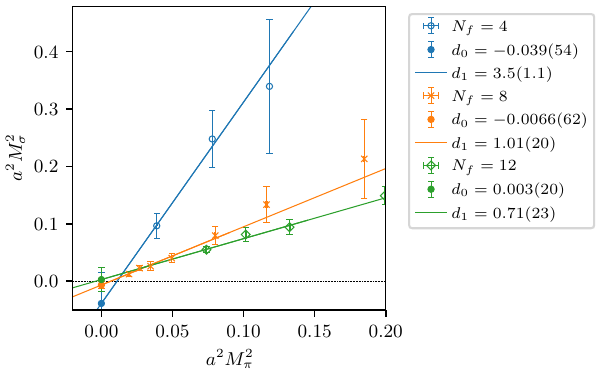}

  \end{center}
\caption{\label{fig:sigma2vspi2-1}Results for $M^2_\sigma$ for $N_f=4, 8, 12$, showing in each case the fit to Eq.~\eqref{eq:msigma-recap}.}
\end{figure}

\begin{figure}[hp]
  \begin{center}
  \includegraphics[width=0.51\textwidth]{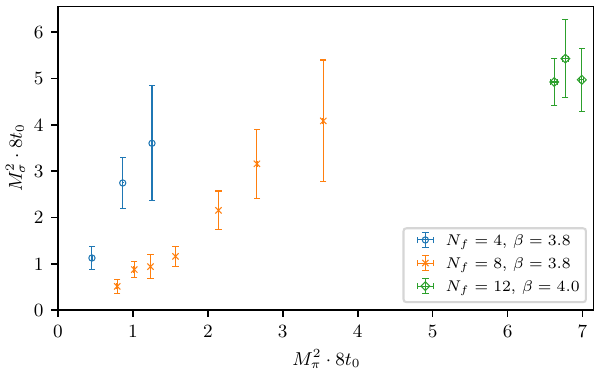}

  \end{center}
\caption{\label{fig:sigma2vspi2-8t0}Results for $M^2_\sigma$ for $N_f=4, 8, 12$ in units of $1/\sqrt{8t_0}$.}
\end{figure}

Here we present summary plots of $M^2_\sigma$ vs $M^2_\pi$ for the three theories considered.
Figure~\ref{fig:sigma2vspi2-1} shows the results in lattice units,
including the fits for $d_0, d_1$,
while Fig.~\ref{fig:sigma2vspi2-8t0} shows the results in units of $1/\sqrt{8 t_0}$.
 
\noindent
  
\noindent

Our new data,
summarized in Tab.~\ref{tab:msigma-fit-results},
indicate that $M_{\sigma}^{2}$ is dominated by the term $d_{1}M_{\pi}^{2}\gg d_{0}$,
such that $M_{\sigma}^{2} \simeq d_{1}M_{\pi}^{2}$ not only for $N_{f}=12$
but also for $N_{f}=4$ and 8
(with the caveat that our data are still some distance from the chiral limit).
As suggested in Sec.~\ref{sec:antiveneziano},
this implies that whether $M_{\sigma}^{2}>M_{\pi}^{2}$ is determined by whether $d_{1}>1$,
which may in turn be linked to whether $\gamma_{m}>1$.
The approximate equality of the power-law behaviour of $m_{f}$   for $\sigma$ as well as  $\pi$
(i.e.~the non-universal ``$\gamma_{m}$'' value~\cite{Aoki:2014oha, Aoki:2016wnc})
 may be the cause of the comparable values of the
$M_{\sigma}$ and $M_{\pi}$ in the $N_{f}=8$ theory.

 \begin{table}[hp]
  \centering
  \caption{\label{tab:msigma-fit-results}
    Summary of the fits of the data for $M_{\sigma}$ to Eq.~\eqref{eq:msigma-recap}
     in lattice units,
    including also the approximate value of $\gamma_{m}$ estimated by comparing
    the fit value for $d_{1}$ with Eq.~\eqref{eq:d1-slope}  and that from hyperscaling, both consistent with each other for $N_f=8$ and $12$.
    (We do not observe hyperscaling in the $N_f=4$ case.)
  }
  \begin{ruledtabular}\begin{tabular}{ccccc}
$N_f$ & $d_0 a^2$ & $d_1$ & $\gamma_m$ & $\left.\gamma_m\right|_{\textnormal{hyperscaling}}$ \\
\hline
4 & $-0.039(54)$ & $3.5(1.1)$ & $2.11(21)$ &  \\
8 & $-0.0066(62)({}^{81}_{144})$ & $1.01(20)({}^{32}_{49})$ & $1.01(20)$ & $\simeq 1$ \cite{Aoki:2012zwc,Aoki:2013xza,Aoki:2016wnc} \\
12 & $0.0030(202)({}^{41}_{27})$ & $0.712(229)({}^{36}_{40})$ & $0.66(31)$ & 0.4--0.5 \cite{Aoki:2012eq, Aoki:2016wnc} \\
\end{tabular}\end{ruledtabular}

\end{table}
\noindent

The fit result $d_0 \simeq 0$ within (large) uncertainties,
both for $N_f=4$ and $N_f=8$,
arises from Eq.~\eqref{eq:msigma-recap},
which is obtained from the WT identity,
and so is valid for the present lattice data with $\chi \gg 1$.
However,
in the region of the chiral limit we would expect
chiral log effects to become more significant.
As discussed in Eq.~\eqref{chirallog} of Sec.~\ref{sec:antiveneziano},
this will increase the chiral limit value of $d_0$,
relative to that obtained from these relatively heavy-mass data,
as $d_0=M_{\sigma}^{2}|_{m_f=0}>0$~\cite{Matsuzaki:2013eva}.
In contrast, the fit result $d_0\simeq 0$ in $N_f=12$   in the conformal phase has no such chiral log  corrections.

Meanwhile,
the fit value of $d_1$ has no such chiral log effects,
neither in the broken nor the conformal phase,
and  hence  the obtained result for $F_\sigma/(F_\pi/\sqrt{2})$
may be identified with that in the chiral limit,
which is of direct relevance to the comparison with
the anticipated properties of a composite Higgs.

In addition to the result of a hyperscaling fit for $N_f=8$ and $12$,
we also present in Table~\ref{tab:msigma-fit-results}
the value of $\gamma_{m}$ obtained by solving Eq.~\eqref{eq:d1-slope}
(the prediction of $d_{1}$ from holographic and linear sigma models)
for the fitted value of $d_{1}$.
For $N_f=8$ and $12$,
this is consistent with the hyperscaling result.
The larger value  $\gamma_m=\GammaMFromDOneNfFour$ in the $N_{f}=4$ case cannot be verified,
as that theory does not exhibit hyperscaling.
However,
it may be suggestive of an alternative description
of infrared QCD with $N_{f} \lesssim 4$, with $\Lambda_{\rm IR} \sim \Lambda_{\rm UV}$:
the gauged NJL model with induced strong four-fermion interaction
and effectively reduced gauge coupling $\alpha_{\rm eff} \ll 1$.
The model in the broken phase
has $\gamma_m \simeq 1+ \sqrt{1-\alpha_{\rm eff}/\alpha_{\rm cr}}  \simeq 2$ for $\alpha_{\rm eff} \ll \alpha_{\rm cr}$,
with the four-fermion operator becoming relevant~\cite{Miransky:1988gk}.
By contrast,
in the $N_f=8$ case,
$\Lambda_{\rm IR} \ll \Lambda_{\rm UV}$ is realized by the essential singularity scaling,
with the gauge coupling staying strong $\alpha_{\rm eff} \simeq \alpha_{\rm cr}$ even at induced four-fermion coupling,
such that $\gamma_m \simeq 1$.
If it is the case, from Eq.~(\ref{d_1}) we may have $F_\sigma/(F_\pi/\sqrt{2}) \simeq \sqrt{3N_f/2}/\sqrt{d_1} \simeq \sqrt{2}$ for $N_f=4$.\footnote{
  Indeed,
  we may expect this to extend to $N_{f}=2$,
  where $\gamma_{m}\simeq 2 \Rightarrow d_{1} \simeq 3$,
  and hence $F_\sigma/(F_\pi/\sqrt{2}) \simeq 1$;
  this would also be expected from linear sigma model predictions,
  as discussed in Footnote~\ref{linearsigmaNf2}.
}

Further,
we may also observe  from Fig.~\ref{fig:sigma2vspi2-8t0},
though with large uncertainty,
that
\begin{equation}
  M_{\sigma}^{\textnormal{(anomalous)}}\sqrt{8t_{0}}\simeq \frac{1}{2
   (3-\gamma_m)}\;,\quad\forall N_{f}\;,
  \label{sigmamass8t0}
\end{equation}
consistently  with the arguments in Sec.~\ref{sec:antiveneziano}, Eqs.~(\ref{gradientflowasIR}), (\ref{sigmamassbroken}), and~(\ref{sigmamass12}).

Considering now each $N_{f}$ in more detail,
in the $N_{f}=4$ case,
$M_{\sigma}$ lies between $\rho$ and $\pi$ in the observed range of $m_{f}$,
similarly to the behavior observed in $N_{f}=2+1$ QCD.
$M_{\sigma}^{2}$ decreases rapidly,
with slope $d_{1}\simeq \ChiralLimitDOneNfFour$,
towards the chiral limit.
$d_{0}$ appears very small,
comparable with the value in the $N_{f}=8$ case,
up to a factor of $1/(3-\gamma_{m})^{2}$
in units of $\Lambda_{\mathrm{IR}}=1/\sqrt{8t_{0}}$.\footnote{
  Eq.~(\ref{sigmamass8t0}) also imply that for $N_f=2$, with $\gamma_m=2$ similarly to $N_f=4$, $M_\sigma^2 \sim (M_\rho/2)^2  + 3 M_\pi^2\simeq  (450\, {\rm MeV})^2$ for the physical point $M_\pi\simeq 140$ MeV, in rough agreement with the reality for $f_0(500)$.}
However,
the relatively large uncertainties on these data
mean that this statement cannot yet be made definitively.

For $N_{f}=8$,
the $\sigma$ mass is comparable with
(with hints of being slightly smaller than)
that of the pseudo-NG boson $\pi$
in the observed range of $m_{f}$.
The values found for $d_{0}$ and $d_{1}$ obtained Eq.~(\ref{sigmadecayconst2})
are consistent with those previously found~\cite{Aoki:2014oha, Aoki:2016wnc},
with
$F_\sigma/(F_\pi/\sqrt{2}) $ also observed to be
consistent with
the holographic and linear sigma models
described in Eq.~(\ref{linearsigmarel}).

We may regard $\sigma$ in the $N_f=8$ theory as
the technidilaton in the
walking technicolor model~\cite{Yamawaki:1985zg,Bando:1986bg} with $N_D$ electroweak doublets,
which would correspond to the physical Higgs boson with
mass $M_{\textnormal{Higgs}}\simeq125\textnormal{ GeV} \simeq v_{\mathrm{EW}}/2$.
 Then
   we would expect
$d_{0} = M_{\textnormal{Higgs}}^{2}
 \simeq   N_D\cdot (F/\sqrt{2})^{2}/4 $,  which reads   $ d_0\simeq  N_D/4\cdot \ChiralLimitFSquaredOverTwo$,
 with our result $F\simeq \ChiralLimitF$ in Eq.~(\ref{eq:F_chiral}).
This is     in rough consistency with  our direct measurement of $d_0$ in Table~\ref{tab:msigma-fit-results} (up to chiral log effects, Eq.~(\ref{chirallog}), and large uncertainties).
Taken together with Eq.~\eqref{MrhovsTzero},
which reads
$d_0 \sim (2\cdot  F/\sqrt{2})^2$
(with $M_\rho/(F/\sqrt{2})\sim 8$, from Table~\ref{tab:spect_chiral:ratio}),
Eq.~(\ref{sigmamass8t0}) would imply
$d_0 \sim (1/4 \cdot 1/\sqrt{8 t_0})^2 \simeq (M_\rho/4)^2$;
that is,
the implied Higgs mass would be  roughly consistent with reality,
up to a factor of 2,
for $N_D=4$.
The result in Eq.~\eqref{sigmadecayconst2}
implies that $v_{\mathrm{EW}}/F_{\sigma}\simeq  \sqrt{N_D}/4$,
which may be compared directly with $\simeq 0.27$ from the LHC experiments
for the signal strength of the 125 GeV Higgs~\cite{Matsuzaki:2015sya}.

For $N_{f}=12$,
$d_{0}$ is also very small, 
consistent with zero  as suggested by Eq.~(\ref{sigmamass12}).
Setting $d_{0}=0$  for simplicity  in Eq.~\eqref{eq:msigma-recap}
and adopting the scale $\Lambda_{\mathrm{IR}}=1/\sqrt{8t_{0}}$
gives
\begin{equation}
  M_{\sigma}^{2}\cdot 8t_{0}=d_{1}M_{\pi}^{2}\cdot 8t_{0}\;;
\end{equation}
hyperscaling implies that
both sides of this equation should be constant in a conformal theory,
independent of the deforming mass $m_{f}$.
This is indeed what we see in Fig.~\ref{fig:sigma2vspi2-8t0}.
In this case,
the reason that $M_{\sigma}<M_{\pi}$ is that $d_{1}<1$
(in turn, via Eq.~\eqref{eq:d1-slope}, because $\gamma_{m}<1$)---
because $\sigma$ in the conformal phase is a pseudo-NG boson,
while $\pi$ is not,
as discussed in Sec.~\ref{sigmamass}.
Using the measured values of $d_{1}$ and $\gamma_{m}$,
we then find
\begin{align}
  M_{\sigma}&\simeq\sqrt{d_{1}}M_{\pi}\simeq\ChiralLimitSqrtDOneNfTwelve M_{\pi}\;,&\frac{F_{\sigma}}{F_{\pi}/\sqrt{2}}\simeq5.6-5.7\;(\textnormal{for }\gamma_{m}=0.4-0.5)\;.
\end{align}

\begin{figure}[hp]
  \centering
  \includegraphics[width=\textwidth]{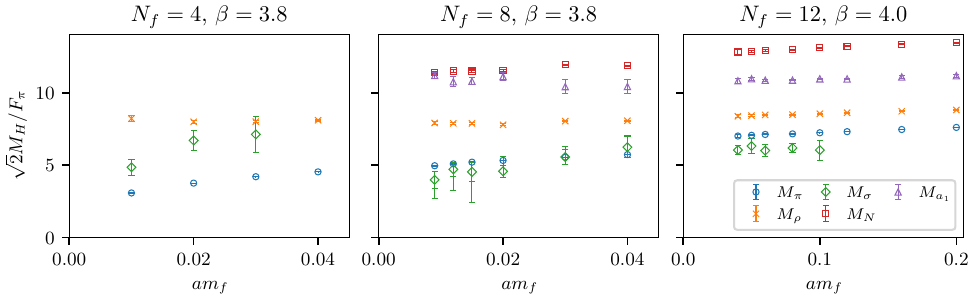}
  \caption{\label{fig:spect_ratio_Fpi}The ratio $\sqrt{2}M_H/F_\pi$ for $H=\pi,\sigma,\rho,a_1$ and $N$ as a function of $m_f$ for $N_f = 4$, 8, and 12 QCD\@,
  using the $N_f=12$ data from Appendix G of Ref.~\cite{Aoki:2016wnc} and the $N_f=4$ data from Table~\ref{tab:nf04:scalar}.}
\end{figure}

\begin{figure}[hp]
  \centering
  \includegraphics[width=\textwidth]{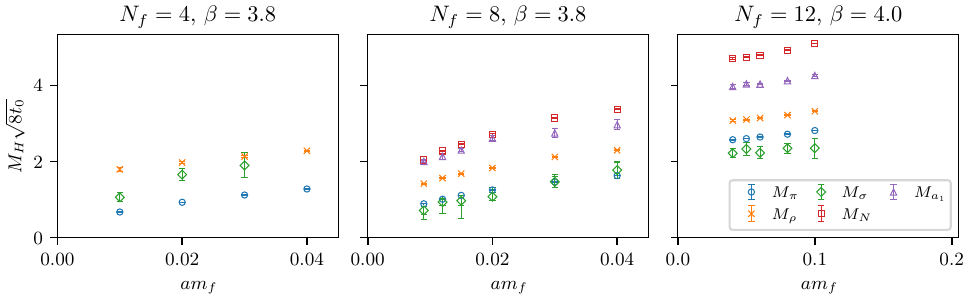}
  \caption{\label{fig:spect_ratio_s8t0}The ratio $M_H\sqrt{8t_{0}}$ for $H=\pi,\sigma,\rho,a_1$ and $N$ as a function of $m_f$ for $N_f = 4$, 8, and 12 QCD\@.}
\end{figure}

We also present summary plots of $\sigma$ mass together with non-singlet spectra
 in units of  $\sqrt{2} F_\pi$ in Fig.~\ref{fig:spect_ratio_Fpi}
and in units of $1/\sqrt{8 t_0}$ in Figs.~\ref{fig:sigma2vspi2-8t0} and~\ref{fig:spect_ratio_s8t0}.
The former are consistent with the picture  that  for $N_f=8$ the non-singlet masses behave in the same hyperscaling as $F_\pi$,  while $\sigma$ and $\pi$ do not, and for $N_f=12$ all the spectra behave in the universal hyperscaling.
In the latter case, the $m_{f}$ dependence is approximately consistent with Eq.~\eqref{MrhovsTzero}.

It is interesting to note that the ratio $\sqrt{2} M_\rho/F_\pi \simeq 8$  is almost independent of $m_f$ for all $N_f=4,8, 12$,
consistent with our previous results~\cite{Aoki:2013xza,Aoki:2016wnc}.
Its value is also close to that of $N_f=2+1$ QCD (experimental value $\approx 8.3$).\footnote{
This is roughly consistent with   the Pagels-Stokar formula
 $  (F_\pi/\sqrt{2})^2
  =  \frac{N_c}{4 \pi^2} \int_0^{\Lambda_{\rm UV}^2} 
dx x \frac{\Sigma(x)^2 - \frac{x}{4} \frac{d  (\Sigma(x)^2)}{d x} }{(x +  \Sigma( x)^2)^2} 
  \simeq (3/4\pi^2)\cdot M_F^2\simeq (M_F/4)^2$ ($M_F=m_f^{(\rm constituent)}$) with the integral dominated by $x<M_F^2$, 
  roughly independently of $N_f$,
  under  simplest ansatz for the mass function   
  $\Sigma(x)\simeq M_F  \, (x<M_F^2)$.
 This is   to be compared with
  $M_\rho \simeq 2 M_F$ ($=2 m_f^{(R)} \simeq M_\pi$ for $N_f=12$) roughly independently of $N_f$ .

}
The chiral extrapolation of this ratio for $N_f=8$, given in Eq.~\eqref{eq:Mrho_F_chiral}, is also consistent.
The Kawarabayashi-Suzuki-Ryazuddin-Fayyazuddin (KSRF) relation~\cite{Kawarabayashi:1966kd,PhysRev.147.1071} suggests that this ratio should be equal to the $\rho$--$\pi\pi$ coupling (up to a factor of $\sqrt{2}$) 
in the broken phase, which would then also be universal---independent both of $m_{f}$ and $n_{f} (< n_f^{\rm cr})$. 
A similar observation has already been reported in \cite{Appelquist:2018yqe,Nogradi:2019iek,Kotov:2021mgp},
where lattice numerical results confirm that the ratio $M_\rho/F_\pi$ remains insensitive to $N_f$ up to $N_f \leq 10$. 
We will further explore this ratio in a companion publication to this work focusing on the $S$ parameter~\cite{Latkmi-Sparameter}.

From a phenomenological perspective, in the $N_f=8$ walking technicolor scenario, 
such non-singlet spectra could appear as resonances detectable in experiments, 
while new particles predicted by many other extensions of the Standard Model have yet to be observed. 
Notably, the $\rho$ meson in the $N_f=8$ theory could lie at the order of a TeV within uncertainties, 
as indicated in Eq.~\eqref{eq:Mrho_F_chiral}.
This opens the possibility that the 2 TeV diboson excess reported by ATLAS and CMS 
could be interpreted as a contribution from technivector mesons~\cite{Fukano:2015zua, Fukano:2015hga}. 
To confirm or exclude this scenario, further progress is needed on two fronts: 
lattice simulations with lighter fermion masses near the chiral limit, 
and more precise experimental studies of TeV-scale resonances in ongoing and future LHC runs.

\clearpage
\newpage
\section{The flavor-singlet pseudoscalar state}\label{sec:pseudoscalar}

In this section, we present lattice results for the flavor-singlet pseudoscalar state for different number of flavors: $N_f=4$, 8 and 12.
Preliminary results have been reported before in conference proceedings~\cite{Aoki:2016fxd,Aoki:2017fnr}.

The flavor-singlet pseudoscalar state has quantum numbers $J^{PC}=0^{-+}$.
In $N_f=2+1$ QCD, this state is known as the $\eta^\prime$ particle, with mass $M_{\eta^\prime}= 958$ MeV.
In the following, for convenience we refer to the flavor-singlet pseudoscalar state in theories with many flavors using the same name as in QCD\@.
Witten~\cite{Witten:1979vv} and Veneziano~\cite{Veneziano:1980xs} showed that the $\eta^\prime$ mass is directly related to the contribution of the axial anomaly as:
\begin{equation}
    M_{\eta^\prime}^2= \left(M_{\eta^\prime}^2\right)^{({\rm anomalous})} +M_\pi^2\;.
\end{equation}
As discussed in Sec.~\ref{sec:antiveneziano},
in the Veneziano limit, where $N_c\rightarrow \infty$ with $N_c \alpha=$ fixed and $ n_f\equiv N_f/N_c ={\rm fixed} \ll 1$, where $\Lambda_{\rm IR}
\sim \Lambda_{\rm UV}$
and gluon loops dominate, we have
\begin{equation}
    \left(M_{\eta^\prime}^2\right)^{({\rm anomalous})}
    /\Lambda_{\rm IR}^2\sim \left(M_{\eta^\prime}^2\right)^{({\rm anomalous})}
     /\Lambda_{\rm UV}^2\sim  n_f
     \ll 1\;.
\end{equation}
Hence the $\eta^\prime$ has a parametrically small mass
 in the chiral limit $M_\pi^2=0$, and behaves as a pseudo NG boson.
 In the realistic $N_f=2+1$, $N_c=3$ QCD with $m_s\gg m_d\gg m_u$, however, we have
 \begin{equation}
     \left(M_{\eta^\prime}^2\right)^{({\rm anomalous})}
      \simeq M_{\eta^\prime}^2 + M_\eta^2- 2M_K^2 \simeq (0.726\, {\rm  GeV})^2\;,
 \end{equation}
 and hence~\cite{Veneziano:1979ec}
 \begin{equation}
     \left(M_{\eta^\prime}^2\right)^{({\rm anomalous})}
      /\Lambda_{\rm IR}^2 \sim \left(M_{\eta^\prime}^2\right)^{({\rm anomalous})}
      /M_\rho^2  \simeq 1\,  (\sim  n_f
      )\;.
 \end{equation}
Thus the $\eta^\prime$ behaves as a pseudo-NG boson only in the idealized case of the Veneziano limit where  $n_f \ll 1$,
even in the chiral limit.

Conversely,
as we described in Sec.~\ref{sec:antiveneziano},
large $N_f=4,8,12 $ QCD are close to  the anti-Veneziano limit, where $ n_f\equiv N_f/N_c=$ fixed $\gg 1$ for $N_c \rightarrow \infty$ with $N_c \alpha=$ fixed,  in which case  $\Lambda_{\rm IR}
\ll \Lambda_{\rm UV}$, and fermion loops dominate over gluon loops in the computation of the chiral anomaly, and we have~\cite{Matsuzaki:2015sya}
\begin{equation}
    \left(M_{\eta^\prime}^2\right)^{({\rm anomalous})}
    /\Lambda_{\rm IR}^2\sim   n_f^2
    \gg 1\;.
\end{equation}
Then we have
as in Eq.~(\ref{eq:R-definition}):
\begin{equation}
  M_{\eta^\prime}^2/\Lambda_{\rm IR}^2
  \simeq ({{M_{\eta^\prime}}^{({\rm anomalous})}} /\Lambda_{\rm IR})^2
  \propto n_f^2 \gg 1\;.
\end{equation}

It is notoriously difficult to compute the mass of the flavor-singlet pseudoscalar, because the pion contribution to the two-point function on the lattice is statistically challenging to remove.
In our calculations we design a gluonic operator (without fermion fields) to construct the correlation function.
This bypasses the challenge, because a gluonic operator does not couple directly to $\pi$ states.
The same method has been adopted in $N_f=2+1$ QCD and has led to results in agreement with experiment~\cite{Fukaya:2015ara}.

In practice, the gluonic interpolating operator used in our calculations is the topological charge density.
On the lattice, the topological charge density operator is defined through the clover-plaquette field strength tensor $G^{\mu \nu}(x)$:
\begin{equation}
  \label{eq:qtopo}
  q(x) = \frac{1}{32\pi} \epsilon_{\mu \nu \rho \sigma} \Tr G^{\mu \nu}(x) G^{\rho \sigma}(x) ,
\end{equation}
The two-point function $\langle q(x) q(y) \rangle$ is computed for all pairs of points $(x, y)$ in the four-dimensional volume $L^3 \times T$.
Using a Fast Fourier Transform (FFT) allows this to be done efficiently, obtaining a full-volume object in a single calculation step.
Moreover, because of translation invariance, the two-point correlator only depends on the distance $r=|x-y|$ and we average all contributions at fixed distance to increase statistics and reduce discretization effects due to the breaking of rotational symmetry.
For a particle freely propagating in four dimensions~\cite{Shuryak:1994rr}, the correlator takes the form
\begin{equation}
  \label{eq:correlator2}
  C(r)=\frac{A}{r^{1.5}}\left(1+\frac{3}{8r}\right)e^{-M_{\eta^\prime}r}
\end{equation}
at large distances $r \rightarrow \infty$.
We fit the lattice data of $C(r)$ to the form in Eq.~\eqref{eq:correlator2}, and we choose a specific fitting window of distances $r \in [r_{\rm min},r_{\rm max}]$ to extract the two parameters $A$ and $M_{\eta^\prime}$ corresponding to the propagating single-particle ground state.
\textit{A posteriori} we check that the fitting window is at large enough $r$ that the form in Eq.~\eqref{eq:correlator2} is valid.
For the fitting procedure we use the \texttt{lsqfit} Python package~\cite{lsqfit}, and we assign large (uninformative) Bayesian priors on the parameters, based on the analysis of effective masses at large $r$.

Additionally, it is well known that ultraviolet fluctuations strongly affect the operator in Eq.~\eqref{eq:qtopo}.
We utilize the gradient flow~\cite{Luscher:2010iy} of the Symanzik action as a smearing technique to remove the UV components and smooth the operator before constructing the two-point function.
In other words, the interpolating operator in Eq.~(\ref{eq:qtopo}) is used to construct the correlator, but $G^{\mu \nu}(x)$ is computed for several values of the flow time $t_w$.
We choose $t_w$ values in units of the lattice spacing $a^2$ lying in the interval $[0,3]$ and in steps of 0.15.
This is a large range and we expect that the operator $q(x)$ at some smearing $t_w$ will have an approximate physical size with good overlap to the ground state.
This physical size can be identified with a smearing scale $s_w = \sqrt{8t_w}$.
As a consequence, we obtain a large number of correlators $C_{t_w}(r) = -\langle q_{t_w}(x) q_{t_w}(y) \rangle$.
The statistical fluctuations are dramatically reduced by the gradient flow smearing, such that correlators at larger $t_w$ can be easily fitted to the exponential form in Eq.~\eqref{eq:correlator2}.
Moreover, the autocorrelation effects on $C_{t_w}(r)$ are negligible.
However, the smearing scale $s_w$ introduces systematic corrections that have to be addressed~\cite{Bruno:2014ova}.
In fact, it turns out that the dominant source of uncertainty in extracting $M_{\eta^\prime}$ comes from systematic effects of the fitting procedure.
There are mainly two competing effects:
\begin{enumerate}[leftmargin=*]
\item Eq.~(\ref{eq:correlator2}) can only be assumed to be valid in a specific region of large $r$, where the ground state dominates: in other words we should make sure that the extracted mass does not depend on the value of $r_{\rm min}$.
\item The correlator at large distances suffers from larger statistical fluctuations and can be extracted only at large values of the smearing scale $s_w$, where smearing artifacts~\cite{Bruno:2014ova} are larger: in other words we should check that the extracted mass does not depend on $r_{\rm max}$.
\end{enumerate}

\begin{figure}[thb]
  \centering
  \includegraphics{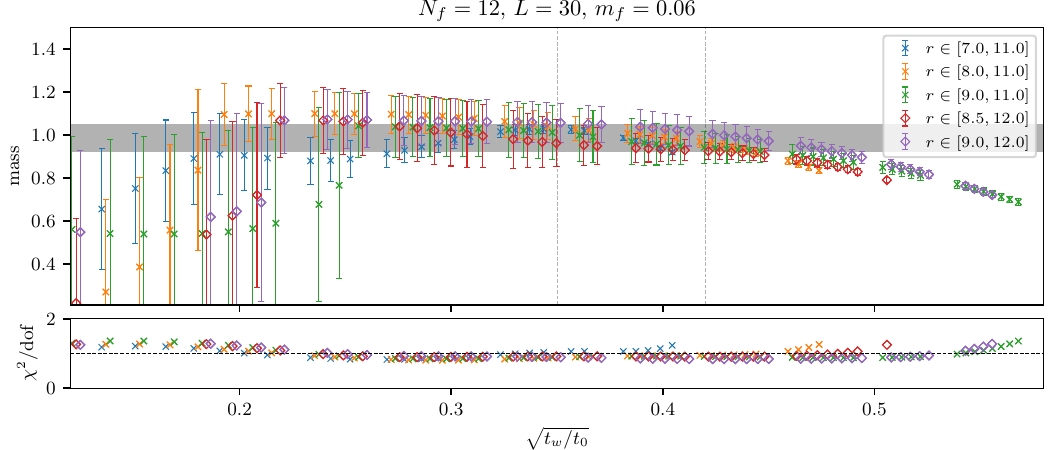}

  \caption{\label{fig:eta-prime-syst}The $\eta^\prime$ mass fitted for different distance regions and smearings, for a specific ensemble of $N_f=12$ QCD. The error bars on the horizontal axis are much smaller than the symbols, and points for different fitting ranges are shifted for clarity. $\sqrt{8t_0}=7.2680(99)$ for this ensemble. The gray band shows the estimate of the systematic error, computed as described in the text.}
\end{figure}

We estimate the systematic uncertainty on $M_{\eta^\prime}$ by looking for a plateau in the fitting range $[r_{\rm min},r_{\rm max}]$, and for a plateau in the smearing range $s_w = \sqrt{8t_w}$.
Representative examples of such plateaux for one ensemble of $N_f=12$ QCD are shown in Fig.~\ref{fig:eta-prime-syst}.
Note that $s_w$ is always smaller than the fit range boundary $r_{\rm min}\geq7$ we typically choose, and that $r_{\rm max}$ is always less than half the lattice spatial extent $L$.
When the smearing range $s_w$ is considered in units of the characteristic radius given by the gradient flow scale $\sqrt{8t_0}$, we seem to find a common region of $\sqrt{t_w/t_0}$ values for all the ensembles at fixed $N_f$ where the fitted mass does not change within the statistical uncertainty.
This region does not appear to depend on the fermion mass nor on the volume: it corresponds to the interval $[0.69,0.81]$ for $N_f=4$, $[0.39,0.51]$ for $N_f=8$, and $[0.35,0.41]$ for $N_f=12$.
The difference in these intervals might be related with the nature of the flavor-singlet pseudoscalar state as $N_f$ is increased because such a smearing scale $s_w$ corresponds to some physical scale for the operator with the best coupling to the ground state.
In the identified region, we take the difference between the largest and the smallest fitted mass as an estimate of the systematic error.
The statistical errors of the individual points are typically smaller than or of the same order as this systematic uncertainty.
Additional plots for the rest of the ensembles are collated in Appendix~\ref{sec:app-etafits}.

\begin{figure}[bht]
  \centering
  \includegraphics[width=\textwidth]{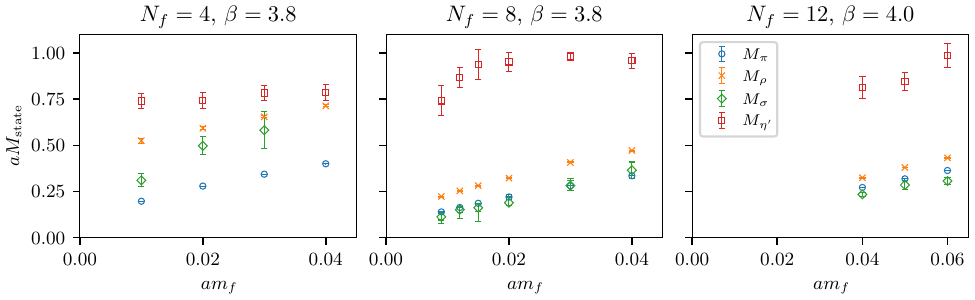}

  \caption{\label{fig:spectrum-all}The flavor-singlet scalar and pseudoscalar spectrum compared to flavor-non-singlet pseudoscalar and vector spectrum for $N_f=4$, 8 and 12. The $\eta^\prime$ mass has error bars that reflect a large systematic uncertainty. For the other states only statistical errors are reported.}
\end{figure}

\begin{figure}[h]
\begin{center}
  \includegraphics[width=0.52\textwidth]{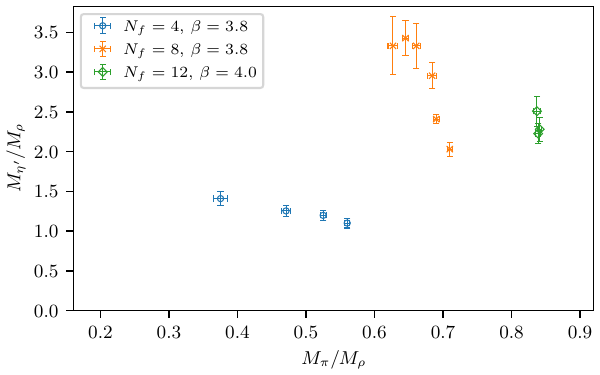}
\end{center}

\caption{\label{fig:nfALL_latkmi_eta_ratio_rho} Comparison of the results for $M_{\eta^\prime}$ between the theories with $N_f=4, 8, 12$ when normalised by $M_\rho$.}
\end{figure}

We can now compare the flavor-singlet pseudoscalar state with the rest of the low-lying spectrum as we change the number of flavors.
The results obtained for $N_f=4$, 8 and 12 QCD are summarized in Fig.~\ref{fig:spectrum-all}
in units of $1/a$.
We identify a notable increase in the gap between the flavor-singlet pseudoscalar and the vector meson,  visible in  Fig.~\ref{fig:nfALL_latkmi_eta_ratio_rho}. In $N_f=4$ QCD the mass ratio $M_{\eta^\prime}/M_{\rho}$ is close to one  and gradually increases towards the chiral limit, while it grows  rapidly from near 2 to $\sim 3-4$ for $N_f=8$, and  it stays around 2.5
for $N_f=12$.
We shall return to this point later.

\begin{figure}[thb]
  \centering
  \includegraphics[width=0.52\textwidth]{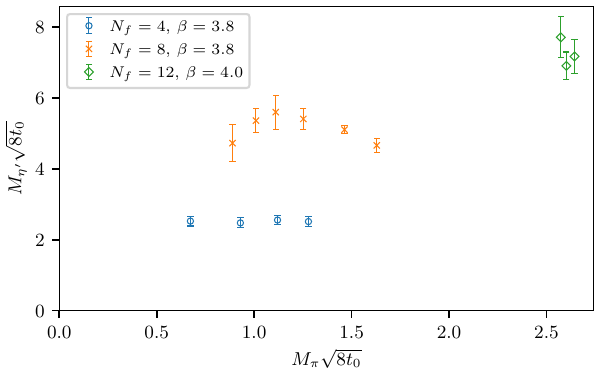}

  \caption{\label{fig:eta-prime-all}Comparison of the flavor-singlet pseudoscalar mass for $N_f=4$, 8 and 12 as a function of the pion mass.
  The hadronic masses are in units of the gradient flow scale $1/\sqrt{8t_0}$ for the various $N_f$ theories.
  Different quark mass regions are explored for different $N_f$ values, in particular for $N_f=12$.}
\end{figure}

\begin{figure}
\begin{center}
  \includegraphics[width=0.52\textwidth]{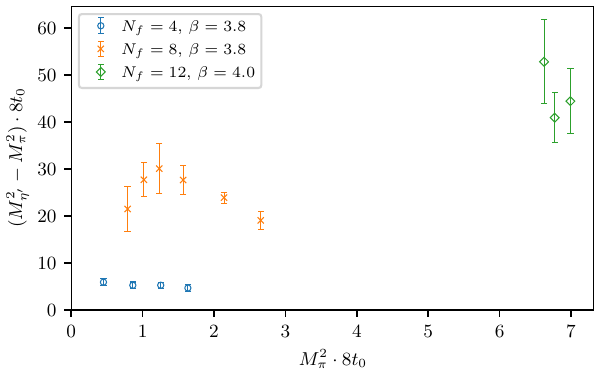}
\end{center}

\caption{\label{fig:nfALL_latkmi_eta_ratio_8t0} Comparison of the difference $M_{\eta^\prime}^2-M_\pi^2$ normalised by the gradient flow scale $\sqrt{8t_0}$.
}
\end{figure}

In Sec.~\ref{sec:antiveneziano} we discussed the expectation that
\begin{equation}
  \frac{{M_{\eta^{\prime}}^{2}}^{\textnormal{(anomalous)}}}{\Lambda_{\mathrm{IR}}^{2}}\sim  n_{f}^{2} \;.
\end{equation}
As anticipated in that discussion,
we do not see this proportionality
when we choose $M_{\rho}$ as the IR scale,
as is shown in Fig.~\ref{fig:nfALL_latkmi_eta_ratio_rho}.
None of the values of $N_{f}$ gives flat behavior,
and the $N_{f}=12$ data are not higher than $N_{f}=8$ as would be predicted.

However,
when $\Lambda_{\mathrm{IR}}=1/\sqrt{8t_{0}}$ is chosen instead,
we obtain the result shown in Fig.~\ref{fig:eta-prime-all}.
In this case,
the resulting behavior for each $N_{f}$ is significantly flatter,
and the values obtained in the three cases display the predicted scaling.
Specifically,
we observe that
\begin{equation}
  M_{\eta^{\prime}}^{2} \cdot 8t_{0}
  \simeq {M_{\eta^{\prime}}^{2}}^{\textnormal{(anomalous)}} \cdot 8t_{0}
  \simeq \begin{cases}
      (2.5)^{2} & N_{f}=4\;, \\
      (5.0)^{2} & N_{f}=8\;, \\
      (7.5)^{2} & N_{f}=12\;,
    \end{cases}
    \label{etaprimet0}
\end{equation}
independently of $m_{f}$, in close agreement with the expectation Eq.~(\ref{eq:R-definition}) in  the anti-Veneziano limit
as shown in Fig.~\ref{fig:nfALL_latkmi_eta_ratio_8t0}.

In view of this result,
we may now gain some further understanding of the gross features of $M_{\eta^\prime} / M_\rho$,
shown in Fig.~\ref{fig:nfALL_latkmi_eta_ratio_rho}.
From Eq.~\eqref{MrhovsTzero},
we expect that in the broken phase,
$M_\rho \sqrt{8t_0}\rightarrow 1$ towards the chiral limit,
and $\rightarrow 2$ towards the heavy-fermion limit,
while we have observed that $M_{\eta^\prime}$  is roughly constant.
For $N_f=4$ and 8,
moving to larger $M_\pi/M_\rho$ in Fig.~\ref{fig:nfALL_latkmi_eta_ratio_rho} corresponds to
dividing the constant $M_{\eta^{\prime}}\sqrt{8t_0}$
by $M_{\rho}\sqrt{8t_0}\rightarrow 2$,
and so giving $M_{\eta^\prime}/M_{\rho} \simeq 1.25$ and $2.5$ respectively,
which is approximately what we observe.
Moving to smaller $M_\pi/M_\rho$ meanwhile takes the opposite limit,
and in Fig.~\ref{fig:nfALL_latkmi_eta_ratio_rho}
we do indeed see the data move in this direction,
although the limit is not reached.
Meanwhile,
the data for $N_f=12$ are constant within uncertainties,
as we would expect.

\clearpage
\newpage
\section{Summary}\label{sec:summary}

We have investigated the mass spectra of the SU(3) gauge theory with $N_f=4$,
$8$, and $12$ fundamental flavors (4-, 8-, and 12-flavor QCD), using a
first-principles lattice gauge theory analysis. This analysis used the
tree-level improved Symanzik gauge action and the Highly Improved Staggered
Quark (HISQ) fermion action.

In the case of $N_f = 8$ QCD, we observe approximate conformality, which we
interpret as indicating the theory is in the hadronic phase just below the
conformal window; i.e. the chiral limit shows walking
dynamics  with a light flavor-singlet scalar ``$\sigma$'' as a pseudo NG boson (pseudo dilaton) of the scale symmetry, broken spontaneously and explicitly by the same origin of the spontaneous mass generation, 
a candidate for the composite Higgs~\cite{Aoki:2016wnc,Aoki:2014oha}.
This property and the $N_f = 8$
flavor structure match the simplest version of the walking technicolor (WTC)
model, which is advocated as a viable candidate for a theory of physics
beyond the standard model (BSM).
Regarding the walking dynamics, the mass
spectra at small fermion mass $m_f$ are of particular interest. These have been
investigated in this paper by updating our previous
results~\cite{Aoki:2016wnc,Aoki:2014oha} with a new lightest fermion mass: $m_f
= 0.009$ (in lattice units). We have compared the $N_f = 8$ spectra with those
in $N_f = 4$ QCD\@; the latter is close to real-world QCD, showing spontaneous
chiral symmetry breaking, but without the walking dynamics. For $N_f =
4$, we have updated our data to include a finer lattice spacing. This has
allowed us to extract the flavor-singlet scalar spectrum in $N_f = 4$. 

We also compared  the spectrum for $N_f=4,8$ on the same footing with the previous results of $N_f=12$ ~\cite{Aoki:2012eq, Aoki:2016wnc}, which was shown to be the
conformal phase without spontaneous chiral symmetry breaking in contrast to $N_f=4,8$.

Moreover,
we have examined the flavor-singlet pseudoscalar spectra  $``\eta^\prime$'' for all $N_f = 4$, 8,
and 12 QCD, and discussed the $N_f$ dependence of the spectra in terms of the
scaling law emerging in the anti-Veneziano limit  described in Sec.~\ref{sec:antiveneziano}.

For $N_f = 8$, we have confirmed that two important properties remains intact in
the updated data: first, the improvement in the HISQ action has continued to
allow the taste violation to be suppressed to 5\% or less of the Goldstone pion
mass,
and second, the ratio $M_\rho/M_{\pi}$ increases with decreasing $m_f$,
consistently with the spontaneous breaking of the chiral symmetry.
We have also
observed this ratio increasing similarly for $N_f = 4$ QCD\@.

In $N_f = 8$, we have evaluated the chiral limits of $F_{\pi}$, $M_{\pi}/m_f$,
$M_{\rho}$, $M_{a_0}$, $M_{a_1}$, $M_{b_1}$, $M_N$, and $M_{N^*_{\bf 1}}$ with a
polynomial ansatz in $m_f$. All of their chiral limit values have been positive
finite with good fit qualities, and are consistent with our previous
studies~\cite{Aoki:2016wnc}. For the chiral condensate, we have investigated
three different observables: $\langle\bar{\psi}\psi\rangle$, $\Sigma$, and $F^2B/2$, as defined in
Eqs.~(\ref{eq:GMOR})--(\ref{eq:Sigma_mf}). With the inclusion of the new data at
the lightest value of $m_f$, the chiral limit values for these three observables
are now consistent when a linear fit ansatz is used, as is expected
theoretically. While previously, the central value of the $\Sigma$ extrapolation was
negative~\cite{Aoki:2016wnc}, the inclusion of the updated spectral data gives a
positive result for the central value. Thus, the update has provided further
evidence of the broken chiral symmetry for $N_f = 8$.

We have observed that the flavor-singlet scalar mass $M_\sigma$ in the new ensemble
for $N_f = 8$ at $m_f=0.009$,
similarly to previous observations at heavier
$m_f$~\cite{Aoki:2016wnc,Aoki:2014oha}, is significantly smaller than the other
hadron masses except the pion.
At this new mass in particular, $M_\sigma$ is  still even smaller
than the pion mass $M_\pi$,
similarly to  the previous result,
and in contrast to the expectation in the chiral limit that  $M_\sigma>M_\pi=0$.
Conversely, $M_\sigma$ in $N_f = 4$ tends to remain
 larger than
$M_\pi$,  though getting closer, with decreasing $m_f$.
The light $M_\sigma$ in $N_f = 8$ indicates a dilatonic feature and a stable
sigma-meson, which contrasts sharply with real-world QCD\@.
We have performed
chiral extrapolations of the light $M_\sigma$ data both with a simple linear ansatz,  $M_\sigma = c_0 + c_1 m_f$,
and
with the Ward-Takahashi (WT) identity for the scale symmetry, $M_\sigma^2=d_0 + d_1 M_\pi^2$, Eq.~(\ref{eq:msigma-summary}).
The results are given together with those for $N_f=4, 12$ in Table~\ref{tab:msigma-fit-results}.
Both extrapolations
for $N_f=8$ resulted in
masses consistent with the discovered Higgs boson mass, 125 GeV  $\simeq v_{\rm EW}/2=\sqrt{N_D/4} \cdot F/\sqrt{2}$.
Owing to the
updated data, the lower systematic errors are improved by more than a factor of
two.

Eq.~(\ref{eq:msigma-summary}) may
also be derived by the dilaton chiral perturbation
theory (dChPT) in the broken phase, up to chiral log effects that are irrelevant to our data points, lying as they do far from the chiral limit.
However,
chiral log effects will raise the value of $M_\sigma$ near the chiral limit,
and as such a value estimated from a fit form excluding them
(and extrapolated from data in the region where they are not present)
will by necessity underestimate the chiral limit value $d_0$.
This is reflected in our data;
the central value for the fit of $d_0$ is negative,
although the large uncertainties still do allow for a positive value.

On the other hand, the fit value of $d_1$ is identified with that in the chiral limit where chiral log effects disappear, and so is the ratio $F_\sigma/(F_\pi/\sqrt{2})$ as  in Eq.~(\ref{d_1}),
which is most relevant to the Higgs phenomenology.
It was
obtained from the observed value of $d_1$
for $N_f=8$ to give Eq.~(\ref{sigmadecayconst2}):
\begin{equation}
 F_\sigma/(F_\pi/\sqrt{2})= \ChiralLimitFourOverSqrtDOneNfEight \quad {\rm for}  \quad \gamma_m=1,
   \end{equation}
  which is consistent with the previous results
$F_\sigma/(F_\pi/\sqrt{2})\simeq  4$~\cite{Aoki:2016wnc,Aoki:2014oha}.
This implies
$v_{\rm EW}/(F_\sigma|_{m_f=0}) \simeq  \sqrt{N_D}/4$ to be compared with  $v_{\rm EW}/F_\sigma|_{m_f=0} \simeq 0.27$, the value from the LHC data when $\sigma$ identified with the Higgs~\cite{Matsuzaki:2015sya}, where
$v_{\rm EW} =250 $ GeV and $N_D$ is the number of weak doublets of the WTC model.

More importantly,  the WT identity formula  Eq.~(\ref{eq:msigma-summary})  is valid in both the broken  (at least for our data points, which lie away from the chiral limit)  and the conformal phase,  thus is our basic framework to compare $N_f=4,8,12$ on the same footing.
The measured values of $d_1$ for $N_f=8$ and
 $N_f=12$, when combined with the measured $\gamma_m$ through hyperscaling, gives values of $F_\sigma/F_\pi/\sqrt{2}$ consistent with those from linear sigma and holographic models, Eq.~(\ref{linearsigmarel});
this in turn gives the relation between
 $d_1=(1+\gamma_m)/(3-\gamma_m)$, Eq.~(\ref{eq:d1-slope}), consistent with the fit values, as summarised in Table~\ref{tab:msigma-fit-results}.
It is remarkable that for $N_f=12$, $\sigma$ with $M_\sigma^2 \simeq 0.7 M_\pi^2$ is definitely lighter than $\pi$ in spite of the P-wave
bound state compared with S wave $\pi$;
this is consistent with the picture discussed in Sec.~\ref{sec:antiveneziano} that $\sigma$ is a pseudo NG boson while $\pi$ is not.
  If this relation is also valid for $N_f=4$ without fit value of $\gamma_m$ due to the lack of
 hyperscaling,
the  large value of $d_1$  would imply $\gamma_m \simeq \GammaMFromDOneNfFour$, suggesting the system to be effectively described by the gauged NJL model with reduced effective gauge coupling in the infrared.
If this relation is valid also in the whole conformal phase where SD equation gives $\gamma_m=1-\sqrt{1-\alpha_*/\alpha_{\rm cr}}$, then we may predict  $M_\sigma^2 \simeq d_1 M_\pi^2 =M_\pi^2\,\, (n_f=n_f^{\rm cr})$ further down to  $ M_\sigma^2 \simeq d_1 M_\pi^2= 1/3\cdot M_\pi^2 \,\, (n_f=16.5/3) $.

Our focus in this work has been on the flavor-singlet pseudoscalar meson
$\eta^{\prime}$. We have extracted the $\eta^{\prime}$ correlator from the two-point function of
the topological charge density, $\langle q(x)q(y)\rangle$, where pion contaminations are
absent. To increase the statistics, the correlator has been expressed as a
function of the distance $r = |x - y|$, and all contributions at a fixed $r$
averaged. We have utilized the gradient flow smearing technique to remove
ultraviolet (UV) noise and achieve enhanced overlap with the ground state.

We have estimated $M_{\eta^{\prime}}$ from a plateau region $[r_{\rm min},r_{\rm max}]$
in the correlator. To avoid lattice spacing, smearing, and finite size
artifacts, the plateau region was restricted to values of $r$ much larger than
the smearing scale (which in turn was much larger than the lattice spacing), and
much smaller than the spatial lattice extent, in all cases. Within these
constraints, $M_{\eta^{\prime}}$ was observed to be independent of the exact choice of
the plateau region and the smearing scale. We have found such a parameter region
for all our ensembles. In particular, a suitable range of smearing scale was
found that was independent of $m_f$ for each $N_f$.

There are theoretical expectations~\cite{Matsuzaki:2015sya} that the mass
$M_{\eta^{\prime}}$ scales with $N_f$, as a consequence of an anti-Veneziano limit where
$n_f\equiv N_f/N_c \gg 1$ is fixed as $N_c\rightarrow \infty$  with $N_c \alpha$ fixed, based on the WT identity for the flavor-singlet axial-vector current, Eq.~(\ref{WTforetaprime}) with Eqs.~\eqref{non},~\eqref{WTI}.
For each $N_f =4,8,12$, $M_{\eta^{\prime}}$ was found to be
heavier than the other hadrons, and less dependent on $m_f$. We have adopted the
gradient flow scale $\sqrt{8t_0}$ as a common measure to compare the theories with
different $N_f$. The normalized mass $M_{\eta^{\prime}}\cdot \sqrt{8t_0} \sim n_f$  was observed to
increase as a function of $n_f$,  Fig.~\ref{fig:eta-prime-all} and Eq.~(\ref{etaprimet0}):
\begin{equation}
  M_{\eta^{\prime}}^{2} \cdot 8t_{0}
  \simeq \begin{cases}
      (2.5)^{2} & N_{f}=4\;, \\
      (5.0)^{2} & N_{f}=8\;, \\
      (7.5)^{2} & N_{f}=12\;,
       \end{cases}
\end{equation}
 independently of $m_f$,
which is suggestive of anti-Veneziano scaling  in Eq.~(\ref{eq:R-definition}).
This is the core result of this paper.

Several subjects remain to be studied in future work. For $N_f=8$ QCD, there are
number of reasons to study a region of $m_f$ even smaller than the $m_f=0.009$
to which this study was extended: to achieve a stable chiral limit value of the
chiral condensate $\Sigma$; to obtain a more precise value for the chiral limit value
of the dilatonic $\sigma$ mass in the chiral limit, in order to confirm (or
otherwise) the consistency with the discovered Higgs mass; and to verify whether
the pion becomes lighter than the Higgs-identified dilaton, which for a walking
theory is expected to happen in the vicinity of the chiral limit. Another
important topic is to explore the mass spectra in the continuum limit. To this
end, the simulations presented here need to be extended to multiple smaller
lattice spacings.
Infra-red dominance and the suppression of taste violations by
the HISQ action means that the majority of the spectra of $N_f = 8$ QCD are not
expected to be very sensitive to the lattice spacing;
in particular, we anticipate this being the case for the $\eta^{\prime}$ mass.
To study the
anti-Veneziano scaling discussed above in detail, continuum and chiral
extrapolations of $M_{\eta^{\prime}}$ are needed for all of $N_f = 4$, 8, and 12.

We believe that our first-principles results indicating the walking dynamics in
$N_f = 8$ QCD will provide a very important guide for BSM model building with
the composite Higgs perspective.

\clearpage
\newpage
\section*{Acknowledgements}
We thank the Lattice Strong Dynamics (LSD) collaboration for sharing their plots and their preliminary unpublished results for the flavor-singlet scalar mass on their ensembles. 
We gratefully acknowledge Dr.~Masafumi Kurachi and Dr.~Kei-ichi Nagai for their collaboration during the early stages of this work.
Numerical calculations have been carried out on the high-performance computing systems at KMI (${\Large\varphi}$), at the Information Technology Center in Nagoya University (CX400), and at the Research Institute for Information Technology in Kyushu University (CX400 and HA8000) both through the HPCI System Research Projects (Project ID: hp140152, hp150157, hp160153) and through general use.\\
This work is supported by the JSPS Grants-in-Aid for Scientific Research (S) No. 22224003, (C) No. 16K05320 (Y.A.) for Young Scientists (A) No.16H06002 (T.Y.), (B) No.25800138 (T.Y.), (B) No.25800139 (H.O.), (B) No.15K17644 (K.M.).
E.B.~acknowledges the support of the UKRI Science and Technology Facilities Council (STFC) Research Software Engineering Fellowship EP/V052489/1,
the EPSRC ExCALIBUR programme ExaTEPP (project EP/X017168/1),
the STFC Consolidated Grant No.\ ST/T000813/1, and the Supercomputing Wales programme, which is part-funded by the European Regional Development Fund (ERDF) via Welsh Government.
K.M. is supported by the OCEVU Labex (ANR-11-LABX-0060) and the A*MIDEX project (ANR-11-IDEX-0001-02), funded by the ``Investissements d'Avenir'' French government program and managed by the ANR.
H.O. is supported in part by the JSPS KAKENHI (Nos.~21K03554, 22H00138).
T.Y. is supported in part by Grants-in-Aid 
for Scientific Research (Nos.~19H01892, 23H01195, 23K25891) and
MEXT as ``Program for Promoting Researches on the Supercomputer Fugaku''
Grant Number JPMXP1020230409.
This work is supported by the JLDG constructed over the SINET6 of NII.\\

\section*{Open access}
For the purpose of open access, the authors have applied a Creative Commons Attribution (CC BY) licence to any author accepted manuscript version arising.

\section*{Data availability statement}
Raw and processed data generated during the preparation of this work are available at Ref.~\cite{datapackage}.

\section*{Software availability statement}

This work was in part based on the MILC collaboration’s public lattice gauge theory code,
version 7~\cite{milc7}.
The workflows used to transform raw data into the results presented
are available at Ref.~\cite{datapackage}.

\bibliography{eta-prime_latkmi}
\appendix
\section{Details of lattice simulations at $m_f=0.009$}\label{sec:app-lattice}

The simulation parameters of the new ensemble are summarized in Table~\ref{tab:spect48:Nconf}.
The measurements of the hadron spectra are carried out with the parameters shown in Table~\ref{tab:hadron_stat}.
The statistical error is estimated by the jackknife analysis.

\begin{table*}[ht]
\caption{\label{tab:spect48:Nconf}
 Parameters of the new ensemble in $N_f=8$.
 $L$ and $T$ for the spatial and temporal size for $L^3\times T$
 lattice, staggered fermion mass $m_f$, molecular dynamics time step
 $\Delta\tau$, number of masses for the Hasenbusch preconditioning
 $N_{m_{\rm H}}$, values of Hasenbusch masses $m_{\rm H}^i$,
 and maximum number of thermalized trajectories $\overline{N}_{\rm Traj}^{\rm max}$ in $\tau = 0.5$ unit
 are shown for each ``stream''.
 $N_{\rm Traj}^{\rm max}$ denotes maximum number of thermalized trajectories
 in $\tau = 1$ unit as in other ensembles~\cite{Aoki:2016wnc}.
}

\begin{ruledtabular}\begin{tabular}{cccccccccccc}
$L$ & $T$ & $m_f$ & $\tau$ & $\Delta\tau$ & $N_{m_{\mathrm{H}}}$ & $m_{\mathrm{H}}^1$ & $m_{\mathrm{H}}^2$ & $m_{\mathrm{H}}^3$ & $m_{\mathrm{H}}^4$ & $\overline{N}_{\mathrm{Traj}}^{\mathrm{max}}(N_{\mathrm{Traj}}^{\mathrm{max}})$ & Stream \\
\hline
48 & 64 & 0.009 & 0.5 & 0.0025 & 4 & 0.2 & 0.4 & 0.6 & 0.8 & 864(432) & 1 \\
 &  &  &  &  &  &  &  &  &  & 10240(5120) & 2 \\
\end{tabular}\end{ruledtabular}

\end{table*}

\begin{table*}[!tbp]
\caption{
Numbers for trajectories ($N_{\rm Traj}$) in $\tau = 1$ unit,
stream ($N_{\rm str}$),
configuration ($N_{\rm conf}$),
for spectrum and singlet scalar ($\sigma$) measurements in
the $N_f = 8$ new ensemble.
The bin size of jackknife analysis ($N_{\rm bin}$) and
number of measurements per configuration ($N_{\rm meas}$)
are also summarized.
}
\label{tab:hadron_stat}

  \begin{ruledtabular}\begin{tabular}{ccccccccc}
Meas & $L$ & $T$ & $m_f$ & $N_{\mathrm{Traj}}$ & $N_{\mathrm{str}}$ & $N_{\mathrm{conf}}$ & $N_{\mathrm{bin}}$ & $N_{\mathrm{meas}}$ \\
\hline
Spectrum & 48 & 64 & 0.009 & 5440 & 2 & 680 & 68 & 8 \\
$\sigma$ & 48 & 64 & 0.009 & 5120 & 1 & 2560 & 160 & 64 \\
\end{tabular}\end{ruledtabular}

\end{table*}

The hadron masses for $\pi$, $\rho$, $N$ and their parity partners, $a_0$, $a_1$, $b_1$, $N^*_{\bf 1}$, at $m_f = 0.009$ are tabulated in Table~\ref{tab:spect48}.
The table also shows the results for the decay constant $F_\pi$ and the chiral condensate $\langle \overline{\psi}\psi\rangle$.
The notations and calculation methods for the hadron spectra are described in our previous paper~\cite{Aoki:2016wnc}.
As in the paper, we observe a small taste symmetry breaking in the pion mass as shown in Table~\ref{tab:taste_pion48}.

\begin{table}[ht]
\caption{Results of hadron spectra and $M_\sigma$ at $m_f = 0.009$ on $L=48$
in $N_f = 8$.}
\label{tab:spect48}

  \begin{ruledtabular}\begin{tabular}{cc}
$M_\pi$ & $0.13950(56)$ \\
$F_\pi$ & $0.03971(20)$ \\
$\langle \bar \psi \psi \rangle$ & $0.0055516(54)$ \\
$M_\rho$ & $0.2225(24)$ \\
$M_{a_0}$ & $0.2355(71)$ \\
$M_{a_1}$ & $0.3144(59)$ \\
$M_{b_1}$ & $0.3196(86)$ \\
$M_N$ & $0.3202(22)$ \\
$M_{N_1^{*}}$ & $0.4156(65)$ \\
$M_{\sigma}$ & $0.112(17)({}^{0}_{33})$ \\
\end{tabular}\end{ruledtabular}

\end{table}

\begin{table*}[ht]
\caption{Mass of the NG pion and the taste partners $M_{\pi_\xi}$ at $m_f = 0.009$ on $L=48$ in $N_f = 8$.}
\label{tab:taste_pion48}

  \begin{ruledtabular}\begin{tabular}{cccccccc}
$\xi_5$ & $\xi_4\xi_5$ & $\xi_i\xi_5$ & $\xi_i\xi_4$ & $\xi_i\xi_j$ & $\xi_4$ & $\xi_i$ & $\xi_I$ \\
\hline
0.13950(56) & 0.14062(56) & 0.14042(54) & 0.14112(59) & 0.14131(58) & 0.14181(60) & 0.14198(57) & 0.14263(54) \\
\end{tabular}\end{ruledtabular}

\end{table*}

\begin{figure}[thb]
  \centering
  \includegraphics[width=\textwidth]{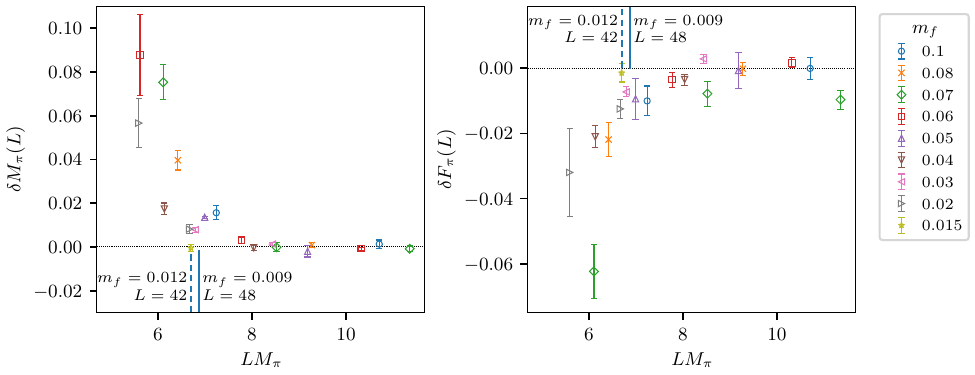}

  \caption{Updated figures for $\delta M_\pi(L)$ and $\delta F_\pi(L)$ defined in Eq.~(\ref{eq:fv_rel}) from the one in the previous paper~\cite{Aoki:2016wnc}. In this paper we do not use fermion masses heavier than $m_f=0.04$ and they are reported here only for comparison with our previous results. Points with error bars show the difference between the specified quantity at the given lattice size $L$ and at the largest volume considered for that value of the bare mass. Different symbols and colors represent different bare masses as indicated in the legend. The two vertical lines in blue represent the two values of mass at which only a single volume was considered: the solid line is $m_f=0.009$, $L=48$, and the dashed line $m_f=0.012$, $L=42$.}
  \label{fig:spect_chiral:vol}
\end{figure}

In the previous study~\cite{Aoki:2016wnc}, we found that the finite volume effects for $M_\pi$ and $F_\pi$ are negligible, when $LM_\pi \gtrapprox 7$ at smaller $m_f$ by using
\begin{equation}
\delta M_\pi(L) = \frac{M_\pi(L)-M_\pi(L_{\rm max})}{M_\pi(L_{\rm max})}
\ \ {\rm and}\ \
\delta F_\pi(L) = \frac{F_\pi(L)-F_\pi(L_{\rm max})}{F_\pi(L_{\rm max})}
\ ,\label{eq:fv_rel}
\end{equation}
with $L_{\rm max}$ being the largest lattice volume at each $m_f$.
Figure~\ref{fig:spect_chiral:vol} is the updated figure from the previous study~\cite{Aoki:2016wnc}.
The figure shows that the value of $LM_\pi$ at $m_f = 0.009$, which is the vertical solid line, is similar to the one at $m_f = 0.012$.
From this observation, we expect that finite volume effects in the new data are similarly suppressed as at higher $m_f$ values.

\section{Details of the new chiral extrapolations with $m_f=0.009$}\label{sec:app-chiral}

\begin{figure}[tp]
   \centering
   \includegraphics[width=0.52\textwidth]{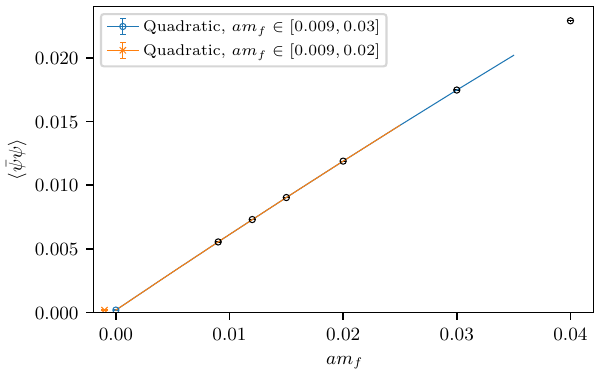}

   \caption{Chiral extrapolations of the chiral condensate $\langle \overline{\psi}\psi\rangle$ in $N_f = 8$. The range of $m_f$ considered for each fit curve
is noted in the legend.}
   \label{fig:spect_chiral:pbp}
\end{figure}

In Fig.~\ref{fig:spect_chiral:pbp} we show the chiral extrapolations for  $\langle \overline{\psi}\psi\rangle$, supplementing the one of the GMOR relation presented in Sec.~\ref{sec:spectrum}.

\begin{figure}[tp]
   \centering
  \includegraphics[width=0.52\textwidth]{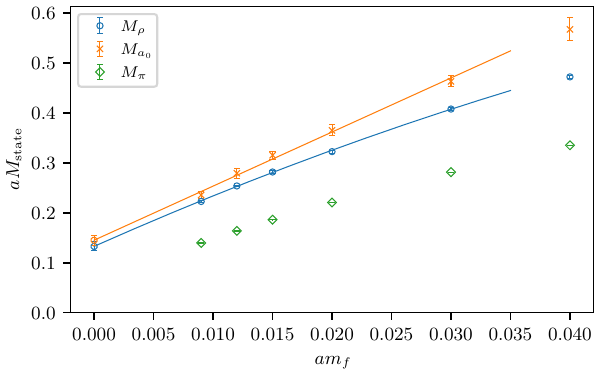}

  \caption{Chiral extrapolations for $M_\rho$ and $M_{a_0}$ in $N_f = 8$. Data for $M_\pi$ are also plotted for comparison.}
   \label{fig:spect_chiral:rho_a0}
\end{figure}

\begin{figure}[tp]
   \centering
  \includegraphics[width=0.52\textwidth]{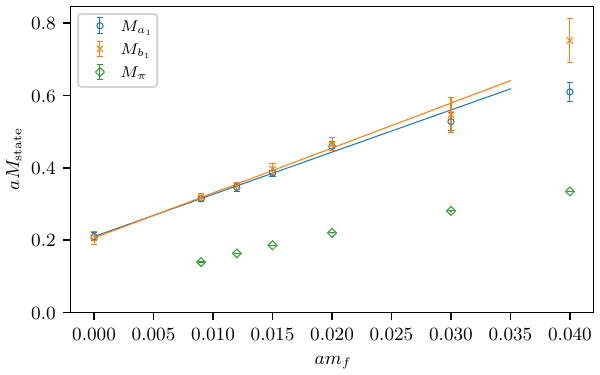}

  \caption{Chiral extrapolations for $M_{a_1}$ and $M_{b_1}$ in $N_f = 8$. Data for $M_\pi$ are also plotted for comparison.}
   \label{fig:spect_chiral:a1_b1}
\end{figure}

\begin{figure}[tp]
   \centering
  \includegraphics[width=0.52\textwidth]{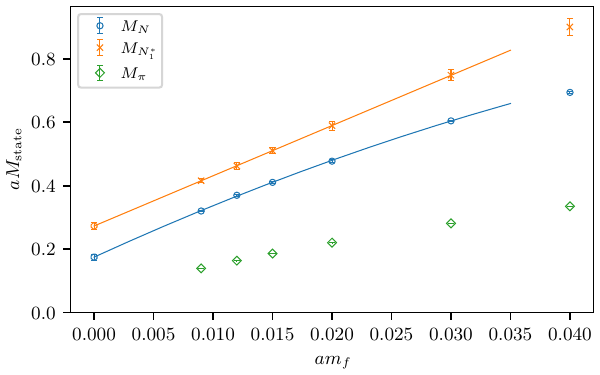}

  \caption{Chiral extrapolations for $M_{N}$ and $M_{N_{\bf 1}^*}$ in $N_f = 8$. Data for $M_\pi$ are also plotted for comparison.}
   \label{fig:spect_chiral:N_Nstar}
\end{figure}

In Figs.~\ref{fig:spect_chiral:rho_a0}-\ref{fig:spect_chiral:N_Nstar} we show the chiral extrapolations of the meson masses including the new $m_f=0.009$ mass point.

\clearpage
\section{Taste-symmetry breaking effects}\label{sec:app-taste}

\begin{figure}[thb]
  \centering
  \includegraphics{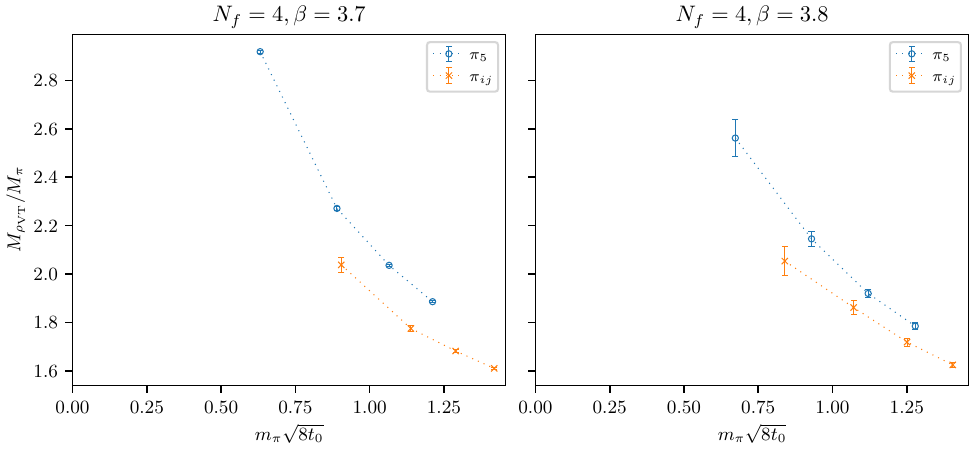}

  \caption{The ratio $M_{\rho}/M_{\pi}$ is shown for $N_f=4$ QCD using different fermion taste combinations for the pseudoscalar meson. Results from two lattice spacings are shown in the two panels. The ratio of masses is shown using the Goldstone pion $\pi_{5}$ and the heaviest pseudoscalar meson $\pi_{ij}$. The horizontal axis uses the pion mass rescaled in units of the gradient flow scale $\sqrt{8t_0}$.}
  \label{fig:nf4-tastesb}
\end{figure}

We report two different lattice spacings for $N_f=4$ and show the ratio $M_{\rho}/M_{\pi}$ using different states for the pseudoscalar meson: the lightest (Goldstone boson) state $\pi_5$ and the heaviest state $\pi_{ij}$ (which comes from operators with tensor taste structure $\xi_i \xi_j$).
Both states are going to be degenerate in the continuum limit, and we can see the reduction in taste symmetry breaking effects when moving from a coarser ($\beta = 3.7$) to a finer ($\beta = 3.8$) lattice spacing.
The results are shown in Fig.~\ref{fig:nf4-tastesb} as a function of the pion mass, $\pi = \pi_5$ or $\pi_{ij}$, in units of the gradient flow scale.

\begin{figure}[tp]
  \centering
  \includegraphics[width=\textwidth]{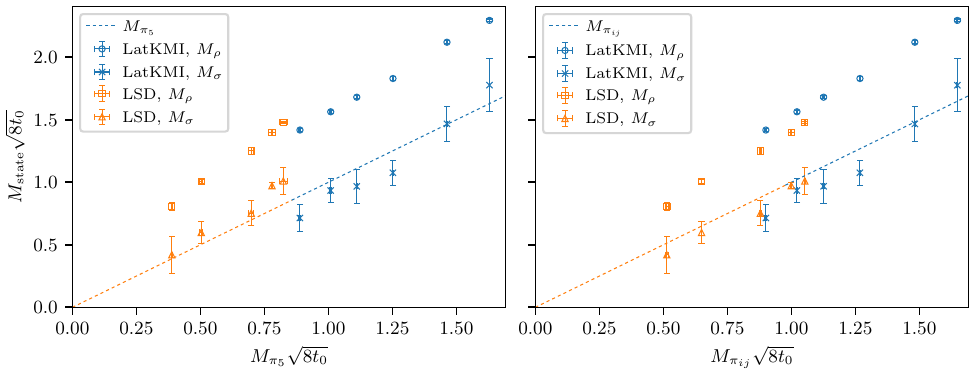}

  \caption{Comparison of the flavor-singlet scalar mass and the vector meson mass between the LatKMI collaboration and the LSD collaboration for $N_f=8$. The pion mass in units of the gradient flow scale is used on the x axis. The left panel uses the Goldstone pion $\pi_5$ while the right panel uses the tensor taste structure pion $\pi_{ij}$. The corresponding pion mass is used to draw the dashed line.}
   \label{fig:comparison_LSD-nf8}
\end{figure}

\begin{figure}[tp]
  \centering
  \includegraphics[width=\textwidth]{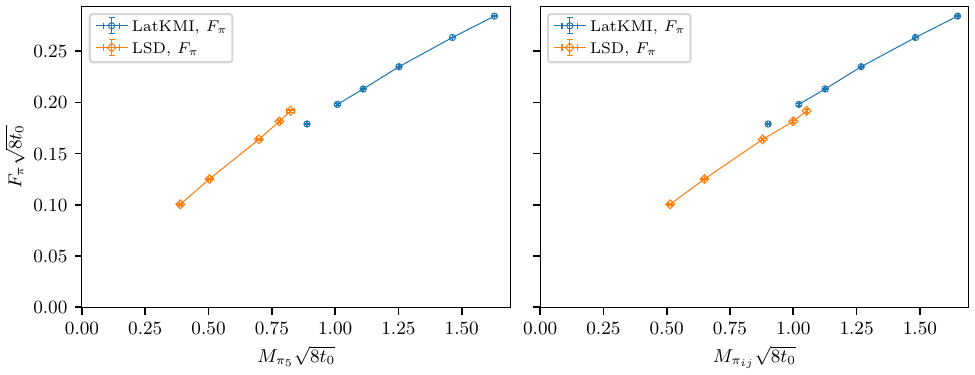}

  \caption{Comparison of the pion decay constant between the LatKMI collaboration and the LSD collaboration for $N_f=8$. The pion mass in units of the gradient flow scale is used on the x axis. The left panel uses the Goldstone pion $\pi_5$ while the right panel uses the tensor taste structure pion $\pi_{ij}$. The decay constant has been normalized to the 131 MeV convention in usual QCD~\cite{Aoki:2016wnc}.}
   \label{fig:comparison_LSD-nf8-decay}
\end{figure}

\begin{figure}[tp]
  \centering
  \includegraphics[width=\textwidth]{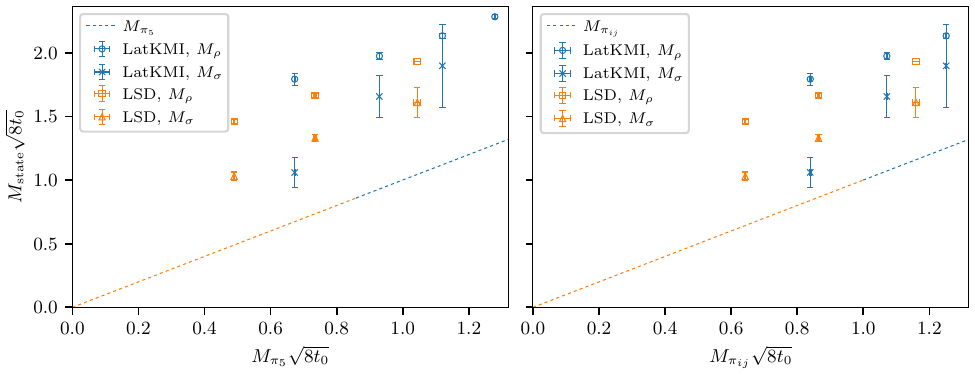}

  \caption{Comparison of the flavor-singlet scalar mass and the vector meson mass between the LatKMI collaboration and the LSD collaboration for $N_f=4$. The pion mass in units of the gradient flow scale is used on the x axis. The left panel uses the Goldstone pion $\pi_5$ while the right panel uses the tensor taste structure pion $\pi_{ij}$. The corresponding pion mass is used to draw the dashed line. The LSD data is only at $\beta=6.6$.}
   \label{fig:comparison_LSD-nf4}
\end{figure}

At the smallest mass considered, the results for $N_f=8$ align more closely with those from the LSD collaboration~\cite{Appelquist:2018yqe}
when compared using a common scale, given by the Nambu-Goldstone pion mass in units of the gradient flow scale $\sqrt{8 t_0}$.
This is illustrated in Fig.~\ref{fig:comparison_LSD-nf8}.
However, the two collaborations use different lattice actions, which can introduce discretization effects, including variations in taste breaking.
We noted in Sec.~\ref{sec:lattice-setup} that the taste splitting in our data remains small.
We may observe this by comparing the pion masses of different taste channels.
For example, if we replace the Goldstone pion mass $M_{\pi_5}$ with the heavier tensor taste pion mass $M_{\pi_{ij}}$, 
our data do not appreciably change.
In contrast, the data from the LSD collaboration shift significantly towards heavier masses, 
as shown in Figs.~\ref{fig:comparison_LSD-nf8} and \ref{fig:comparison_LSD-nf8-decay}, 
making the compatibility between the two datasets even more striking.
The results for $N_f = 4$ are also shown in Fig.~\ref{fig:comparison_LSD-nf4}.
Based on the scale of $t_0$, it appears that the LSD results for $N_f=4$ correspond to a finer lattice spacing, 
which likely reduces taste-breaking effects.


\clearpage
\section{Fits of the flavor-singlet psudoscalar}\label{sec:app-etafits}
\subsection{$N_f=4$}
\begin{figure}[thb]
  \centering
  \includegraphics{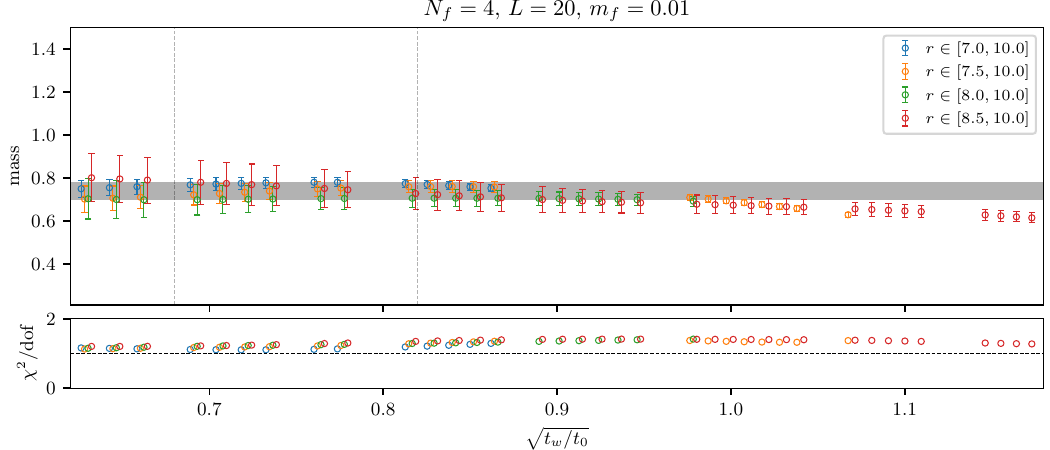}

  \caption{The $\eta^\prime$ mass fitted for different distance regions and smearings, for a specific ensemble of $N_f=4$ QCD with $L=20$ and $m_f=0.01$.}
  \label{fig:nf4_L20_mf0-01_compare_smearing_rmax10-00_with_sys}
\end{figure}

\begin{figure}[thb]
  \centering
  \includegraphics{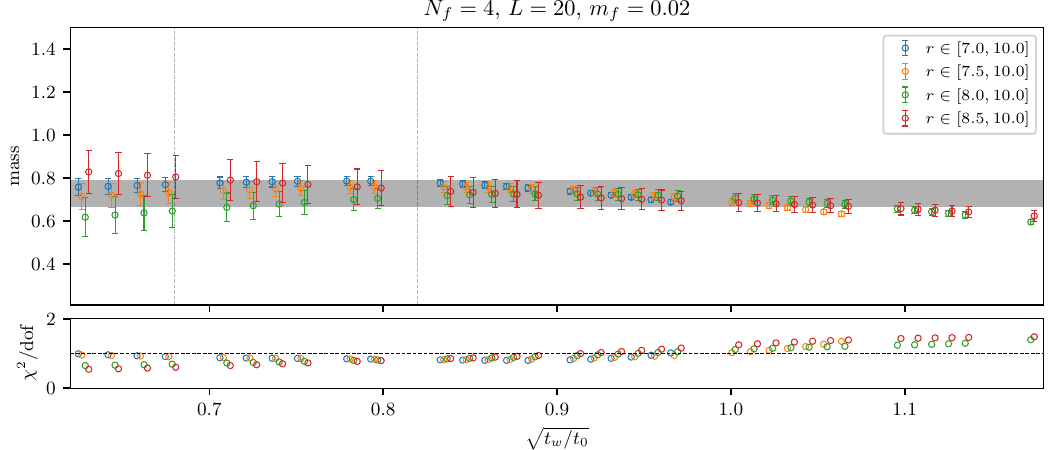}

  \caption{The $\eta^\prime$ mass fitted for different distance regions and smearings, for a specific ensemble of $N_f=4$ QCD with $L=20$ and $m_f=0.02$.}
  \label{fig:nf4_L20_mf0-02_compare_smearing_rmax10-00_with_sys}
\end{figure}

\begin{figure}[thb]
  \centering
  \includegraphics{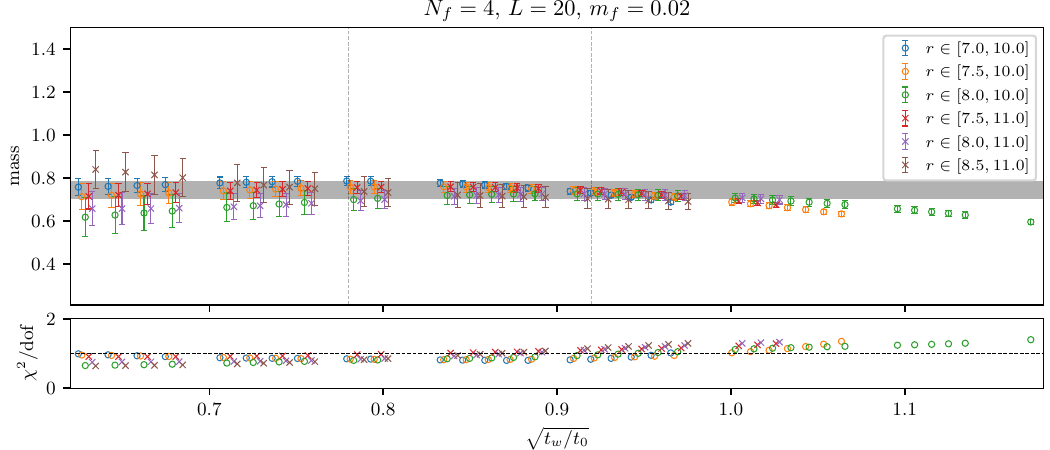}

  \caption{The $\eta^\prime$ mass fitted for different distance regions and smearings, for a specific ensemble of $N_f=4$ QCD with $L=20$ and $m_f=0.02$. Different $r_{\rm max}=11$.}
  \label{fig:nf4_L20_mf0-02_compare_smearing_rmax11-00_with_sys}
\end{figure}

\begin{figure}[thb]
  \centering
  \includegraphics{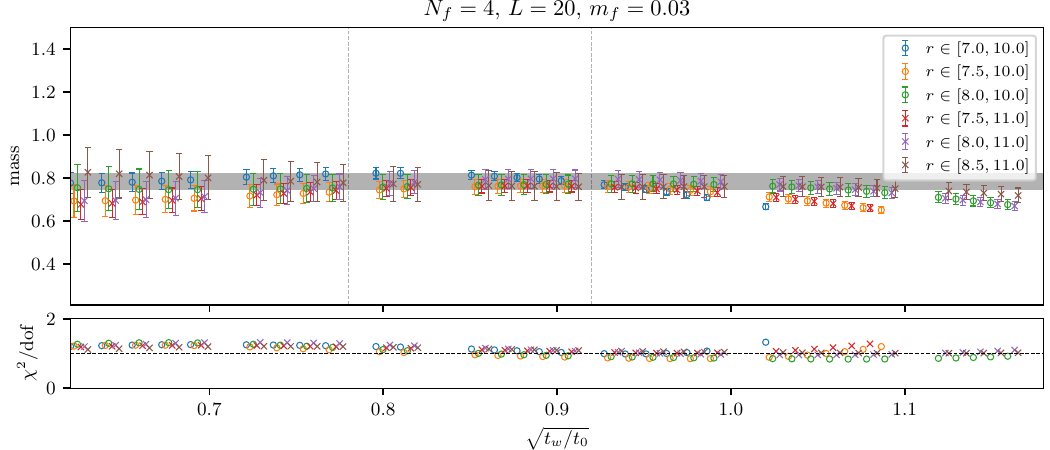}

  \caption{The $\eta^\prime$ mass fitted for different distance regions and smearings, for a specific ensemble of $N_f=4$ QCD with $L=20$ and $m_f=0.03$.}
  \label{fig:nf4_L20_mf0-03_compare_smearing_rmax11-00_with_sys}
\end{figure}

\begin{figure}[thb]
  \centering
  \includegraphics{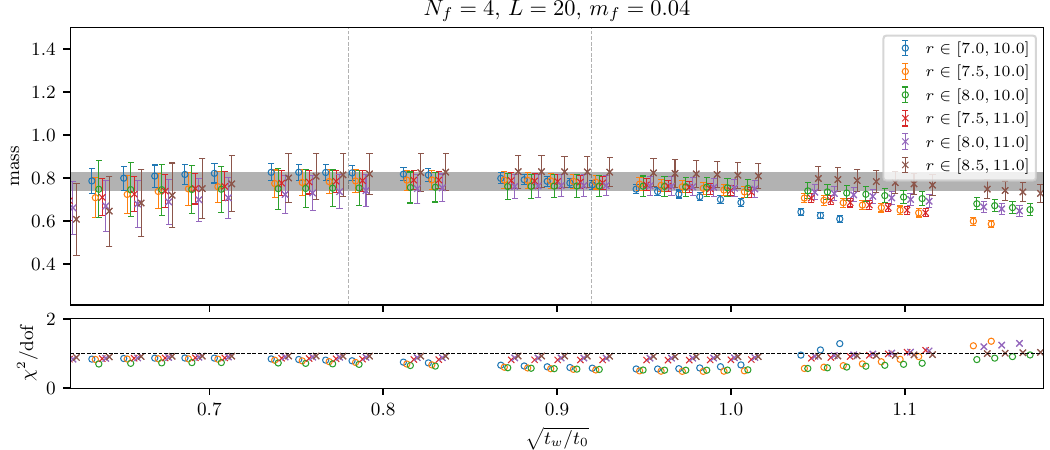}

  \caption{The $\eta^\prime$ mass fitted for different distance regions and smearings, for a specific ensemble of $N_f=4$ QCD with $L=20$ and $m_f=0.04$.}
  \label{fig:nf4_L20_mf0-04_compare_smearing_rmax11-00_with_sys}
\end{figure}

The plateaux for the fits for each ensemble studied in the $N_f=4$
theory are presented in Figs.~\ref{fig:nf4_L20_mf0-01_compare_smearing_rmax10-00_with_sys}, 
\ref{fig:nf4_L20_mf0-02_compare_smearing_rmax10-00_with_sys}, 
\ref{fig:nf4_L20_mf0-02_compare_smearing_rmax11-00_with_sys},
\ref{fig:nf4_L20_mf0-03_compare_smearing_rmax11-00_with_sys}, and 
\ref{fig:nf4_L20_mf0-04_compare_smearing_rmax11-00_with_sys}.

\clearpage
\subsection{$N_f=8$}
\begin{figure}[thb]
  \centering
  \includegraphics{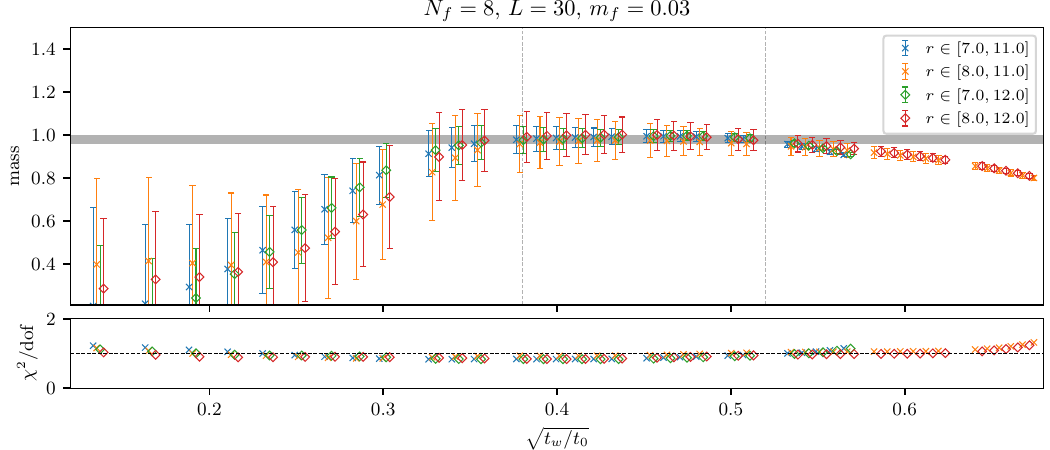}

  \caption{The $\eta^\prime$ mass fitted for different distance regions and smearings, for a specific ensemble of $N_f=8$ QCD with $L=30$ and $m_f=0.03$.}
  \label{fig:nf8_L30_mf0-03_compare_smearing_rmax10-00_with_sys}
\end{figure}

\begin{figure}[thb]
  \centering
  \includegraphics{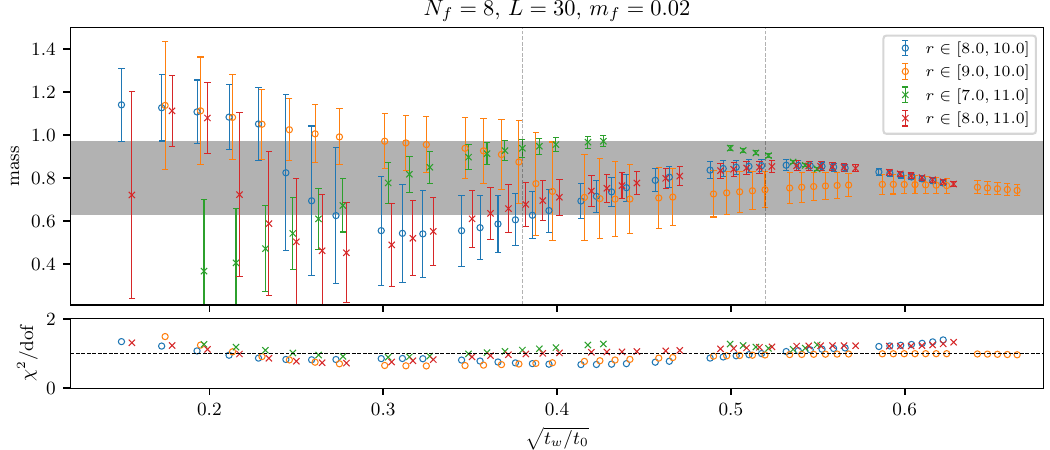}

  \caption{The $\eta^\prime$ mass fitted for different distance regions and smearings, for a specific ensemble of $N_f=8$ QCD with $L=30$ and $m_f=0.02$.}
  \label{fig:nf8_L30_mf0-02_compare_smearing_rmax11-00_with_sys}
\end{figure}

\begin{figure}[thb]
  \centering
  \includegraphics{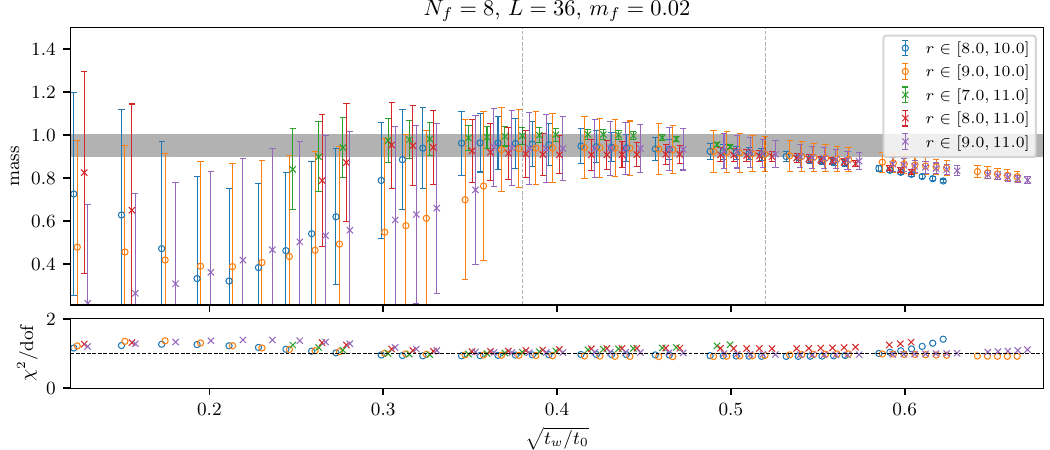}

  \caption{The $\eta^\prime$ mass fitted for different distance regions and smearings, for a specific ensemble of $N_f=8$ QCD with $L=36$ and $m_f=0.02$.}
  \label{fig:nf8_L36_mf0-02_compare_smearing_rmax11-00_with_sys}
\end{figure}

\begin{figure}[thb]
  \centering
  \includegraphics{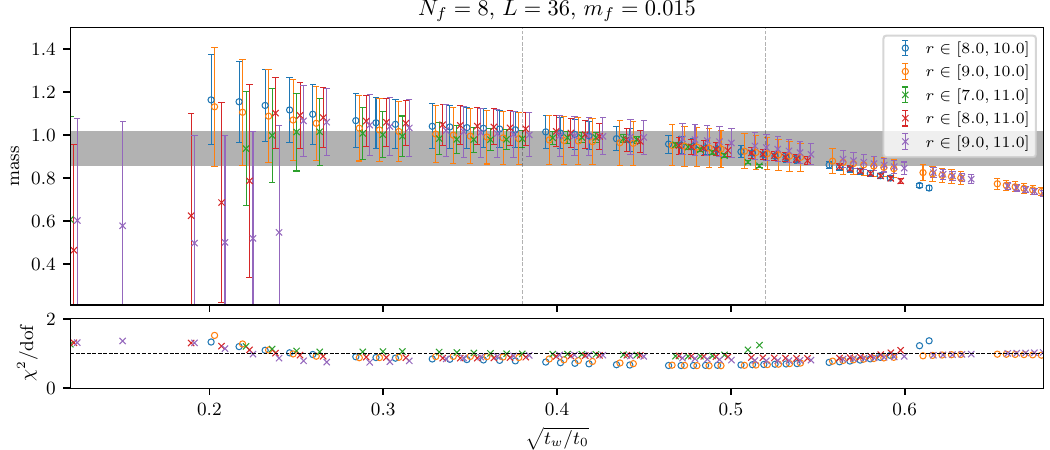}

  \caption{The $\eta^\prime$ mass fitted for different distance regions and smearings, for a specific ensemble of $N_f=8$ QCD with $L=36$ and $m_f=0.015$.}
  \label{fig:nf8_L36_mf0-015_compare_smearing_rmax11-00_with_sys}
\end{figure}

\begin{figure}[thb]
  \centering
  \includegraphics{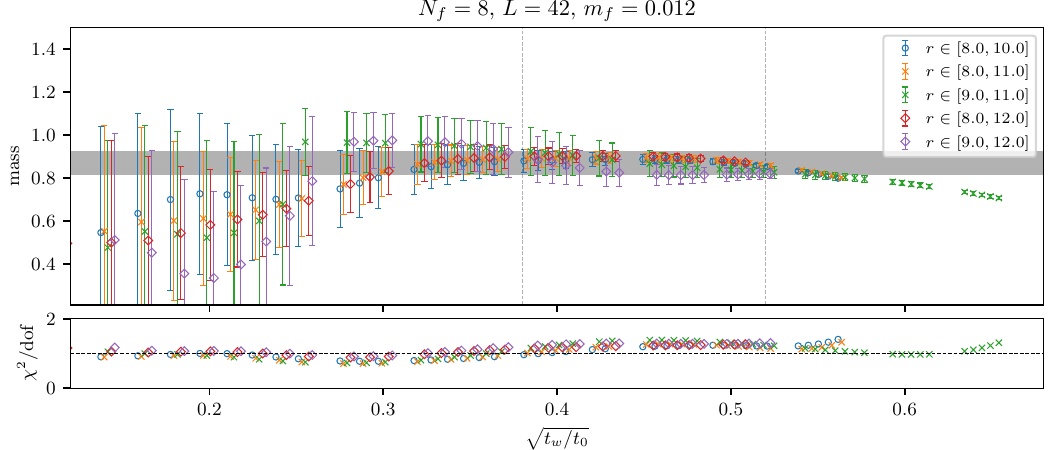}

  \caption{The $\eta^\prime$ mass fitted for different distance regions and smearings, for a specific ensemble of $N_f=8$ QCD with $L=42$ and $m_f=0.012$.}
  \label{fig:nf8_L42_mf0-012_compare_smearing_rmax12-00_with_sys}
\end{figure}

\begin{figure}[thb]
  \centering
  \includegraphics{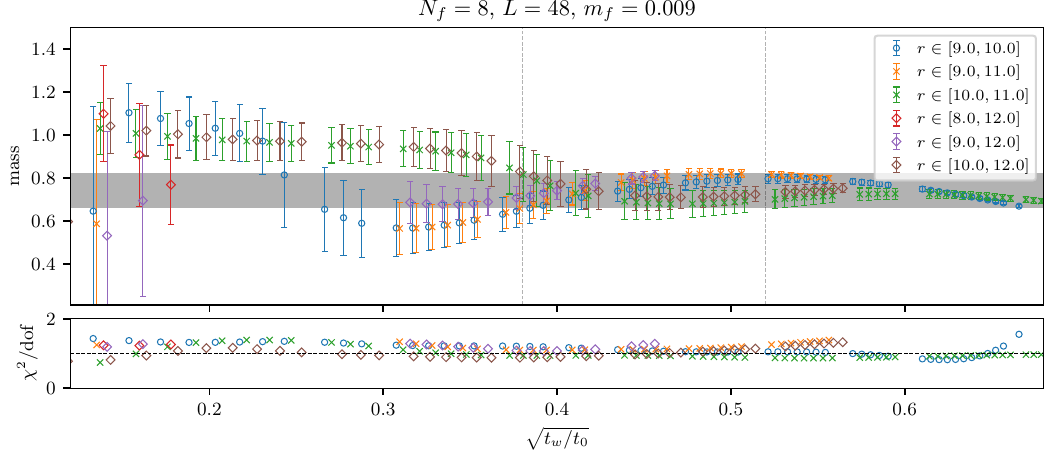}
  \caption{The $\eta^\prime$ mass fitted for different distance regions and smearings, for a specific ensemble of $N_f=8$ QCD with $L=48$ and $m_f=0.009$.}
  \label{fig:nf8_L48_mf0-009_compare_smearing_rmax12-00_no_sys}
\end{figure}

The plateaux for the fits for each ensemble studied in the $N_f=8$
theory are presented in Figs.~\ref{fig:nf8_L30_mf0-03_compare_smearing_rmax10-00_with_sys}, 
\ref{fig:nf8_L30_mf0-02_compare_smearing_rmax11-00_with_sys}, 
\ref{fig:nf8_L36_mf0-02_compare_smearing_rmax11-00_with_sys},
\ref{fig:nf8_L36_mf0-015_compare_smearing_rmax11-00_with_sys},
\ref{fig:nf8_L42_mf0-012_compare_smearing_rmax12-00_with_sys}, and 
\ref{fig:nf8_L48_mf0-009_compare_smearing_rmax12-00_no_sys}.

\clearpage
\subsection{$N_f=12$}
\begin{figure}[thb]
  \centering
  \includegraphics{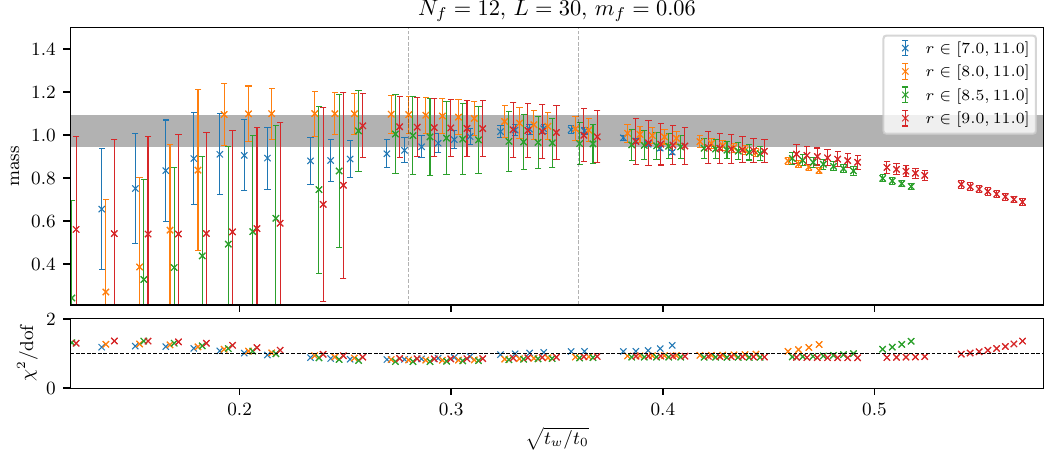}

  \caption{The $\eta^\prime$ mass fitted for different distance regions and smearings, for a specific ensemble of $N_f=12$ QCD with $L=30$ and $m_f=0.06$.}
  \label{fig:nf12_L30_mf0-06_compare_smearing_bw1_rmin7p0-rmax11p0_with_sys}
\end{figure}

\begin{figure}[thb]
  \centering
  \includegraphics{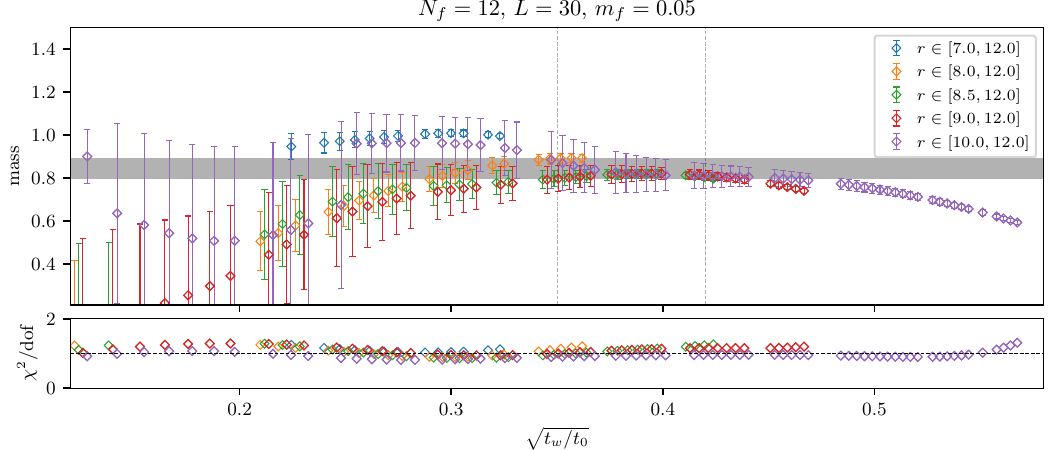}

  \caption{The $\eta^\prime$ mass fitted for different distance regions and smearings, for a specific ensemble of $N_f=12$ QCD with $L=30$ and $m_f=0.05$.}
  \label{fig:nf12_L30_mf0-05_compare_smearing_bw1_rmin7p0-rmax11p0_with_sys}
\end{figure}

\begin{figure}[thb]
  \centering
  \includegraphics{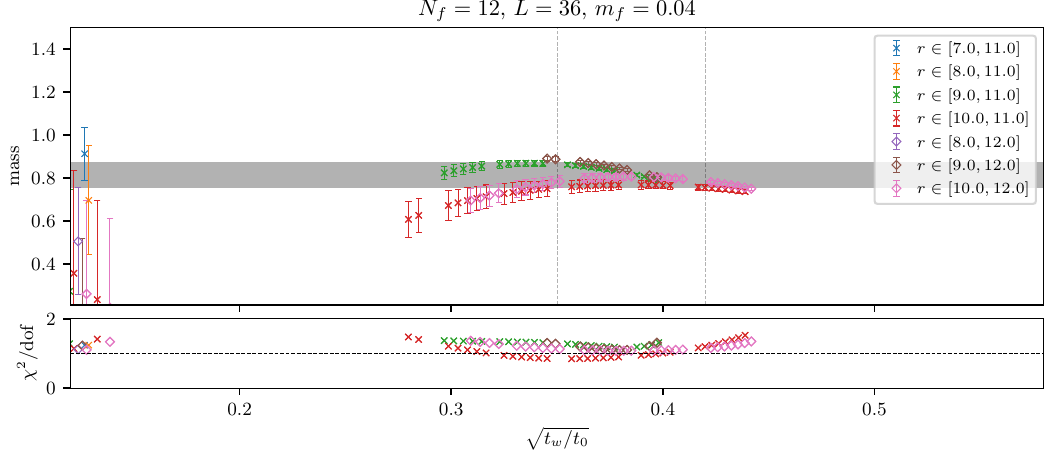}

  \caption{The $\eta^\prime$ mass fitted for different distance regions and smearings, for a specific ensemble of $N_f=12$ QCD with $L=36$ and $m_f=0.04$.}
  \label{fig:nf12_L36_mf0-04_compare_smearing_bw1_rmin7p0-rmax12p0_with_sys}
\end{figure}

The plateaux for the fits for each ensemble studied in the $N_f=12$
theory are presented in Figs.~\ref{fig:nf12_L30_mf0-06_compare_smearing_bw1_rmin7p0-rmax11p0_with_sys}, 
\ref{fig:nf12_L30_mf0-05_compare_smearing_bw1_rmin7p0-rmax11p0_with_sys}, and 
\ref{fig:nf12_L36_mf0-04_compare_smearing_bw1_rmin7p0-rmax12p0_with_sys}.

\end{document}